\documentclass[
 reprint,
 superscriptaddress,
 amsmath,amssymb,
 aps,
]{revtex4-2}

\usepackage{dcolumn}%
\usepackage{bm}%
 \usepackage{float}
 \usepackage{comment}
\usepackage{graphicx}

\usepackage{algorithm}
\usepackage{algpseudocode}
\usepackage{float}

\usepackage{xcolor}

\begin{document}

\preprint{APS/123-QED}

\title{Topological Origin of the Diversity of Timescales in Recurrent Neural Circuits}%

\author{Marco Zenari}
\email{marco.zenari.2@phd.unipd.it}
 \affiliation{Padova Neuroscience Center, University of Padova, Italy.}%
\author{Luca Taffarello}%
 \affiliation{Padova Neuroscience Center, University of Padova, Italy.}

\author{Luca Mazzucato}
\affiliation{Institute of Neuroscience, Departments of Biology, Mathematics \& Physics, University of Oregon, Eugene, Oregon 97403, United States
}%

\author{Amos Maritan}
\affiliation{Department of Physics and Astronomy ``Galileo Galilei'', University of Padova, Italy and INFN, Padova Division, Italy}

\author{Samir Suweis}
\affiliation{Department of Physics and Astronomy ``Galileo Galilei'', University of Padova, Italy and INFN, Padova Division, Italy}
 \affiliation{Padova Neuroscience Center, University of Padova, Italy.}

\date{\today}%

\begin{abstract}
Structural and functional heterogeneities are hallmarks of cortical circuits; for example, broad degree distributions characterize neuronal connectivity, while intrinsic timescales vary across neurons within the same circuit.
However, a mechanistic link explaining how  functional diversity emerges from heterogeneity in connectivity is lacking. To bridge this gap, we introduce a random recurrent network in which connectivity is generated by a configuration model with tunable degree heterogeneity, and synaptic weights exhibiting varying levels of correlation. Using generating-functional methods, we derive a heterogeneous dynamical mean-field theory (hDMFT) that yields degree-conditioned effective stochastic dynamics.
The theory shows that partial symmetry in the weights induces a non-Markovian memory term in the form of an emergent self-coupling, whose degree-dependent strength, arising from degree heterogeneity, produces a broad distribution of activity timescales.
We obtain analytic stability criteria demonstrating that degree heterogeneity lowers the critical gain and localizes unstable modes onto hubs. The resulting rich dynamical landscape includes silent, chaotic, and multistable regimes, which we uncover via spectral, replica, and Lyapunov exponent analyses. We highlight the computational benefits of the observed timescale heterogeneity by revealing that, under an external input drive featuring a broadband spectrum, hub neurons with slow timescales act as effective integrators, demixing the slow input components. Finally, instantiating the model with the empirically measured topology from the MICrONS cubic-millimeter mouse connectome explains the broad range of single-neuron timescales and their positive correlation with in-degree observed in resting-state recordings. Our results provide a mechanistic link between connectome topology, neural dynamics, and computation, identifying hubs in partially symmetric networks as a natural substrate for multiplexed processing across timescales.
\end{abstract}

\maketitle

\section{INTRODUCTION}
Structural and functional heterogeneities are a defining feature of nervous systems, spanning levels of organization from ion channels and dendrites to synapses, cell types, microcircuits, and large-scale networks. Across these scales, variability is increasingly understood not as incidental biological “noise,” but as a generative resource that shapes stability, sensitivity, and computational repertoire. In this sense, the functional organization of neural systems is not reducible to uniform components: collective behavior emerges from structured diversity.

At the functional level, heterogeneity restructures the spectrum of network dynamics by linearizing otherwise strongly nonlinear regimes, broadening the dynamic range, and enabling selective entrainment across frequencies \cite{kusmierz2020edge,kusmierz2025hierarchy,landau2016impact,marti2018correlations,dahmen2025heterogeneity}. These reconfigurations thereby alter which computations are possible and how efficiently they can be implemented.

Within this view, diversity of intrinsic timescales plays a central functional role. At the ethological level, the wide range of timescales and non-Markovian structure observed in naturalistic behavior \cite{berman2016predictability} may require a wide variety of neuronal timescales as its substrate \cite{mazzucato2022neural}. In recurrent neural networks (RNNs), distributed timescales improve temporal credit assignment and support multiplexed computations: long constants of integration stabilize memory and context, while short timescales preserve responsiveness and precision \cite{bernacchia2011reservoir,iigaya2019deviation}. Such heterogeneity enhances reservoir richness and separability, mitigates interference, and can expand the operating region between quiescence and runaway activity, complementing related advantages attributable to heavy-tailed synaptic statistics, structured connectivity, and balance constraints \cite{kusmierz2020edge, landau2016impact, marti2018correlations,dahmen2025heterogeneity}.

Several mechanisms have been proposed to account for the emergence of broad timescale distributions in RNNs, yet a precise mechanistic link to the underlying structural heterogeneity is lacking. Timescale heterogeneity has been reported in network models operating at the edge of instability \cite{chen2024searching}, or when synaptic couplings exhibit heavy-tailed distributions \cite{shi2025brain}.
However, in both cases, when the couplings are independent, the resulting timescale heterogeneity diminishes with increasing network size and does not persist in the thermodynamic limit, as we show numerically in Appendix~\ref{app:timescales_scaling} and argue heuristically in Appendix~\ref{app:iid}. Heavy-tailed couplings combined with partial symmetry, instead, sustain it over the range of $N$ tested (Appendix~\ref{app:timescales_scaling}).
Timescale diversity can also arise at the level of collective modes, depending on which directions in neuron space are activated by the input, even in otherwise homogeneous networks with local connectivity motifs~\cite{dahmen2019second, shao2025impact}; here we focus instead on the intrinsic timescales of single neurons in spontaneous activity.

Alternatively, timescale heterogeneity emerges in recurrent circuits whose synaptic weights exhibit a distribution of strong self-couplings~\cite{stern2023reservoir}. While this provides a  generative model for heterogeneity that is not fine tuned and survives the scaling limit, it leaves open a key question: How do heterogeneous self-couplings emerge from the biological ingredients of neural circuits? To achieve this goal, one needs to identify two ingredients: first, which features of connectivity may induce effective self-couplings in the dynamics and, second, how to generate heterogeneity in the effective self-coupling distribution.

The first ingredient is provided by the overrepresentation of reciprocal connections  (partially symmetric weights \cite{clark2023dimension, clark2025symmetries}, potentially induced by Hebbian plasticity \cite{clark2024theory}), which gives rise to effective self‐coupling terms in the dynamics. However, the resulting distribution of self-couplings is homogeneous, leaving open the second question about how heterogeneity may arise.

Here, we demonstrate that heterogeneous distributions of effective self-couplings arise from heterogeneity in connectivity degree, in the presence of partial symmetry. Empirical analyses point to broad, heavy‐tailed (often approximately lognormal) degree and strength distributions across species and brain areas, implying that neurons have widely heterogeneous number of synaptic connections \cite{piazza2025physical}. Recent studies in theoretical ecology \cite{park2024incorporating, poley2025interaction, aguirre2024heterogeneous} have highlighted that heterogeneity in connectivity can reshape the dynamical regime of a network, generating multiscale fluctuations and long memory traces \cite{metz2025sparsedmft, azaele2024generalized}.

We show that partial symmetry generates an effective self-coupling whose strength is proportional to the degree of each neuron, so that degree heterogeneity turns it into a distribution of self-couplings across neurons. A heterogeneous connectivity distribution thereby induces a diversity of neuronal timescales, with ``hub'' neurons of larger degree exhibiting slower fluctuations. Our approach unifies the ingredients of partial symmetry \cite{clark2025symmetries, marti2018correlations, sherf2025complexity} and  degree heterogeneity \cite{piazza2025physical} to provide a biologically plausible and mechanistic explanation for the observed diversity of neuronal timescales.

We test our theoretical predictions of the link between degree, symmetry, and timescales using empirical data from the MICrONS dataset, which provides the structural connectome and physiological responses of neurons in a cubic millimeter of the mouse visual cortex.
When endowed with the synaptic couplings with statistics matching the MICrONS connectome, our model generated time-varying neural activity whose diversity of timescale is consistent with the empirical one. Our results show that the coexistence of degree heterogeneity and partial symmetry within the same circuit is sufficient to explain the experimentally observed diversity of temporal dynamics in cortical circuits.

\section{MODEL}
\subsection{Topology of neural circuit connectivity in the mouse visual cortex}

We consider a recurrent neural network (RNN) described by pre-activation variables \(x_i\) for \(i = 1, \ldots, N\), whose dynamics are governed by
\begin{equation}
    \frac{dx_i}{dt} = -x_i + \sum_{j} W_{ij}\,\phi(x_j) + I_i(t),
    \label{eq:RNN}
\end{equation}
where \(W_{ij}\) denotes the synaptic connectivity from pre-synaptic neuron \(j\) to post-synaptic neuron \(i\), and \(\phi(x)\) is a nonlinear transfer function. The external input \(I_i(t)\) is initially set to zero to study the network in the spontaneous activity regime, while the effects of a broadband input are discussed in Sect.~\ref{sect:broadband}. The standard approach to studying neural network dynamics~\cite{sompolinsky1988chaos} assumes that the elements of $W$ are drawn independently and identically from a Gaussian distribution, thereby introducing {\it quenched disorder} in the couplings. In RNNs with this class of i.i.d. connectivity, all neurons exhibit the same timescale in the large-$N$ limit, which is inconsistent with experimental observations~\cite{stern2023reservoir}.

\begin{figure*}[ht]
\centering
\includegraphics[width=1\linewidth]{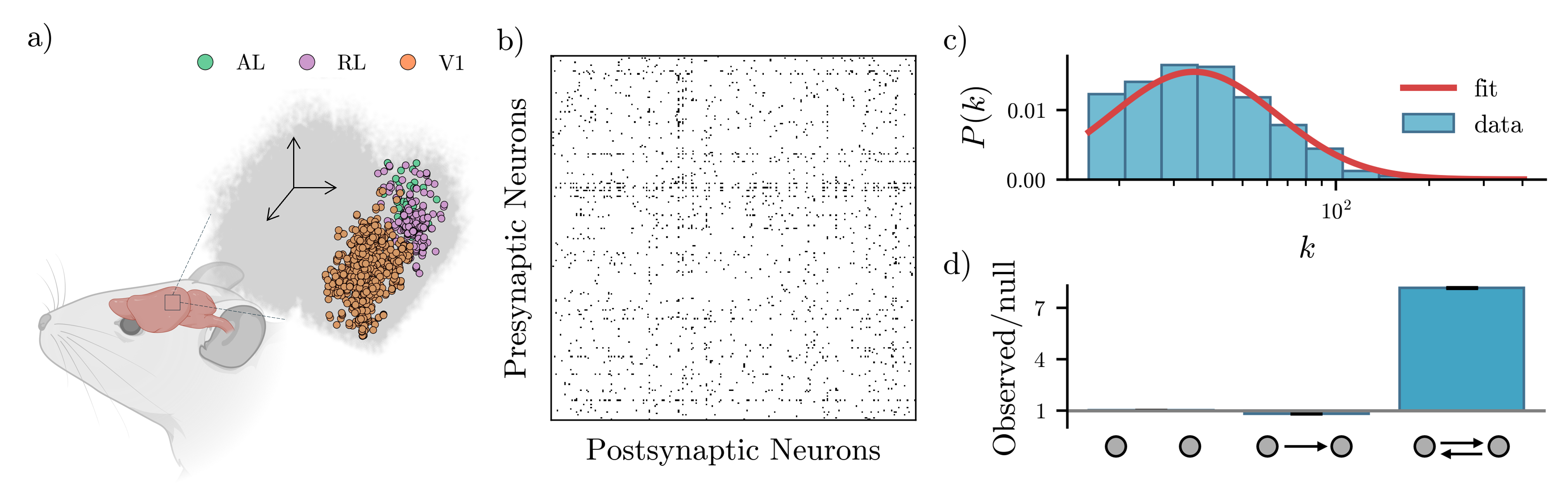}
\caption{\textbf{Nontrivial topology of neural circuits in mouse visual cortex \cite{microns2025functional}.} (a) Spatial distribution of neurons within the MICrONS Cubic Millimeter volume. Colored markers indicate the subset of proofread neurons analyzed in this study, highlighted according to visual cortical area: primary visual cortex (V1), anterolateral visual area (AL), and rostrolateral visual area (RL). Gray points show the surrounding neuronal population within the same volume. (b) Corresponding binary adjacency matrix of synaptic connectivity among the $N=1965$ proofread neurons shown in (a). (c) In-degree distribution $p(k)$ of the neurons together with a lognormal fit (red curve; shape parameter $\sigma=0.69$, scale parameter $\mu=3.83$), indicating a heavy-tailed connectivity profile. (d) Pairwise connectivity between neurons compared to an Erd\H{o}s--R\'enyi null model with connection probability matching empirical observation ($p=2.67\%$). Bars report the observed counts of non-connected, unidirectional, and reciprocal pairwise connections (95.25\%, 4.17\%, and 0.58\% of all neuron pairs, respectively) normalized by their expectation under the null model (horizontal gray line at 1). Reciprocal connections are strongly overrepresented, while unidirectional connections are underrepresented relative to the null model (Monte Carlo test with $10000$ random networks, $p<10^{-4}$). Quantifying reciprocity by the Pearson correlation between opposite directed edges gives $\gamma_s = 0.197$ (see Appendix \ref{app:microns}).}
\label{fig:microns}
\end{figure*}

However, biological neural circuits exhibit complex connectivity patterns that substantially deviate from this i.i.d. Gaussian assumption. To gain insight into the topological structure of real-world neural networks, we analyzed the MICrONS dataset~\cite{microns2025functional}, which provides the structural connectome of a cubic millimeter of the mouse visual cortex (Fig.~\ref{fig:microns}a-b), integrated with large-scale physiological recordings from the same cortical patch. This combined structural and functional dataset allows, for the first time, the determination of how the dynamical features of neurons arise from the underlying network topology.

Our analysis of the circuit's structural connectivity (Fig.~\ref{fig:microns}b) revealed two highly nontrivial topological features (Fig.~\ref{fig:microns}c-d): a heterogeneous degree distribution exhibiting a lognormal organization of in- and out-degrees across cortical neurons~\cite{piazza2025physical}; and a strong correlation between reciprocal pairs~\cite{song2005highly}, which controls the degree of partial symmetry in the couplings. Restricting the analysis to the $N=1965$ proofread neurons, the in-degree distribution is well described by a lognormal with $\mu=3.83$ and $\sigma=0.69$ (Fig.~\ref{fig:microns}c), and reciprocal pairs are overrepresented with respect to an Erd\H{o}s--R\'enyi network of the same density, with a structural reciprocity $\gamma_s=0.197$ (Fig.~\ref{fig:microns}d). The details of the analysis are given in Sec.~\ref{sec:microns} and Appendix~\ref{app:microns}.

\subsection{Recurrent neural networks with nontrivial topology} \label{sec:model}
To account for a nontrivial network topology, the couplings $W_{ij}$ between pre- (j) and post- (i) synaptic neurons are generated by two independent random processes, which separately determine the network structure and the coupling strengths. Specifically, we construct the synaptic matrix as $W_{ij} = A_{ij} J_{ij}$, where $A_{ij}$ specifies the connectivity structure and $J_{ij}$ determines the synaptic weight. The structural component $A_{ij}$ is generated from a Bernoulli process, $A_{ij} \sim \mathrm{Be}(p_{ij})$. Throughout this work, we assume an undirected connectivity architecture generated using the configuration model. A degree sequence $\mathbf{k} = \{k_1,\ldots,k_N\}$ is drawn from a target distribution $P(k)$, and edges are instantiated independently with probability $p_{ij} = k_i k_j / (N K)$, where $K$ denotes the mean degree.
Self-connections are excluded, $A_{ii}=0$, so that the network contains no autapses. The undirected construction is a modelling choice made for tractability: it assigns a single
structural label, the degree $k_i$, to each neuron, so that the mean-field theory of
Sec.~\ref{sec:hDMFT} reduces to a one-parameter family of effective single-neuron processes
indexed by $k$.

In our model, the choice of the distribution $P(k)$ reflects the structure of the couplings and extends the fully connected case, which has been the focus of most studies in the literature~\cite{sompolinsky1988chaos, crisanti2018path, helias2020statistical}. This framework provides a natural way to investigate how the architecture of the connectome shapes the function of neural circuits.  In the MICrONS dataset, we found that $P(k)$ closely follows a lognormal distribution, a pattern consistently observed across species and consistent with previous reports~\cite{piazza2025physical}. Motivated by these findings, in this paper we study the effects of structural heterogeneity by adopting a lognormal distribution as a proxy for biologically plausible connectomes, and compare the results with those obtained using a Poisson distribution as a null model for homogeneous networks.

Given the adjacency matrix, the synaptic strengths $J_{ij}$ are drawn from a Gaussian ensemble with zero mean, $\mathrm{mean}(J_{ij}) = 0$, and standard deviation $\mathrm{std}(J_{ij}) = g / \sqrt{K}$, and with a prescribed correlation between reciprocal pairs, $\mathrm{corr}(J_{ij}, J_{ji}) = \gamma$, which controls the degree of partial symmetry in the couplings~\cite{marti2018correlations}. This construction decouples structural heterogeneity from synaptic-level symmetry, allowing us to systematically investigate their combined impact on recurrent dynamics. The parameter \( g \) represents the standard gain of the network and is scaled as \(1/\sqrt{K}\), ensuring that the typical interaction strength remains of order one, \(\mathcal{O}(1)\), in the large-network limit.

The statistics of the effective couplings $W_{ij}$ emerge from the combination of the adjacency matrix $A_{ij}$ and the synaptic strengths $J_{ij}$, yielding mean, variance, and correlations
$$
\mathbb{E}[W_{ij}] = 0, \quad \mathrm{Var}[W_{ij}] = p_{ij}\,g^2/K, \quad \mathrm{Corr}[W_{ij}, W_{ji}] = \gamma \ .
$$
Thus, the effective statistics of $W_{ij}$ are fully determined by the network topology $p_{ij}$ and the Gaussian strength parameters $g$ and $\gamma$.
In cortical circuits reciprocity has a structural component, the over-representation of bidirectionally connected pairs, and a synaptic one, the correlation between the strengths of reciprocal synapses. The former can be interpreted as an effect of developmental processes~\cite{markram1997physiology, song2005highly, wang2006heterogeneity}, the latter as the outcome of Hebbian learning~\cite{clark2023dimension, clark2024connectivity, clark2024theory, clark2025symmetries}. If the two are independent, in sparse networks they combine multiplicatively into an effective symmetry $\gamma_{\rm eff}\simeq\gamma_{\rm s}\gamma_{\rm w}$, where $\gamma_{\rm s}$ and $\gamma_{\rm w}$ are the correlations of reciprocal edges and of reciprocal weights; neither is sufficient on its own. Our undirected model has $\gamma_{\rm s}=1$, and $\gamma$ should be read as this effective symmetry. In Appendix~\ref{app:sim_directed} we show that directed networks with different $(\gamma_{\rm s},\gamma_{\rm w})$ at equal $\gamma_{\rm eff}$ yield the same timescale statistics.
We will see below that partial symmetry influences the dynamics in a manner similar to the introduction of self-couplings~\cite{stern2014dynamics}.

In what follows, we analyze the dynamics of our model using the generating functional approach for random neural networks ~\cite{sompolinsky1988chaos}. We uncover a rich phase diagram emerging by varying the network parameters $g,\gamma$ and the connectivity degree. Crucially, our model predicts the emergence of a heterogeneous distribution of timescales in the chaotic regime, where higher degree neurons exhibit slower timescales. We then demonstrate the functional benefits of heterogeneous timescales in demixing broadband input signals. Finally, we test the relationship between connectivity topology and timescales using the MICrONS dataset ~\cite{microns2025functional}.

\section{HETEROGENEOUS DYNAMICAL MEAN FIELD THEORY} \label{sec:hDMFT}

We derive a heterogeneous dynamical mean-field theory (hDMFT), in the spirit of the heterogeneous mean-field theory of dynamical processes on networks~\cite{pastor2001epidemic}, for Eq.~(\ref{eq:RNN}) that reduces structured recurrent networks to an effective single-unit stochastic dynamics with topology-dependent statistics and a symmetry-controlled memory kernel. We obtain the hDMFT by generalizing the  Martin-Siggia-Rose-De Dominicis-Janssen (MSRJD) generating functional approach~\cite{sompolinsky1988chaos, crisanti2018path, helias2020statistical,de1978dynamics, martin1973statistical, janssen1976lagrangean} to structured interactions, as previously done for networks with multiple populations or structured connectivity~\cite{kadmon2015transition, aljadeff2015transition, mastrogiuseppe2018linking, stern2023reservoir}, leveraging tools from theoretical ecology~\cite{park2024incorporating, poley2025interaction, aguirre2024heterogeneous}. The details of the derivation can be found in appendix \ref{app:DMFT}.
Briefly, we define the generating functional of the dynamics:
\begin{widetext}
\begin{equation}
    Z[\mathbf{\psi}] = \int \mathcal{D}\mathbf{x} \, \mathcal{D}\mathbf{\hat{x}} \,
    \exp \Bigg[ i \sum_i \int dt \, \hat{x}_i(t) \Big( \dot{x}_i(t) + x_i(t) - \sum_j A_{ij} J_{ij} \phi(x_j(t)) \Big)
    + i \sum_i \int dt \, x_i(t) \psi_i(t) \Bigg] \prod_{i=1}^NP(x_i(0)),
    \label{zofpsi}
\end{equation}
\end{widetext}
where we have introduced the source field $\mathbf{\psi}$. We then average the functional
$\langle\langle Z[\mathbf{\psi}]\rangle_A\rangle_J$ over the quenched disorder of \(A_{ij}\) and \(J_{ij}\).
In the large-$N$ limit, the effective functional can be evaluated at its saddle point, yielding the generating functional of the effective dynamics for a neuron with degree $k$, described by the stochastic differential equation
\begin{equation}
    \label{eq:DMFT}
    (\partial_t + 1) x_k(t) = g \sqrt{\frac{k}{K}} \, \eta(t) + \gamma g^2 \frac{k}{K} \int_0^t \mathrm{d}t' \, G(t, t') \, \phi(x_k(t')),
\end{equation}
for all $k$ in the support of the degree distribution $P(k)$.
The zero-mean Gaussian noise $\eta(t)$ must be determined self-consistently, with
\begin{equation}\label{eq:corr}
    \langle \eta(t) \eta(t') \rangle = \sum_{k} P(k) \frac{k}{K} C_k(t, t'),
\end{equation}
where $C_k(t, t') = \langle \phi(x_k(t)) \phi(x_k(t')) \rangle$,
together with the response function
\begin{equation}\label{eq:response}
    G(t, t') = \sum_{k} P(k) \frac{k}{K}  G_k(t, t'),
\end{equation}
where $G_k(t,t') = \left\langle \frac{\delta \phi(x_k(t))}{\delta h(t')} \right\rangle\big|_{h=0}$, and $h(t)$ is an infinitesimal external input added to the right-hand side of Eq.~(\ref{eq:DMFT}), so that $G_k$ is the linear response of a unit with degree $k$ to a perturbation of its input at time $t'$.
Here, the averages $\langle \cdots \rangle$ are computed over both the ensemble of disordered couplings $J$ and the ensemble of adjacency matrices $A$.

Note that in this derivation, all degree-degree correlations between nodes are neglected by construction of the ensemble. The effective dynamics is obtained at leading order in $1/K$, which treats the input to each neuron as Gaussian, with corrections of order $1/k$ that are largest for low-degree neurons (Appendix~\ref{app:DMFT}).

The hDMFT, Eq.~(\ref{eq:DMFT}), describes the effective dynamics of a neuron with degree $k$. The influence of network interactions is captured by the Gaussian noise term, which is weighted by the square root of the number of presynaptic neurons, $k$.

The last term on the right-hand side in (\ref{eq:DMFT}) represents a non-Markovian memory effect arising from correlations between the couplings, producing recurrent contributions expressed as a memory kernel. The impact of this term on the dynamics is discussed in detail in what follows. Intuitively, this memory kernel represents the reverberation of a neuron's own activity returning through the network. This reverberation becomes stronger for neurons with larger degree $k$, and it only arises when $\gamma > 0$.
Partial symmetry thus generates the self-coupling, while the degree sets its strength. In a homogeneous network, $\gamma$ slows down all neurons equally. With a heterogeneous degree distribution, high-degree nodes (hubs) experience a larger effective self-coupling, proportional to their degree, while low-degree nodes remain fast, so that the same mechanism produces a distribution of self-couplings across neurons.
This degree-dependent slowing is the key physical mechanism underlying the heterogeneous timescales in the network.

\section{PHASE STRUCTURE AND DYNAMICAL REGIMES}
We map the dynamical regimes induced by degree heterogeneity and partial symmetry by determining when the silent fixed point destabilizes and how the spectrum reorganizes across parameters. We first obtain the stability boundary of the silent phase from a linear analysis of the hDMFT~(\ref{eq:DMFT}). We then quantify hub-driven destabilization effects by analyzing the Jacobian of Eq.~(\ref{eq:RNN}) in the asymmetric case ($\gamma=0$). Finally, we characterize the resulting disordered phase using Largest Lyapunov Exponents and, where relevant, replica-based diagnostics.

The stationary fixed-point solution $x_k^*$ of the dynamical Eqs.~(\ref{eq:DMFT}) reads
\begin{equation}
    x_k^* - \gamma g^2 \frac{k}{K} \, \chi \, \phi(x_k^*) = g \sqrt{\frac{k}{K}} \, \eta^*,
    \label{eq:stat_DMFT}
\end{equation}
where $\eta^*$ is a Gaussian random variable with variance $q^2 = \sum_k P(k) (k/K) \, \langle \phi(x_k^*)^2 \rangle$, and $\chi = \int d\tau \, G(\tau)$, with $G(\tau) = G(|t-t'|)$ evaluated at stationarity.

Eq.~(\ref{eq:stat_DMFT}) is formally equivalent to the stationary equations of recurrent neural networks with an effective self-coupling $s_k = \gamma g^2 \frac{k}{K} \chi$ ~\cite{stern2014dynamics, stern2023reservoir}.
To investigate the stability of the fixed point, we perform a linear stability analysis of the DMFT Eq.~(\ref{eq:DMFT}) linearized around it. Within this framework, the fixed-point ansatz provides a self-consistent prediction of its own breakdown. Details of the computation are provided in appendix \ref{app:LSA}.

The critical condition reads
\begin{equation}
    1 = g^2 \sum_k P(k) \frac{k^2}{K^2}
    \left\langle \big(\phi'(x_k^*)\big)^2
    \left( \frac{1}{1 - \gamma g^2 \chi \frac{k}{K} \phi'(x_k^*)} \right)^2 \right\rangle,
\label{eq:stabilitygamma}\end{equation}
where the parameter $\chi$ must be computed self-consistently, and the average $\langle \cdots \rangle$ is taken over realizations of the stochastic process defined by Eq.~(\ref{eq:stat_DMFT}).

This average is generally difficult to compute analytically because the nonlinear transfer function $\phi(x)$ precludes an exact determination of the stationary distribution $P(x^*)$. For the specific choice $\phi(x) = \tanh(x)$ and zero-mean coupling strengths $J_{ij}$, the fixed-point regime coincides with the classical silent phase. In this regime, there exists a unique fixed point $x_k^* = 0$ for all degrees $k$, which is stable for sufficiently small gain $g$ and for any value of $\gamma$. Consequently, the stationary distribution reduces to a Dirac delta centered at zero, $P(x^*) = \delta(x^*)$, since the self-consistent parameter $q^2$ vanishes. The critical condition for the onset of instability therefore simplifies to
\begin{equation} \label{eq:crit_cond_silent}
    1 = g^2 \sum_k P(k) \frac{k^2}{K^2}
    \left( \frac{1}{1 - \gamma g^2 \chi \frac{k}{K}} \right)^2.
\end{equation}

\subsection{High-degree neurons destabilize the silent fixed point phase}
We start by considering the asymmetric case, $\gamma = 0$, to isolate the impact of structural heterogeneity on the network dynamics. In this limit, the critical condition~(\ref{eq:crit_cond_silent}) reduces to
\begin{equation}
    1 = g^2 \, \frac{\mathbb{E}[k^2]}{K^2},
    \label{eq:crit_cond_gamma0}
\end{equation}
indicating that the stability of the silent phase depends directly on the second moment of the degree distribution, $\mathbb{E}[k^2]$, and hence on the presence of highly connected hubs. For a fixed mean degree $K$, a larger $\mathbb{E}[k^2]$ lowers the critical gain $g_c$, making networks with more heterogeneous degree distributions less stable. Highly connected nodes, whose degrees deviate strongly from the mean, promote the transition of the system away from the silent state.

\begin{figure*}[ht]
    \centering
    \includegraphics[width=1\linewidth]{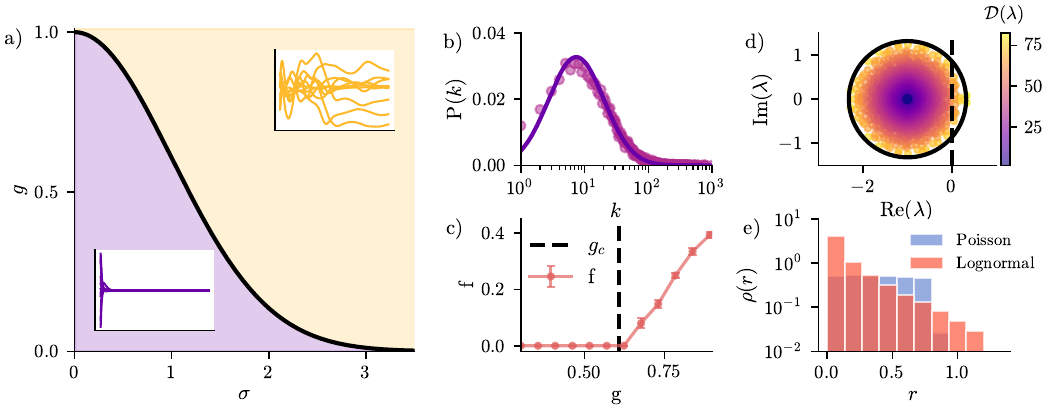}
     \caption{\textbf{
    Effects of heterogeneous degree distribution at $\gamma=0$ (asymmetric case).}
    (a) Phase diagram for $\gamma = 0$ with a lognormal degree distribution. The solid black line corresponds to the theoretical prediction, Eq.~(\ref{eq:crit_cond_LN_gamma0}), for the critical gain, separating the trivial fixed-point phase from the chaotic phase. Insets show examples of dynamics in the two regions.
     (b) Example of a synthetic degree distribution for a network with $N = 2000$ units, drawn from a lognormal distribution with $\mu = 3$ and $\sigma = 1$. The solid line represents the theoretical degree distribution, and the dots are the realized degrees obtained through the configuration model
    sampling procedure (finite-size and binomial sampling of the edges account for the
    deviation).
     (c) Numerical validation of the critical gain prediction for a network with a lognormal degree distribution ($\mu = 3$, $\sigma = 1$) and no cross-correlations ($\gamma = 0$). The parameter $f$ quantifies the temporal volatility of neuronal activities at stationarity and is evaluated over $20$ network realizations with $1000$ neurons each.
    (d) Eigenvalue spectrum of the Jacobian matrix computed at the trivial fixed point for a lognormal degree distribution with $\mu = 3$ and $\sigma = 1$ for a network realization with $N = 2000$ nodes and $g = 0.8$. Scatter points are colored based on the degree score defined in Eq.~(\ref{eq:degree_score}).
      (e) Radial density distribution of the eigenvalue disk for networks with $N = 2000$ units and Poisson ($K = 100$, $g = 0.8$) or lognormal ($\mu = 3$, $\sigma = 1$, $g = 0.8$) degree distributions.
    }
    \label{fig:panel1}
\end{figure*}

In the case of an Erd\H{o}s--R\'enyi network, which can serve as a proxy for a homogeneous network with a Poisson degree distribution, the second moment is $\mathbb{E}[k^2] = K + K^2$. The corresponding prediction for the critical gain is
\begin{equation}
    g_c^{\text{ER}} = \sqrt{\frac{K^2}{K + K^2}},
      \label{eq:gc_ER}
\end{equation}
which gives $g_c \lesssim 1$ for $K \gg 1$, in agreement with~\cite{kadmon2015transition}. This represents only a small correction compared to the critical value for the fully connected case~\cite{sompolinsky1988chaos}.

In the case of a lognormal degree distribution with parameters $\mu$ and $\sigma$, the moments of the distribution are given by $K = \exp(\mu + \sigma^2 / 2)$ and $\mathbb{E}[k^2] = \exp(2\mu + 2\sigma^2)$. These expressions are exact for real-valued random variables drawn from a lognormal distribution. In our network construction, however, degrees are integers obtained by discretizing such samples, so the corresponding moments are approximate. The approximation becomes increasingly accurate as $\mu$ and $\sigma$ grow, i.e., in the regime $K \gg 1$ considered here. In this limit, the ratio of moments entering the critical condition is independent of $\mu$, and the expression for the critical gain simplifies to
\begin{equation}
g_c^{\text{LN}} = e^{-\sigma^2/2}.
\label{eq:crit_cond_LN_gamma0}
\end{equation}

In the lognormal case, the phase diagram shown in Fig.~\ref{fig:panel1}a reveals that network instability grows with increasing heterogeneity of the degree distribution, controlled by $\sigma$. A corresponding example of a sampled degree distribution is presented in Fig.~\ref{fig:panel1}b.

We verify the theoretical prediction through numerical simulations by computing the parameter $f = \big\langle \sqrt{\langle x^2 \rangle - \langle x \rangle^2 } \big\rangle_T$, where $\langle \cdot \rangle$ denotes an average over neurons, and $\langle \cdot \rangle_T$ indicates a temporal average computed in the stationary regime. The parameter $f$ quantifies the overall activity fluctuations in the network: it vanishes in the silent phase and becomes positive in the unstable regime. The comparison between the theoretical prediction and numerical simulations is presented in Fig.~\ref{fig:panel1}c.

The effects of structural heterogeneity on the network dynamics can be further characterized by analyzing the eigenvalue spectrum of the stability matrix associated with the full model~(\ref{eq:RNN}) at the silent fixed point, $x_i^* = 0$ for all $i$:
\begin{equation}
    M_{ij} = -\delta_{ij} + (1 - \delta_{ij}) W_{ij},
    \label{eq:stability_matrix}
\end{equation}
where $\delta_{ij}$ is the Kronecker delta.

Following  the modified circular law for networks with heterogeneous degree distributions~\cite{baron2022rmt}, the eigenvalues of the stability matrix for $\gamma=0$ are confined within a disk of radius

\begin{equation}
    \mathcal{R} = g \frac{\sqrt{\mathbb{E}[k^2]}}{K}.
    \label{eq:radius_eigenspectrum}
\end{equation}

Considering the diagonal term in Eq.~(\ref{eq:stability_matrix}), the requirement that the real parts of all eigenvalues of the Jacobian are negative, i.e., that the spectral radius~(\ref{eq:radius_eigenspectrum}) satisfies $\mathcal{R} < 1$, reproduces the critical condition obtained from the linear stability analysis, Eq.~(\ref{eq:crit_cond_gamma0}).

The comparison between the theoretical prediction and the simulations is shown in Fig.~\ref{fig:panel1}d.

We find that the eigenvalue spectrum is not uniformly distributed within the disk. To quantify this, we analyze the radial spectral density, $\rho(r)$. Unlike homogeneous networks (e.g., fully connected or Erdős–Rényi), where eigenvalues are uniformly spread, degree heterogeneity leads to a radial distribution concentrated near the center and decreasing with the distance $r$, as shown in Fig.~\ref{fig:panel1}e.

To quantify this effect, we introduce the degree score of the eigenvalues $\lambda_j$, defined as
\begin{equation}
    \mathcal{D}(\lambda_j) = \sum_i k_i |\psi_i^{(j)}|^{2} ,
\label{eq:degree_score}
\end{equation}
where $\psi^{(j)}$ is the eigenvector corresponding to $\lambda_j$, and $k_i$ is the degree of node $i$. Each eigenvector $\psi^{(j)}$ is normalized to have unit norm, $\|\psi^{(j)}\| = 1$.
This measure captures the extent to which an eigenmode is localized on high- or low-degree nodes. As shown in Fig.~\ref{fig:panel1}d, $\mathcal{D}(\lambda)$ increases with the radius $r$, indicating that eigenvectors associated with eigenvalues near the spectral edge are increasingly localized on high-degree nodes. This demonstrates that, in heterogeneous networks, hubs play a dominant role in the dynamics by controlling the unstable eigenmodes near the critical point.

\subsection{Effects of partial symmetry}
\begin{figure*}
    \centering
    \includegraphics[width=1\linewidth]{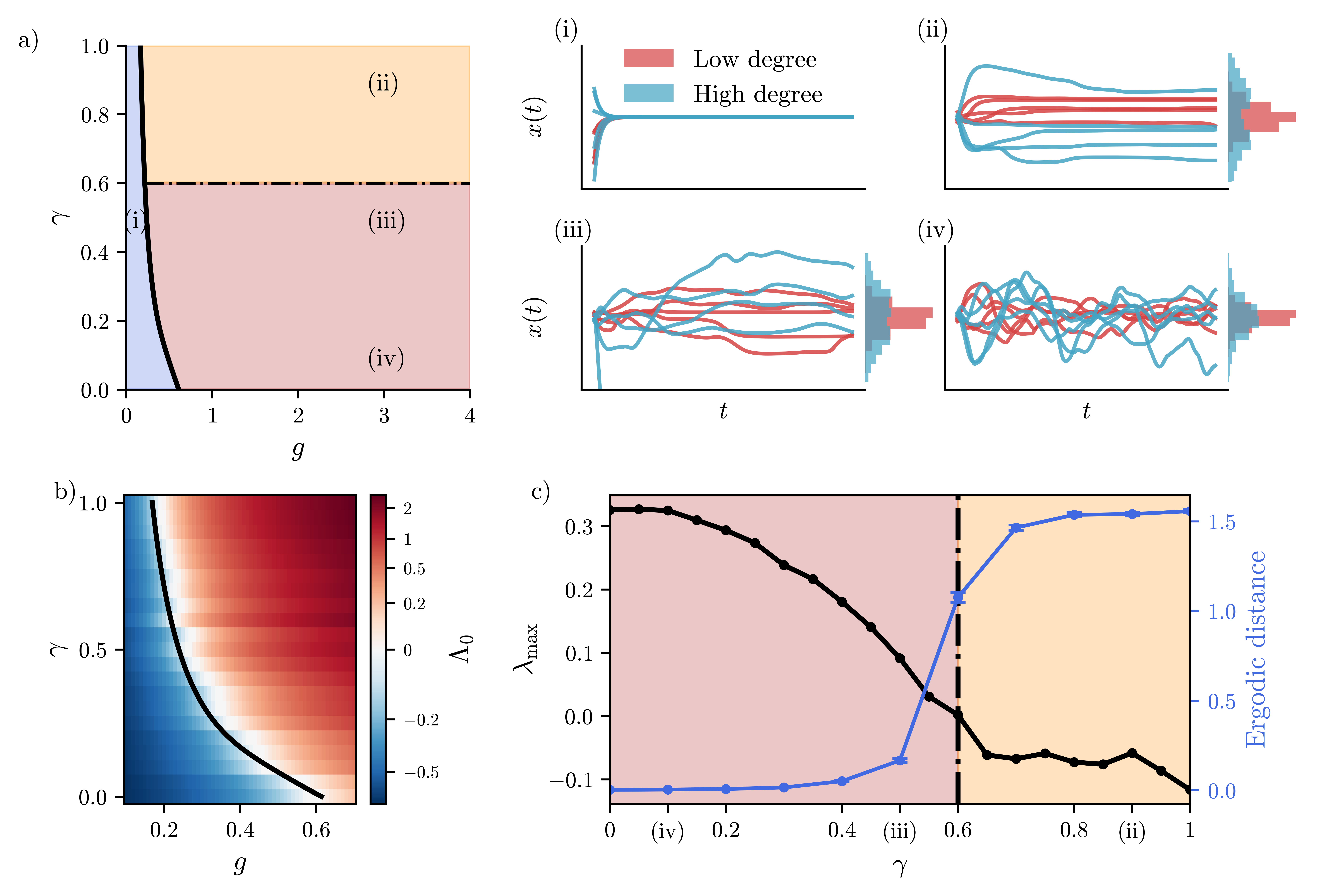}
\caption{\textbf{Phase diagram with heterogeneous and reciprocal connections.}
a) Phase diagram for a network with a lognormal connectivity distribution,
$P(k) = \mathcal{LN}(\mu = 3, \sigma = 1)$, showing phase boundaries.
The black solid line is the theoretical critical line, Eqs.~(\ref{eq:crit_cond_silent}) and (\ref{eq:chi}); the dash-dotted line marks the numerically estimated onset of ergodicity breaking (panel c), drawn as a guide to the eye.
Insets (i)–(iv) illustrate representative dynamical regimes within different regions of the phase diagram. Histograms display the stationary activity distributions of neurons, while trajectories are colored according to the degree percentile (blue for high-degree neurons, red for low-degree neurons). High-degree neurons exhibit larger activity fluctuations and, in some regimes, bistability in their stationary distributions.
b) Numerical validation of the critical condition obtained by solving Eqs.~(\ref{eq:crit_cond_silent}) together with (\ref{eq:chi}), shown as the black solid line.  The heatmap represents the local largest Lyapunov exponent at the trivial fixed point, $\Lambda_0 = \max_j \mathrm{Re}\,\lambda_j(M)\big|_{x^*=0}$, i.e.\ the largest real part of the eigenvalues of the stability matrix, Eq.~(\ref{eq:stability_matrix}), averaged over $10$ network realizations with $N = 4000$ neurons. $\Lambda_0$ changes sign where the silent fixed point loses linear stability. The color scale is logarithmic away from zero, and values with $|\Lambda_0| < 0.03$ are shown in white.
c)  Ergodic distance between replicas (blue, mean $\pm$ s.e.m.\ over $100$ realizations) and global largest Lyapunov exponent $\lambda_{\max}$, computed with the Jacobian method~\cite{engelken2023lyapunov}  (black) of networks with $g = 3$, $\mu = 3$, $\sigma = 1$, and varying $\gamma$. The dashed-dotted line indicates the  numerically estimated onset of ergodicity breaking at $\gamma \simeq 0.6$ (color-coded shades as in panel  a).}
\label{fig:phase_diagram_gamma}
\end{figure*}

Having analyzed the effects of structural connectivity on network dynamics in the asymmetric case ($\gamma=0$), we now explore how these effects interact with those arising from partial symmetry ($\gamma>0$).

Previous work~\cite{marti2018correlations, crisanti1987dynamics} showed that introducing coupling correlations in networks with homogeneous degree distribution has two effects: i) a slow down of network dynamics, signaled by  increasing single neuron autocorrelation times; and ii) a decrease of the critical gain $g$. These findings are consistent with predictions from the elliptic law~\cite{girko1985circular, sommers1988spectrum}, which indicate that the real axis of the elliptic eigenvalue spectrum is stretched by a factor of $(1+\gamma)$, so that in the fully connected case the silent phase becomes unstable at $g_c = (1+\gamma)^{-1}$, i.e.\ the critical gain decreases monotonically with the degree of partial symmetry.

Here we compute the critical line corresponding to the transition from the silent phase to the unstable phase, starting from the critical condition obtained from linear stability and evaluated at the trivial fixed point, given by Eq.~(\ref{eq:crit_cond_silent}). The dynamic susceptibility $\chi = \int d\tau G(\tau)$ is given by the weighted average $\chi = \sum_k P(k) \frac{k}{K}\chi_k$, with
 \begin{equation}
\chi_k = \left\langle \frac{\delta \phi(x_k^*)}{\delta h^*} \right\rangle = \frac{1}{1-\gamma g^2 \chi k/K},
\end{equation}
where $h^*$ is a small constant input added to the right-hand side of the fixed-point equation~(\ref{eq:stat_DMFT}),
yielding
\begin{equation}
\chi = \sum_k P(k) \frac{k}{K} \left(1-\gamma g^2 \chi\frac{k}{K}\right)^{-1}.
\label{eq:chi}
\end{equation}

Together, Eqs.~(\ref{eq:crit_cond_silent}) and (\ref{eq:chi}) yield the critical condition in the plane $(\gamma, \chi)$ in implicit form, which can be solved iteratively.
We validate this prediction numerically by computing the local largest Lyapunov exponent at the trivial fixed point, $\Lambda_0=\max_j \mathrm{Re}\,\lambda_j(M)\big|_{x^*=0}$, where $M$ is the stability matrix~\eqref{eq:stability_matrix}. $\Lambda_0$ is the asymptotic growth rate of infinitesimal perturbations of the silent state and changes sign exactly where this state loses linear stability (Fig.~\ref{fig:phase_diagram_gamma}b). It must be distinguished from the global largest Lyapunov exponent $\lambda_{\max}$, computed along trajectories with a Jacobian-based method~\cite{engelken2023lyapunov}, which characterizes the dynamics in the unstable phase (Fig.~\ref{fig:phase_diagram_gamma}c and Appendix~\ref{app:lyapunov}, Fig.~\ref{fig:supp_phase_diagrams}). Close to the transition $\lambda_{\max}$ grows slowly with $g-g_c$ (Fig.~\ref{fig:supp_lyapunov_exponents}a), so that for finite $N$ and simulation time it is numerically indistinguishable from zero in the vicinity of the critical line.

The numerical analysis reveals a rich dynamical landscape. At small $g$, a silent fixed-point phase exists for all values of $\gamma$. Increasing the coupling strength destabilizes this phase, giving rise to two distinct regimes.  For small $\gamma$ and away from the critical line, a positive global LLE, $\lambda_{\max}>0$, indicates a chaotic phase (Fig.~\ref{fig:phase_diagram_gamma}c and Appendix~\ref{app:lyapunov}, Fig.~\ref{fig:supp_phase_diagrams}). For large $\gamma$, a multistable phase emerges, characterized by negative global Lyapunov exponents and a large number of fixed points. This multistable regime has been previously observed in networks with self-couplings~\cite{stern2014dynamics,stern2023reservoir,clark2024theory}, consistent with the fact that partial symmetry induces an effective self-coupling in the dynamical equations~\eqref{eq:DMFT}.

Examples of network dynamics are shown in Fig.~\ref{fig:phase_diagram_gamma}a, along with the corresponding stationary distributions of neural activities for different groups of units, sorted by the quantiles of their degree. This representation highlights that the dynamics are strongly influenced by hubs with large degree, which exhibit larger variance and increasing bistability compared to nodes with fewer connections.

For large $g$, increasing $\gamma$ leads to a crossover in the  global LLE $\lambda_{\max}$ (Fig.~\ref{fig:phase_diagram_gamma}c) from positive values, indicative of chaotic dynamics, to small but negative values (at least for finite $N$). At large $\gamma$, the  emergence of multiple fixed points  leads to ergodicity breaking, a behavior reminiscent of the glassy phase in spin systems and consistent with recent observations in balanced networks with symmetric couplings~\cite{berlemont2022glassy, tixidre2025cortical}.

This behavior is illustrated in Fig.~\ref{fig:phase_diagram_gamma}c by computing the ergodic
distance between  replicas~\cite{berlemont2022glassy, tixidre2025cortical}, defined as the mean
squared difference between the time-averaged activities of corresponding neurons in two
copies of the same system initialized with different conditions.
We find that the  ergodic distance asymptotically vanishes for small $\gamma$, whereas it converges to a large finite value for large $\gamma$, signaling the breakdown of ergodicity.
\section{EMERGENCE OF HETEROGENEOUS TIMESCALE DISTRIBUTION}\label{sec:emergence}

Partial symmetry ($\gamma>0$) generates a memory term in the hDMFT, governed by the response function $G(\tau)$ that acts as an effective self-coupling. Consequently, the dynamics becomes similar to Ref. \cite{stern2014dynamics,stern2023reservoir}, including bistability at sufficiently large $\gamma$ (trajectory (iii) in Fig. \ref{fig:phase_diagram_gamma}a). This mapping implies that topology and reciprocity jointly induce a distribution of effective self-couplings, and therefore a distribution of intrinsic timescales, across units \cite{stern2023reservoir}.

This similarity can be understood, to a first approximation, by noting that at stationarity the response function $G(t,t')$ depends only on the time difference $\tau = t - t'$, and decays much faster than the autocorrelation times.  Within this regime, where the typical decay time of $G(\tau)$ is much shorter than that of the correlation function $C(\tau)$, we can neglect the temporal memory and, to a first approximation, assume that $G(\tau) \simeq \chi\,\delta(\tau)$ (see Appendix \ref{app:num_DMFT}, Fig. \ref{fig:DMFT_solution}).
The resulting dynamical model can thus be written as
\begin{equation} \label{eq:hDMFT}
    \dot{x}_k = -x_k + g\sqrt{\frac{k}{K}}\,\eta(t) + g^2 \gamma \frac{k}{K} \chi\, \phi(x_k).
\end{equation}

Compared to the model in~\cite{stern2023reservoir}, where the self-coupling distribution can be arbitrary, here the distribution of self-couplings $s_k = g^2 \frac{k}{K} \gamma \chi$ emerges from the microscopic features of the network, namely, the interactions of the degree distribution $P(k)$ and the  partial symmetry parameter $\gamma$.

Intuitively, the microscopic origin of the effective self-couplings is that a neuron with degree \( k \) excites or inhibits the neurons to which it is connected and, due to the correlation of the couplings \( \gamma \), recurrently receives a proportional excitation or inhibition in return. These effective recurrent self-interactions are also proportional to the number of connections of the neuron.
As a result, the effective self-coupling is proportional to the degree, $s_k \propto k$ [Eq.~(\ref{eq:hDMFT})]: high-degree nodes (hubs) receive a larger self-coupling, which pushes them towards bistability, while low-degree nodes remain fast.

The degree, however, also sets the intensity of the recurrent noise, $D_k \propto k$. On its own this does not change the timescale: for $\gamma = 0$ the effective dynamics is linear in $x_k$, the degree only rescales the input, and the normalized autocorrelation is the same for all neurons (Appendix~\ref{app:num_DMFT}). For $\gamma > 0$ the self-coupling creates two symmetric wells once $s_k$ exceeds a threshold, which hubs reach first, and the timescale becomes dominated by the dwell time between the wells (Appendix~\ref{app:bistability}). The two degree-dependent terms then compete: the noise drives the escapes from a well, while the self-coupling sets its depth. Since the barrier grows faster than linearly in $s_k \propto k$, whereas the noise intensity grows only linearly in $k$, their ratio increases with the degree, and hubs switch less often. This degree-dependent slowing is the key physical mechanism.

Numerical solutions of the heterogeneous DMFT equations (Appendix~\ref{app:num_DMFT}) support this interpretation. For a network with a discrete degree distribution \(k \in \{100,1000\}\), in the fully asymmetric case (\(\gamma=0\)) low- and high-degree nodes have essentially the same autocorrelation time. Introducing partial symmetry slows down the dynamics of all units and progressively separates the two classes, with hubs becoming the slowest. At weak symmetry the stationary distribution of hubs broadens while remaining unimodal; at stronger symmetry it becomes bimodal, first for hubs, and the autocorrelation time is then set by the switching between the two wells (Appendix~\ref{app:bistability}). The response kernel \(G(\tau)\) decays faster than the autocorrelation functions, which motivates the approximation \(G(\tau) \simeq \chi\,\delta(\tau)\) used above.

In general, the memory kernel \(G(\tau)\) also encodes recurrent self-interactions arising from correlations of a node's couplings with its second-, third-, and higher-order neighbors---effects that are neglected when \(G(\tau)\) is replaced by a Dirac delta. Comparing this approximation with the full solution of the hDMFT and with direct simulations (Appendix~\ref{app:num_DMFT}, Figs.~\ref{fig:kernels_memory} and \ref{fig:acf_memory_steps}) shows that the instantaneous self-coupling captures the mechanism, namely a memory term whose strength grows with the degree and orders the timescales by degree, but underestimates the timescales of all units: part of the slowing down originates from the finite temporal extent of the memory kernel.
\subsection{Heterogeneous degree distribution}
\begin{figure}[h]
    \centering
\includegraphics[width=\linewidth]{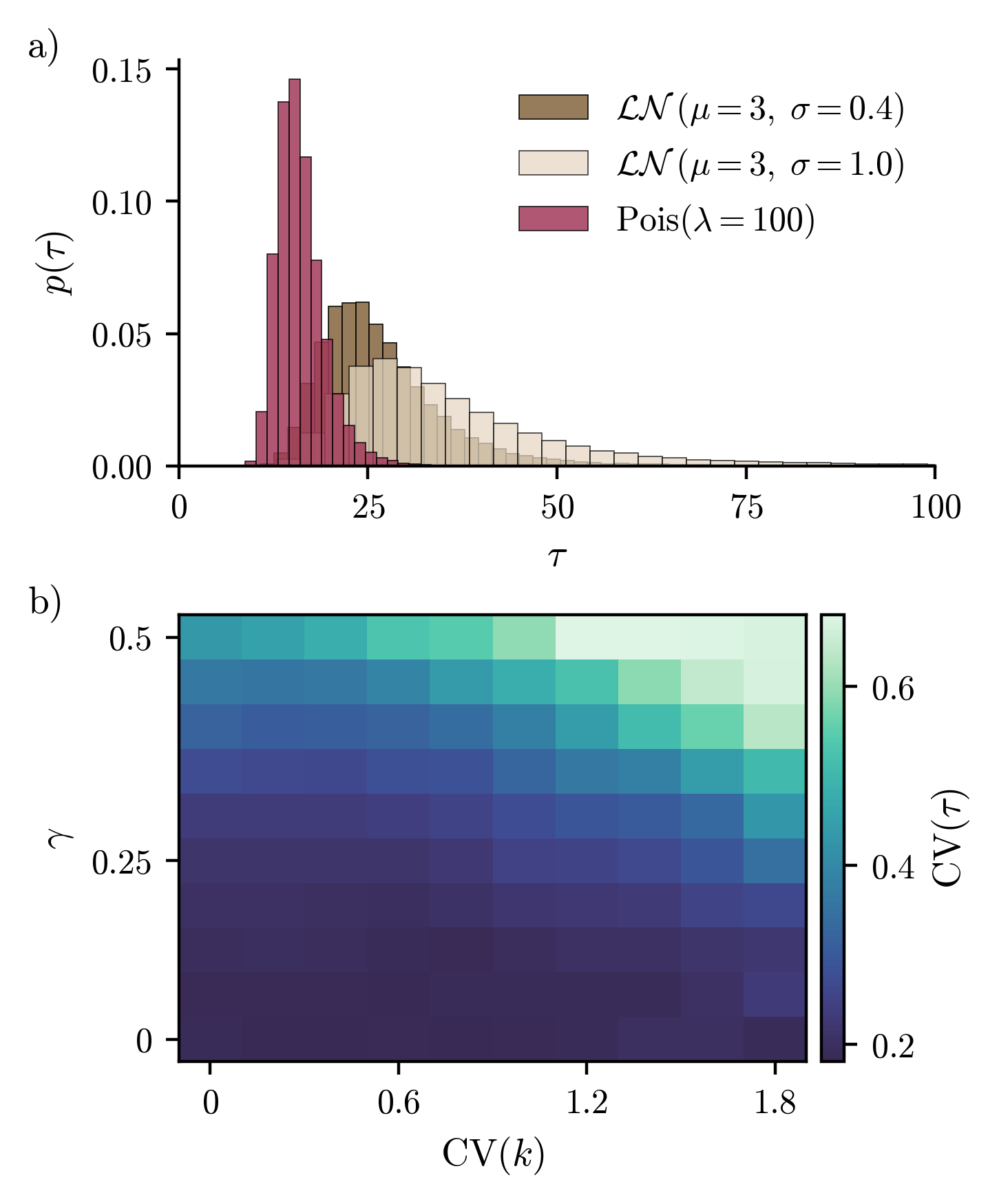}
   \caption{\textbf{Emergence of heterogeneous timescales.} (a) Distribution of neuronal timescales for Poisson and lognormal degree distributions at gain $g = 3$ and symmetry parameter $\gamma = 0.4$. The neuronal timescale $\tau$ is defined as the half-width at half-maximum (HWHM) of each neuron’s autocorrelation function. Parameters of the degree distributions are reported in the legend.
    (b) Coefficient of variation of neuronal timescales, $\mathrm{CV}(\tau)$, as a function of the symmetry parameter $\gamma$ and the coefficient of variation of the lognormal degree distribution $ \mathrm{CV}(k)=\sqrt{e^{\sigma^2}-1}$. Increasing heterogeneity in the degree distribution and larger symmetry both lead to a progressive broadening of the timescale distribution, indicating enhanced heterogeneity in neuronal dynamics. For each degree distribution, dynamics are simulated over 20 independent network realizations with $N = 4000$.}
\label{fig:panel_timescale}
\end{figure}

We now turn to the statistics of single-neuron dynamics at the network level. Heterogeneous degree distributions, such as lognormal distributions, induce a broad distribution of effective self-couplings, which in turn generates a
reservoir of intrinsic timescales, as illustrated in Fig.~\ref{fig:panel_timescale}.

In this setting, we define the timescale of a neuron as the half-width at
half-maximum (HWHM) of the autocorrelation function of its activity. We use the HWHM as a model-free summary of the decay: it assumes no functional form for the
autocorrelation, is not dominated by the tail, and is monotonic in the decay rate, which is what our rank-based comparison across neurons requires.
Figure~\ref{fig:panel_timescale}a shows histograms of neuronal timescales for different
choices of the degree distribution: a Poisson distribution, used as a proxy for a
homogeneous network, and two lognormal distributions inspired by experimentally
observed connectomes, with different values of the scale parameter $\sigma$.

We fix the network parameters by choosing the location parameter $\mu$ (the mean of $\ln k$) of the lognormal degree distribution and the gain $g$, while varying the symmetry parameter $\gamma$ and the scale parameter $\sigma$. For all parameter choices, we verify that the dynamics
remain in the chaotic regime and do not converge to fixed points, ensuring that
autocorrelation functions are well defined. Using extensive numerical simulations, we
then analyze how structural heterogeneity translates into dynamical heterogeneity in
recurrent neural circuits. In particular, we compute the coefficient of variation of
the neuronal timescales, $\mathrm{CV}(\tau)$, defined as the ratio of their standard
deviation to their mean, as a function of the heterogeneity of the degree distribution,
$\mathrm{CV}(k)$, and the symmetry parameter $\gamma$. The results for lognormal degree
distributions are shown in Fig.~\ref{fig:panel_timescale}b.

The results show that for small values of $\gamma$, the effect of introducing
heterogeneity in the connectivity is weak and the overall coefficient of variation of
the timescales remains small. Increasing the degree of partial symmetry, however,
leads to a regime in which the diversity introduced by the connectivity structure is
amplified by symmetry, resulting in network dynamics characterized by a broad range
of intrinsic timescales. The shape of this distribution follows from that of the degree distribution: since the timescale increases monotonically with the degree, $P(\tau)$ is obtained from $P(k)$ by a change of variables, and its tail is set by the ratio between the rate at which high-degree neurons become rare and the rate at which they become slow. An exponential degree distribution can thus produce a power-law tail of timescales, whereas a lognormal one gives a broad tail without a single exponent (Appendix~\ref{app:tail}).

We examine the large $N$ scaling of several alternative models to identify the ones where timescale heterogeneity is preserved in the thermodynamic limit. To this end, we define as a criterion for heterogeneity that $\mathrm{CV}(\tau)$ remains finite at large $N$. If $\mathrm{CV}(\tau)$ vanishes at large $N$, then the heterogeneity in the corresponding model does not survive the thermodynamic limit.
We tested this criterion by computing $\mathrm{CV}(\tau)$ as a function of $N$, up to $N=10^4$, for our model and for other alternative models associated with heterogeneous timescales (Appendix~\ref{app:timescales_scaling}, Fig.~\ref{fig:timescales_scaling}). For both our  model, as well as the alternative model with heterogeneous self-couplings~\cite{stern2023reservoir}, $\mathrm{CV}(\tau)$ stays approximately constant over the whole range of $N$. On the other hand, in models of fully connected networks close to criticality~\cite{chen2024searching} and in models with i.i.d.\ heavy-tailed couplings~\cite{shi2025brain}, both of which display heterogeneous timescales at small $N$, $\mathrm{CV}(\tau)$ decreases monotonically with $N$, suggesting that the heterogeneity does not survive in the thermodynamic limit. In an alternative model where heavy-tailed couplings are combined with partial symmetry, however, the partial symmetry rescues the scaling and $\mathrm{CV}(\tau)$ stays finite al large $N$. A heuristic argument (Appendix~\ref{app:iid}) suggests the origin of this difference: a heterogeneity that persists at large $N$ requires a quenched single-neuron variable that enters the dynamics as a self-feedback, and not only as an input amplitude. For i.i.d.\ couplings the reciprocal terms that would provide this feedback vanish as $N$ grows, whereas partial symmetry keeps them finite and the degree sets their strength, $s_k\propto k$; heterogeneous self-couplings provide the same feedback by construction.

The emergence of heterogeneous timescales can thus be understood as the combined effect
of partial symmetry and heterogeneity in the structural connectivity. Structural
heterogeneity introduces variability in the network architecture that shapes the
dynamics and leads to degree-dependent neuronal behavior. In the absence of partial
symmetry, this effect remains weak.

Partial symmetry provides the self-feedback through which the degree enters the dynamics: it slows down all units, and the hubs more strongly, so that the heterogeneity of the degrees is transferred to the timescales. Because the self-coupling is set by the degree, a quenched single-neuron property, these differences do not average out as the network size grows (Appendix~\ref{app:timescales_scaling}).

To provide a biologically plausible interpretation of our mechanism in a model that obeys Dale's, we examined our analysis to the case of a recurrent network of excitatory and inhibitory spiking neurons in the balanced regime (Appendix~\ref{app:spiking}). We found that, in networks with heterogeneous in-degrees but without reciprocity (namely, i.i.d. connectivity), the timescale of single neurons is independent of the degree. Crucially, we found that endowing the network with reciprocity in the excitatory-to-excitatory connectivity yields timescales that increase with the degree as reciprocity grows, while degree heterogeneity broadens the distribution of timescales (Fig.~\ref{fig:spiking}h,i). As in the rate model, partial symmetry and heterogeneous degree act jointly. Our results are specific to a regime where  reciprocity is strong and the degree spread moderate, while compensating the differences in mean excitation that in-degree heterogeneity produces under Dale's law. These results indicate that the proposed mechanism carries over to the biologically plausible case of balanced networks of spiking neurons.

\section{Network response to broadband input}\label{sect:broadband}
To investigate how structural heterogeneity and partial symmetry shape information
processing, we analyze the response of our network to a broadband input signal, following the approach in~\cite{stern2023reservoir}. We hypothesized that intrinsic timescales act as a substrate for frequency-selective information routing in recurrent neural circuits. Specifically, we quantified the extent to which a recurrent network with heterogeneous self-couplings can demix broadband inputs, such that units with large degree preferentially entrain low-frequency input components.

In our network, we consider two distinct populations of neurons: hub neurons with high degree ($k=1000$) and low-degree neurons ($k=100$). The input is a broadband periodic drive composed of $M=11$ sinusoidal components with distinct frequencies, representing a simplified model of complex, time-varying stimuli. The input is applied uniformly across all neurons and can be written as
\begin{equation}
I_i(t) = A \sum_{m=1}^{M} \sin\!\big(2 \pi f_m t + \phi_i\big), \quad i = 1, \dots, N,
\end{equation}
where $I_i(t)$ is the input current to neuron $i$, $A$ is the input amplitude, $f_m$ is the frequency of the $m$-th sinusoidal component, and $\phi_i$ is a neuron-specific random phase drawn uniformly from $[0,2\pi)$. This formulation allows us to study how intrinsic network properties, including degree heterogeneity and partial symmetry, shape the response of each subpopulation. The random phases are chosen so that the input does not favour any direction in neuron space; mode-selective inputs can generate additional, input-dependent timescale diversity~\cite{dahmen2019second, shao2025impact}

To quantify the degree-dependent modulation of the response, we define a modulation index (MI)~\cite{stern2023reservoir}. Let $\mathrm{PSD}_\mathrm{hub}(f)$ and $\mathrm{PSD}_\mathrm{low}(f)$ denote the power spectral densities (PSDs) of the activity $x_i$ at frequency $f$, averaged over the neurons of the hub and of the low-degree population, respectively. The modulation index is then
\begin{equation}
\mathrm{MI}(f) = \frac{PSD_\mathrm{hub}(f) - PSD_\mathrm{low}(f)}{PSD_\mathrm{hub}(f) + PSD_\mathrm{low}(f)} \,,
\end{equation}
which captures the relative dominance of high- versus low-degree neurons at each frequency.

\begin{figure}[h]
\centering
    \includegraphics[width=1\linewidth]{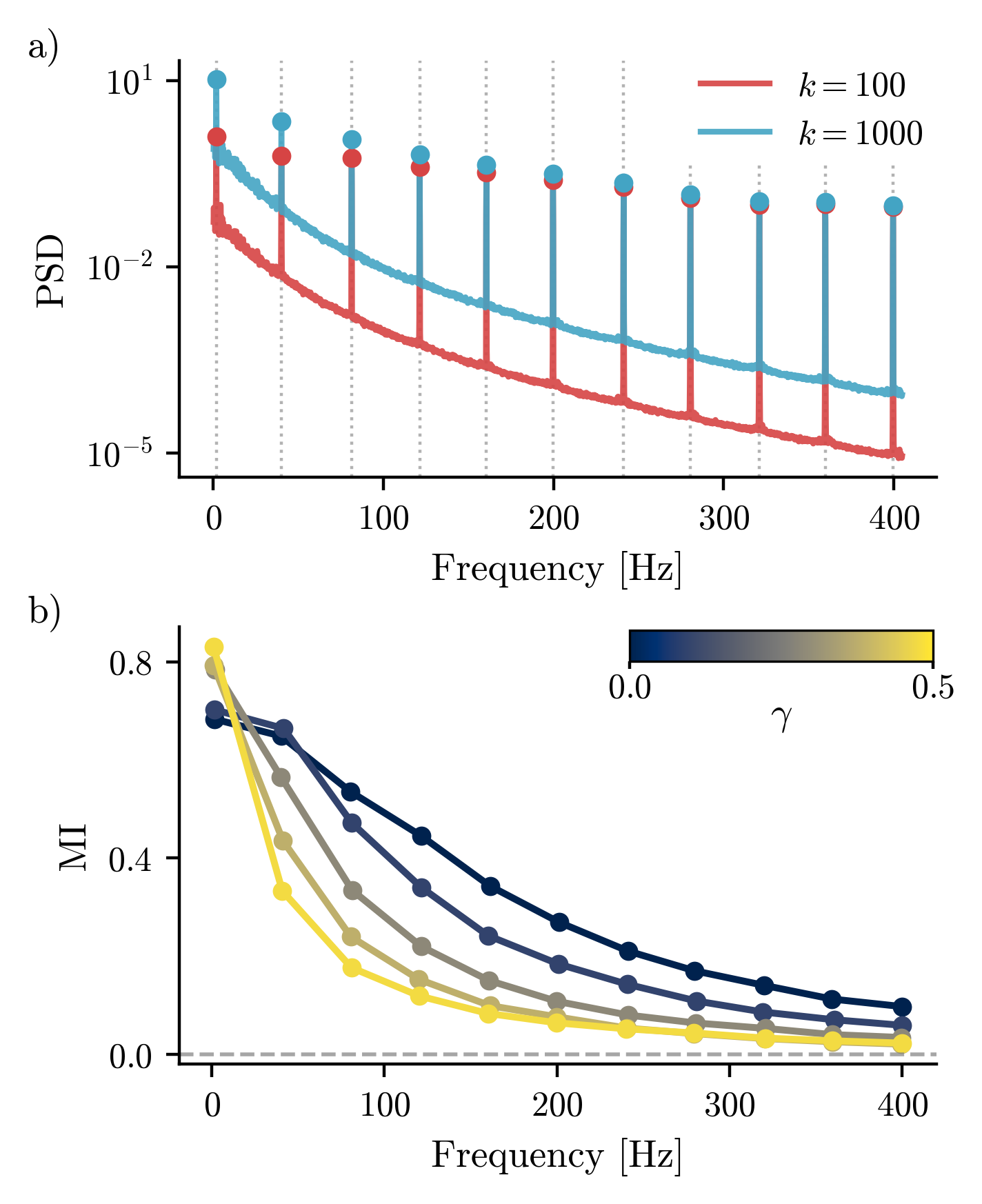}
    \caption{\textbf{Network response to broadband input.} (a) Population-averaged power spectral density (PSD) of the network activity under a broadband periodic drive composed of the superposition of $M=11$ sinusoidal components with different frequencies (vertical dashed lines; input amplitude $A=0.5$). Red and blue curves show the PSD averaged over nodes in a network with $\gamma=0.3$, degree distribution $k \in \{100,1000\}$ and probabilities $P_k=\{0.9,0.1\}$, respectively. Circles indicate the PSD amplitudes evaluated at the driven frequencies. (b) Modulation index ($\mathrm{MI}$) between the high- and low-degree PSD responses as a function of the driving frequency for different values of the symmetry parameter $\gamma$ (color-coded; $\gamma=0.0,\,0.1,\,0.3,\,0.4,\,0.5$). Increasing $\gamma$ enhances the frequency selectivity of the degree-dependent response, making $\mathrm{MI}$ increasingly concentrated at low frequencies, consistent with a stronger dominance of high-degree nodes for the slow input components.  Network parameters: $g=3.0$, $N=4000$.}
    \label{fig:broadband_input}
\end{figure}

We found that while high-degree neurons preferentially entrain lower
frequencies (Fig.~\ref{fig:broadband_input}), at higher frequencies both high- and
low-degree neurons respond similarly; a difference that is enhanced at increasing partial symmetry. This result stands in slight contrast with the heterogeneous self-coupling case \cite{stern2023reservoir}, where units with stronger self-couplings preferentially entrain slower frequency components, while units with weaker self-couplings respond more to faster components. The difference stems from the fact that highly connected neurons act as integrators of a large number of inputs from
presynaptic neurons, rather than as units directly entraining the input, as in networks
with heterogeneous self-couplings.

\section{STRUCTURAL-FUNCTIONAL ANALYSIS OF MOUSE VISUAL CORTEX} \label{sec:microns}

We tested the main predictions of our model, namely that high-degree neurons exhibit longer intrinsic timescales,  using the Cubic Millimeter dataset of the MICrONS project~\cite{microns2025functional}. This dataset provides both detailed structural connectivity of the mouse visual cortex (Fig.~\ref{fig:microns}) and large-scale functional recordings of neural activity obtained through calcium imaging. An example calcium trace is shown in Appendix~\ref{app:microns}, Fig.~\ref{fig:microns_supplementary}. The dataset includes recordings obtained under different behavioral conditions; here we focus on spontaneous activity during rest. Our analysis is restricted to the subset of proofread neurons, for which synaptic connections have been manually validated, ensuring higher confidence in the reconstructed adjacency matrix.

Functional activity was extracted from calcium imaging recordings during resting-state conditions. This choice is motivated by our theoretical framework, which predicts that heterogeneous timescales emerge as an intrinsic property of the network structure, independent of stimulus-driven responses. The timescale of each neuron is defined as the half-width at half-maximum (HWHM) of the autocorrelation function of its calcium trace.

Since the indicator kinetics broaden the autocorrelation and the time unit of the model is arbitrary, this measure estimates relative rather than absolute timescales; this suffices here, as our analysis relies on rank statistics across degrees, which are invariant under the
monotonic distortions involved.

\begin{figure*}[!ht]
    \centering
\includegraphics[width=1\linewidth]{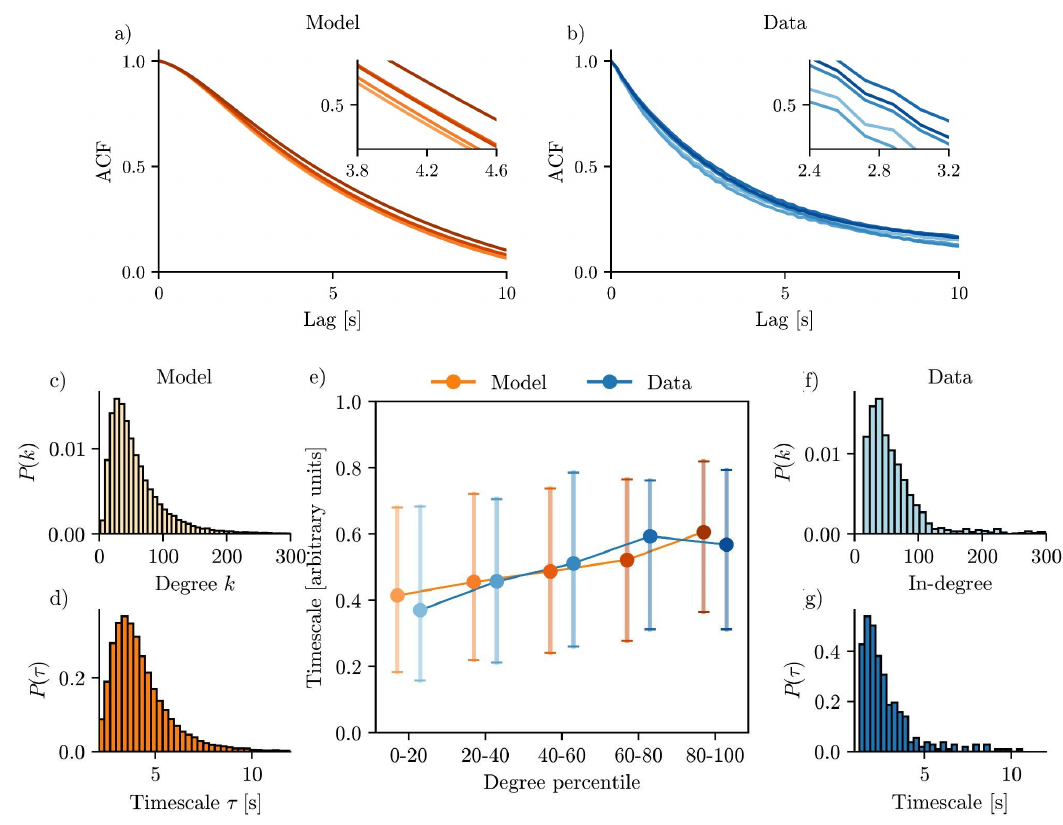} \caption{\textbf{Correlation between degree and intrinsic timescales.} (a) Autocorrelation functions of a recurrent neural network with a lognormal in-degree distribution ($\mu = 3.83$, $\sigma = 0.69$), partial symmetry $\gamma = 0.35$, and gain $g = 4$. The inset shows a magnified view around the half-width at half-maximum (HWHM). Curves represent the mean autocorrelation of units grouped into five percentile bins of the in-degree distribution.  (b) Autocorrelation functions of calcium imaging traces from the MICrONS  dataset~\cite{microns2025functional}, averaged within the same five percentile bins of in-degree. (c) Synthetic in-degree distribution obtained by sampling from a lognormal distribution with $\mu = 3.83$ and $\sigma = 0.69$. (d) Distribution of intrinsic timescales obtained from 100 simulations of a network
with $N = 2000$ units, gain $g = 4$, and partial symmetry $\gamma = 0.35$. (e) Comparison between the correlation of in-degree and intrinsic timescales in the model (panels c--d) and in the MICrONS data (panels f--g). Degrees are grouped by percentiles, and timescales are reported in arbitrary units to enable qualitative
comparison. Medians are shown as central markers, with interquartile ranges (25th--75th percentiles) indicated by vertical bars. Spearman correlation coefficients
are $\rho_{\mathrm{data}} = 0.164$ ($p = 10^{-4}$) for the MICrONS dataset and $\rho_{\mathrm{model}} = 0.159$ for the model.
(f) In-degree distribution of proofread neurons in the MICrONS dataset~\cite{microns2025functional}.
(g) Distribution of intrinsic autocorrelation timescales of calcium traces recorded
at rest in the MICrONS functional dataset~\cite{microns2025functional}. The simulation time constant $\tau_{\mathrm{sim}}$ in (a)-(d) is chosen to match the observed autocorrelation times, $\tau_{\mathrm{sim}} = 2.4\,\mathrm{s}$.}
    \label{fig:microns_matched}
\end{figure*}

For the structural analysis, we extract the in-degree of each proofread neuron from the reconstructed adjacency matrix of the MICrONS dataset~\cite{microns2025functional} (Fig.~\ref{fig:microns}). The resulting in-degree distribution exhibits substantial heterogeneity and is well described by a lognormal distribution with parameters $\mu = 3.83$ and $\sigma = 0.69$. In addition, we observe a significant level of partial symmetry in the connectivity, with a structural reciprocity $\gamma_s = 0.197$, consistent with previous experimental reports~\cite{piazza2025physical,song2005highly}.

We used the empirical degree distribution $P(k)$ to
inform the construction of an RNN model that predicts a distribution of intrinsic
timescales. Since the model
parameter $\gamma$ also depends on the correlation between reciprocal synaptic
strengths (Sec.~II~B), which cannot be uniquely inferred from synapse counts and sizes,
we treat it as an effective parameter, choosing $\gamma = 0.35$ to match the measured
degree--timescale correlation. Values in the range $0.3 \leq \gamma \leq 0.4$ yield
qualitatively similar results (Appendix ~\ref{app:sim_details}).

Within the chaotic regime, the predictions are not sensitive to the
remaining RNN parameter $g$, which controls the overall strength of synaptic couplings.
We use $g = 4$; deep in the chaotic regime the results are insensitive to $g$ (Appendix \ref{app:sim_details}).

The RNN model predicts that neurons with higher in-degree exhibit longer intrinsic timescales. Comparison with the empirical data shows that the model captures the observed variability of timescales for a given degree, closely matching the quantile ranges measured experimentally (Fig.~\ref{fig:microns_matched}e). The Spearman correlation between degree and timescale is $\rho_{\mathrm{data}} = 0.164$ ($p = 10^{-4}$) for the MICrONS dataset and $\rho_{\mathrm{model}} = 0.159$ for the model. The details of the analysis are provided in Appendix~\ref{app:microns}.

This result provides empirical support for the mechanism revealed by our model. The heterogeneity in network connectivity, amplified by partial symmetry due to the overrepresentation of reciprocal synapses, gives rise to a broad distribution of neuronal activity timescales. Overall, the model matches the relationship between degree and timescales observed in the data.

\section{DISCUSSION}

A central question in systems neuroscience is how diverse intrinsic neuronal timescales emerge from collective circuit dynamics. Several mechanisms have been proposed \cite{stern2023reservoir,chen2024searching,chaudhuri2014diversity,chaudhuri2015large,shi2025brain}, yet direct links between measurable circuit topology and within-area timescale diversity remain limited. Here we identify a specific topological mechanism: degree heterogeneity interacting with partial symmetry in synaptic couplings  produces degree-dependent emergent self-couplings and hence heterogeneous intrinsic timescales. We test this prediction in the MICrONS connectome and matched physiology.

\subsection{Empirical evidence for diversity of timescales}
A substantial body of experimental work indicates that intrinsic timescales are organized at the macroscale, forming systematic gradients across cortical areas. In non-human primates, spike autocorrelation times across multiple cortical regions exhibit a clear hierarchical ordering, with shorter timescales in sensory areas and progressively longer timescales toward prefrontal cortex, broadly consistent with anatomical hierarchies defined by long-range projections \cite{murray2014hierarchy}. More recently, this organizing principle has been tested at scale across species and behavioral conditions~\cite{song2024hierarchical}. Together, these results suggest that temporal specialization is a robust feature of mammalian cortical organization, plausibly reflecting conserved circuit-level constraints of intracortical processing.

Importantly, timescale heterogeneity has been reported not only across the cortical hierarchy, but also across neurons within the same area. Prior studies reported broad within-area distributions of intrinsic timescales in diverse systems—from the oculomotor circuitry of fish \citep{miri2011spatial} to primate brainstem \citep{joshua2013diversity} and frontal cortex \citep{cavanagh2016autocorrelation,bernacchia2011reservoir}. In large-scale mouse electrophysiology datasets, individual neurons exhibit markedly different autocorrelation shapes. These shapes are better captured by mixtures of exponentials, consistent with multiple effective timescales rather than a single constant \cite{shi2025brain}. This coexistence of within-area dispersion and measurable connectomic heterogeneity motivates a topology-based mechanism, which we formalized in our model.

\subsection{Heterogeneous dynamic mean field theory}

We found that, in the MICrONS dataset, neurons in the mouse visual cortex also exhibit a broad  distribution of intrinsic timescales. We therefore asked whether this dispersion can arise from measurable connectomic statistics. MICrONS reveals two salient signatures: elevated reciprocity (modeled as partial symmetry) and a broad degree distribution.
To isolate the joint effect of these ingredients, we derived a heterogeneous dynamical mean-field theory (hDMFT) for a configuration model with lognormal degree distributions and reciprocal-pair correlations captured by the symmetry parameter $\gamma$. The hDMFT yields a degree-conditioned effective single-neuron dynamics with a non-Markovian memory term. At long times, the response kernel reduces to a quasistatic gain, mapping partial symmetry to an emergent self-coupling that scales with degree, $s_k=\gamma g^2 (k/K)\chi$ (Eq.~\ref{eq:hDMFT}). This degree-dependent self-coupling yields unit-specific timescales (Appendix \ref{app:num_DMFT}, Fig.~\ref{fig:DMFT_solution}), and does not rely on fine-tuning or proximity to criticality, as the response kernel $G(\tau)$ decays much faster than the activity autocorrelations $C(\tau)$, supporting the effective self-coupling reduction  (Appendix \ref{app:num_DMFT}, Fig.~\ref{fig:DMFT_solution}).
Linear stability and spectral analyses further show how topology controls dynamical regimes: for $\gamma=0, g_c^2=K^2/\mathbb{E}[k^2]$, implying $g_c=e^{-\sigma^2/2}$ for lognormal $P(k)$, and unstable modes localize on hubs (Fig. \ref{fig:panel1}d–e). For $\gamma > 0$, linear stability analysis provides an exact condition for the transition from the silent phase to the unstable phase. The resulting phase diagram, characterized by the largest Lyapunov exponent and replica analysis, exhibits a rich structure comprising silent, chaotic, bistable, and heterogeneous fixed-point regimes (Fig.~\ref{fig:phase_diagram_gamma}).

\subsection{Empirical validation}
Using MICrONS, we find that the in-degree of a neuron is positively associated with its intrinsic timescale. The model parameterized by the empirical degree distribution and an effective symmetry parameter $\gamma$ reproduces both the broad timescale distribution and the positive degree-timescale association (Spearman $\rho\approx 0.16$ in data and model; Fig.~\ref{fig:microns_matched}), with $\gamma$ chosen to match simulations and data and results insensitive to the gain $g$ (Appendix~\ref{app:sim_details}). The effect size is modest (Fig.~\ref{fig:microns_matched}e), but the correlation is stable across successive releases of the connectome (Appendix~\ref{app:microns}). A key limitation is that MICrONS provides anatomical measures of synaptic connectivity, such as synapse counts and sizes, but not the
effective couplings entering the rate model, so that their reciprocal correlation
$\gamma$ cannot be measured directly and enters as an effective parameter.

Our results integrate prior approaches to timescale heterogeneity by providing a microscopic origin for heterogeneous self-couplings \cite{stern2014dynamics,stern2023reservoir}: partial symmetry generates effective self-feedback, and degree heterogeneity converts this feedback into a distribution of strengths. This generalizes symmetry-induced slowing in dense networks to degree-heterogeneous graphs and links topology to both stability thresholds and mode localization \cite{marti2018correlations}. In contrast to approaches that posit intrinsic single-cell heterogeneity \cite{tomita2025dynamical}, our mechanism produces timescale diversity even with identical neurons. Methodologically, hDMFT adapts tools from structured ecological interactions and yields closed-form stability criteria; spectrally, heterogeneity compresses bulk eigenvalues and concentrates unstable directions on hubs, sharpening the connection between connectomic statistics and computational timescales.

The theory yields testable predictions that connect structure to dynamics. First, timescale distributions should track the dispersion of in-degree across layers and areas: at fixed $\gamma>0$, increasing $CV(k)$ should increase $CV(\tau)$. Second, manipulations that alter bidirectionality, including developmental changes, plasticity rules that increase reciprocity, or circuit-specific perturbations, should lengthen timescales as $\gamma$ increases and shift networks toward bistability in high-degree subpopulations.
Third, because the emergent self-coupling scales with degree, stimulus rise/decay and recovery times should systematically vary across degree quintiles, a prediction that could be probed in stimulus-response data from the same preparations.

\subsection{Alternative mechanisms and relation to prior theory}
Previous modeling studies addressing potential explanations for the observed diversity of timescales relied on either single cell or network effects. It has been hypothesized that across-cell heterogeneity in bio-physical properties, such as membrane or synaptic time constants  \citep{gjorgjieva2016computational}, or developmental changes in conductances \citep{gjorgjieva2014intrinsic}, might lead to timescale diversity. However, subsequent modeling work \cite{stern2023reservoir} suggested that, although the heterogeneity in single-cell time constants may affect the dynamics of single neurons in isolation or within feedforward circuits \citep{gjorgjieva2014intrinsic}, strong recurrent dynamics overcomes single cell differences and restores a single timescale for all cells in the same circuit.

Although single-cell properties may not explain the diversity of timescales, a variety of alternative network mechanisms have been proposed. A combination of structural heterogeneities and long-range connections arranged along a spatial feedforward gradient \citep{chaudhuri2014diversity,chaudhuri2015large} was shown to reproduce the population-averaged hierarchy of timescales observed across primate cortical areas \citep{murray2014hierarchy,chaudhuri2015large}. However, it is not known whether such  feedforward spatial structure is present within a local cortical circuit, which would be necessary to explain a diversity of timescales within a single area.

In linear networks, fine-tuning the coupling parameters close to criticality can lead to a heterogeneity of timescales \cite{chen2024searching,shi2025brain}. Slow collective modes and long correlation times emerge as the stability spectrum approaches zero \cite{chen2024searching,shi2025brain}. In this picture, extended timescales arise from a near-marginal spectrum and are therefore strongest in a neighborhood of the instability point; away from that regime the slow tail is reduced, and the timescale heterogeneity depends on how close the system is poised to criticality. A second class of mechanisms instead encodes multiscale dynamics in the statistics of synaptic interactions by departing from Gaussian weight ensembles. Heavy-tailed (e.g., $\alpha$-stable) coupling distributions can generate heavy-tailed timescale statistics in nonlinear networks over a broad parameter range, without requiring fine tuning of the gain \cite{shi2025brain}. However, we found that for independent heavy-tailed couplings this mechanism is a finite-size effect that does not survive the thermodynamic limit (Appendices~\ref{app:timescales_scaling} and~\ref{app:iid}). When heavy-tailed couplings are combined with partial symmetry, instead, the timescale heterogeneity does not decrease with network size over the range of $N$ tested (Fig.~\ref{fig:timescales_scaling}), consistent with the mechanism identified here. In nonlinear networks, a diversity of timescales can be generated via self-tuned criticality maintained by anti-hebbian plasticity \citep{magnasco2009self}; or in networks whose coupling matrix exhibit multiple block-structured connectivity \citep{aljadeff2015transition}.

More recently, Stern et al.~\cite{stern2023reservoir} identified a general mechanism linking the autocorrelation timescale of a network's node to its own self-coupling, a mechanism requiring strong recurrent dynamics for its activation. When applied to a recurrent neural circuit, the authors showed that a distribution of self-couplings across rate units leads to a broad distribution of intrinsic timescales, along with a range of emergent properties, including multistability and spatial demixing~\cite{stern2014dynamics,stern2023reservoir}. How do the effective self‐couplings of \cite{stern2023reservoir} arise in empirical neural circuits? The authors proposed a phenomenological interpretation, supported by a spiking network formulation, whereby a rate unit was interpreted as a neural cluster with potentiated within-cluster couplings. We directly tested this alternative mechanism in the MICrONS data. We first inferred the presence of neuronal clusters and estimated their features using a stochastic block model. When we examined the relationship between cluster size and neuronal timescales, we did not find any evidence for it.  Our main contribution in this study is to provide an alternative explanation of the emergence of such broad self-coupling distributions grounded in circuit connectivity (degree and reciprocity) directly estimated from mouse visual cortex.

As a first exploration, we compared the two mechanisms in balanced networks of spiking neurons obeying Dale's law (Appendix~\ref{app:spiking}). Degree heterogeneity combined with reciprocal connections reproduces the main features of the rate model: single-neuron timescales increase with the in-degree when reciprocity is present, and their distribution broadens with degree heterogeneity, showing that the mechanism can operate in spiking circuits, at least in the regime explored here. A network of heterogeneous assemblies~\cite{litwin2012slow, mazzucato2015dynamics, stern2023reservoir} produces a similar dependence of the timescale on its own structural variable, the assembly size. The two mechanisms differ in which variable is slow, the collective activity of an assembly or the activity of single neurons, and in the structural correlate of the timescale, assembly membership or in-degree; the latter is directly measurable in connectomes such as MICrONS, which makes the two mechanisms experimentally distinguishable.

\subsection{Mechanisms underlying slow dynamics}

In our model, the main driver of slow timescales is the neuronal bistability, consistent with previous studies in networks of rate \cite{stern2014dynamics,stern2023reservoir} or spiking neurons \cite{litwin2012slow,mazzucato2015dynamics,papadopoulos2024modulation}. Bistability is induced by self-coupling terms in the dynamical equations, which can emerge collectively due to a clustered network architecture \cite{amit1997dynamics}, or due to partial symmetry in the connectivity \cite{marti2018correlations}. However, slow timescales may emerge from other mechanisms, such as tuning recurrent couplings close to critical points \cite{toyoizumi2011beyond,chen2024searching,shi2025brain} or from global attractor dynamics in the presence of noise \cite{recanatesi2022metastable}. A clear signature of bistability-induced slow dynamics is a bistable distribution of single-cell input currents, which is potentially accessible to intracellular recordings or voltage sensitive dye imaging. We leave the empirical test of bistability signatures to future work.

\subsection{Partial symmetry regulates network dynamics}
We also found that the amount of partial symmetry in synaptic couplings, determined by the $\gamma$ parameter, regulates the network dynamical regime, toggling between chaotic and multistable phases. In brain circuits,  $\gamma$ is not fixed, but rather may evolve via different plasticity mechanisms. On fast timescales, from subseconds to seconds, it could be up- or down-regulated via short-term facilitation or depression \cite{zucker2002short}. On longer timescales of minutes or hours, it may be sculpted by other plasticity mechanisms involved in learning, such as Hebbian plasticity  \cite{clark2024theory}. One possibility is that circuit plasticity and neuromodulation control $\gamma$, positioning high-degree subpopulations near the edge of instability while keeping the bulk stable to flexibly adapt to varying computational demands and changing environments.

\subsection{Limitations and future directions}

Our model intentionally adopts a minimal random-network setting to isolate the joint impact of degree heterogeneity and partial symmetry. This choice comes with clear limitations. In random model of rate units, we neglect key biological constraints on synaptic organization, such as distance-dependent wiring rules and Dale’s law of cell-type-specific connectivity.
To explore the effects of Dale's law, we extended our theory to the case balanced excitatory and inhibitory spiking networks. Here, we found our mechanism carries over in the presence of Dale's law (Appendix~\ref{app:spiking}). Extending these simulations to the full range of reciprocity and degree heterogeneity measured in cortex, including correlated synaptic strengths, and developing a mean-field description of the spiking case are natural next steps.
Incorporating excitatory–inhibitory structure and type-specific degree distributions into hDMFT would strengthen the link to cortical circuits and clarify how reciprocity and hub structure interact with balanced dynamics. Second, we treat synaptic weights as Gaussian (up to reciprocal correlations) and ignore additional sources of heterogeneity known to shape slow dynamics, such as  structured motifs beyond pairwise reciprocity and degree–weight correlations; although we examined the effects of heavy-tailed distributions in Appendices H-I. Extending the full hDMFT to include non-Gaussian and correlated weight ensembles would allow a more systematic comparison between “topological” and “weight-statistical” routes to timescale diversity and would help delineate which features are necessary to reproduce dataset-specific timescale distributions.
Third, we restrict the connectivity statistics to reciprocal pairs. Among second-order motifs (reciprocal, convergent, divergent and chain), the reciprocal pair is the only one that closes a loop on the same neuron, and it is therefore the lowest-order connectivity statistic that generates an effective self-coupling. Longer loops built from pairwise correlations are already contained in the full memory kernel $G(\tau)$ (Sec.~\ref{sec:emergence}), whereas loops that require correlations among three or more synapses, such as over-represented directed triangles, enter at higher order and would add further, possibly degree-dependent, contributions to the effective self-coupling; including them in the hDMFT is a natural extension of the present work. Similarly, combining degree heterogeneity with the full set of second-order motifs, for which a DMFT has recently been developed \cite{yang2026nonlinear}, could provide further routes to timescale diversity.

A further limitation is that our mean-field treatment relies on an uncorrelated network ensemble, treated at leading order in $1/K$, and neglects higher-order degree–degree correlations and community structure, which may be relevant in real connectomes.

Similarly, the theory is derived for an undirected topology with a single degree per neuron: structural reciprocity in directed networks is treated here only numerically (Appendix~\ref{app:sim_directed}), while networks with distinct in- and out-degrees and their mean-field description are left for future work.
Developing a quenched or partially quenched extension of hDMFT, and quantifying the role of modularity and hierarchical organization, would be important steps toward more realistic circuit descriptions. Finally, while we validate the mechanism against resting-state activity, the framework makes sharper predictions in driven settings: because effective self-feedback scales with degree, stimulus rise/decay times, frequency selectivity, and memory kernels should systematically vary across degree quantiles. Testing these predictions using stimulus-evoked recordings and perturbations that selectively alter reciprocity or hub participation would provide a direct experimental probe of the proposed mechanism.

Conceptually, our results suggest that two ubiquitous anatomical features—broad degree distributions and elevated reciprocity—provide a principled route from connectome statistics to heterogeneous effective self-couplings, yielding a robust substrate for multiplexed temporal processing.

\begin{acknowledgments}
We are grateful to Merav Stern and Ashok Litwin-Kumar for useful discussions. LM is supported by NSF Awards 2617523 and 2238247 and NIH Awards R01MH141929, R01MH127375, R01DA055439.
\end{acknowledgments}

\bibliography{references}%

\clearpage
\onecolumngrid

\appendix

\section{HETEROGENEOUS DYNAMICAL MEAN FIELD THEORY}\label{app:DMFT}
In this appendix, we derive the equations for the effective dynamics of model~(\ref{eq:RNN}). For an introduction to the methods, see~\cite{helias2020statistical}. We derive the Dynamical Mean-Field Theory (DMFT) of a recurrent neural network with pre-activation variables $x_i$, for $i = 1, \dots, N$, obeying the dynamical equations
\begin{equation}
    \frac{dx_i}{dt} = -x_i + \sum_{j} W_{ij} \, \phi(x_j),
    \label{eq_app:RNN}
\end{equation}
where the non-linear transfer function $\phi$ is chosen to be $\phi(x) = \tanh(x)$.
We consider random couplings of the form
\begin{equation}
    W_{ij} = A_{ij} J_{ij},
\end{equation}
where $A_{ij}$ is sampled according to a Bernoulli process:
\begin{equation}
    A_{ij} =
    \begin{cases}
        1 & \text{with probability } p_{ij},\\[2mm]
        0 & \text{with probability } 1 - p_{ij},
    \end{cases}
\end{equation}
with
\begin{equation}
    p_{ij} = \frac{k_i k_j}{N K},
\end{equation}
and $J_{ij}$ is a Gaussian random variable with
\begin{align}
    \mathrm{mean}(J_{ij}) &= 0,\\
    \mathrm{std}(J_{ij}) &= \frac{g}{\sqrt{K}},\\
    \mathrm{corr}(J_{ij}, J_{ji}) &= \gamma.
\end{align}
The scaling of the couplings with the average degree $K$ ensures that, on average, the sum in Eq.~(\ref{eq_app:RNN}) remains of order one.
The factorized form of $p_{ij}$ defines the ensemble considered here, that of uncorrelated networks, i.e.\ without degree--degree correlations, in which the statistics of $A_{ij}$ are determined solely by the degree sequence~\cite{park2024incorporating}. The average over $A$ below is exact for this ensemble; the approximation is made in Eq.~(\ref{eq:A16}), where the expansion is truncated at leading order in $1/K$.

We follow the prescription of generating functional analysis~\cite{sompolinsky1988chaos, crisanti2018path, helias2020statistical}, based on the MSRJD path integral~\cite{de1978dynamics, martin1973statistical, janssen1976lagrangean}, and generalized to account for the heterogeneity of the degree
distribution~\cite{park2024incorporating, poley2025interaction, aguirre2024heterogeneous}.

We define the generating functional as
\begin{equation}
    Z[\mathbf{\psi}] = \int \mathcal{D}\mathbf{x} \, \mathcal{D}\mathbf{\hat{x}} \,
    \exp \Bigg[ i \sum_i \int dt \, \hat{x}_i(t) \Big( \dot{x}_i(t) + x_i(t) - \sum_j A_{ij} J_{ij} \phi(x_j(t)) \Big)
    + i \sum_i \int dt \, x_i(t) \psi_i(t) \Bigg]  \prod_{i=1}^NP(x_i(0)),
\end{equation}
where we have introduced the source field $\mathbf{\psi}$, which will be set to zero at the end of the calculations. The measure $\mathcal{D}[\mathbf{x}, \hat{\mathbf{x}}]$ includes a Jacobian term that ensures the normalization condition $Z[\mathbf{\psi}=0] = 1$. Its
contribution does not affect the resulting stochastic single-neuron equations and can therefore be ignored in the derivation of the dynamics.

The next step is to compute the averages of the generating functional over the quenched disorder. In this case, two independent averages must be performed: one over the disorder of $A_{ij}$ and one over the disorder of $J_{ij}$. We start with the former. The term of the functional affected by the disorder is
\begin{equation}
    \left\langle \left\langle
    \exp\Big[-i \sum_i \int dt \, \hat{x}_i(t) \sum_j A_{ij} J_{ij} \phi(x_j(t)) \Big]
    \right\rangle_A \right\rangle_J.
\end{equation}
Since $A_{ij}$ takes values $0$ or $1$ with probability $p_{ij}$, and since the adjacency matrix is undirected, a single Bernoulli variable $A_{ij}=A_{ji}$ controls both directed edges of a pair, so that the average factorizes over unordered pairs $i<j$ (self-connections being excluded, $A_{ii}=0$), we have
\begin{align} \label{eq:A10}
    &\left\langle \left\langle
    \exp\Big[-i \sum_i \int dt \, \hat{x}_i(t) \sum_j A_{ij} J_{ij} \phi(x_j(t)) \Big]
    \right\rangle_A \right\rangle_J \notag \\
    &= \Big\langle \prod_{\text{$i<j$}} \Big[ (1-p_{ij}) + p_{ij}\,
    \text{}\,
    \text{$\exp\big(Q_{ij}\big)$} \Big] \Big\rangle_J \\
    &= \exp \Bigg( \sum_{i<j} \log \Big[ 1 + p_{ij} \left\langle \exp Q_{ij} - 1 \right\rangle_J \Big] \Bigg) \\
    &\approx \exp \Bigg( \sum_{i<j} p_{ij} \left\langle \exp Q_{ij} - 1 \right\rangle_J \Bigg),
\end{align}
where we have introduced the Gaussian quantity collecting the two directed edges of the pair,
\begin{equation}
    Q_{ij} = -i \int dt \, \hat{x}_i(t) J_{ij} \phi(x_j(t)) - i \int dt \, \hat{x}_j(t) J_{ji} \phi(x_i(t)),
\end{equation}
and used the approximation $\log(1+x) \approx x$.

We now insert $p_{ij} = k_i k_j / (N K)$, and perform the average over the disorder of $J_{ij}$, obtaining
\begin{align}
    &\exp \Bigg( \sum_{i<j} p_{ij} \left\langle \exp Q_{ij} - 1 \right\rangle_J \Bigg) \notag \\
    &= \exp \Bigg( \sum_{i<j} \frac{k_i k_j}{N K} \left[ \exp\Big( \langle Q_{ij} \rangle_J + \frac{1}{2} \langle\langle Q_{ij}^2 \rangle\rangle_J \Big) - 1 \right] \Bigg) \\
    &= \exp \Bigg( \sum_{i<j} \sum_{n=1}^\infty \frac{k_i k_j}{N K} \Big( \langle Q_{ij} \rangle_J + \frac{1}{2} \langle\langle Q_{ij}^2 \rangle\rangle_J \Big)^n \Bigg).
\end{align}

Keeping only the leading order in $K^{-1}$, we obtain

\begin{multline} \label{eq:A16}
\exp \Bigg( \sum_{i<j} p_{ij} \left\langle \exp Q_{ij} - 1 \right\rangle_J \Bigg) = \exp \Bigg( \sum_{i \neq j} \frac{k_i k_j}{N K} \Bigg[ \frac{g^2}{2K} \int dt \, dt' \, \hat{x}_i(t) \hat{x}_i(t') \phi(x_j(t)) \phi(x_j(t'))\\
+\frac{\gamma g^2}{2K} \int dt \, dt' \, \hat{x}_i(t) \phi(x_i(t')) \phi(x_j(t)) \hat{x}_j(t')
+ \mathcal{O}(K^{-2}) \Bigg] \Bigg).
\end{multline}

The neglected terms are the higher cumulants of the input $h_i=\sum_jA_{ij}J_{ij}\phi(x_j)$. Its second and fourth cumulants are $g^2(k_i/K)\,q$ and $3g^4k_i\,q_4/K^2$, with $q_4=\sum_j (k_j/NK)\,\phi(x_j)^4$, so that its excess kurtosis is $3q_4/(k_iq^2)=\mathcal{O}(1/k_i)$. The input is therefore Gaussian at leading order, with corrections that are largest for the lowest-degree neurons.

We continue the calculation by introducing the macroscopic order parameters
\begin{align}
    b(t,t') &= \frac{1}{N} \sum_i \frac{k_i}{K} \, \hat{x}_i(t) \hat{x}_i(t'),\\
    q(t,t') &= \frac{1}{N} \sum_i \frac{k_i}{K} \, \phi(x_i(t)) \phi(x_i(t')),\\
    g(t,t') &= \frac{1}{N} \sum_i \frac{k_i}{K} \, \hat{x}_i(t) \phi(x_i(t')),
\end{align}
which can be introduced into the generating functional together with their conjugate fields by exploiting the integral representation of the Dirac delta. For example, for the field $q(t,t')$ we have
\begin{align}
    1 &= \int \mathcal{D}[q] \prod_{t,t'} \delta \Bigg[ N q(t,t') - \sum_i \frac{k_i}{K} \phi(x_i(t)) \phi(x_i(t')) \Bigg] \\
      &= \int \mathcal{D}[q, \hat{q}] \exp \Bigg( i \int dt \, dt' \, \hat{q}(t,t') \Big[ N q(t,t') - \sum_i \frac{k_i}{K} \phi(x_i(t)) \phi(x_i(t')) \Big] \Bigg).
\end{align}

The disorder-averaged generating functional can then be written as
\begin{equation}
    \bar{Z}[\mathbf{\psi}] = \int \mathcal{D}[\Xi, \hat{\Xi}] \,
    \exp \Big( \Phi[\Xi, \hat{\Xi}] + \Psi[\Xi] + \log Z_0[\hat{\Xi}] \Big),
\end{equation}
where we have introduced the notation $\Xi = (b,q,g)$, $\hat{\Xi} = (\hat{b}, \hat{q}, \hat{g})$, and defined
\begin{align}
    \Phi &= N i \int dt \, dt' \, \Big[ \hat{b}(t, t') b(t, t') + \hat{q}(t, t') q(t, t') + \hat{g}(t, t') g(t, t') \Big],\\
    \Psi &= \frac{N i g^2}{2} \int dt \, dt' \, \Big[ b(t, t') q(t, t') + \gamma g(t, t') g(t', t) \Big],\\
    Z_0 &= \int \mathcal{D}\mathbf{x} \, \mathcal{D}\mathbf{\hat{x}} \,
    \exp \Bigg[ i \sum_i \int dt \, \hat{x}_i(t) \big( \dot{x}_i(t) + x_i(t) \big) + i \sum_i \int dt \, x_i(t) \psi_i(t) \Bigg] \notag \\
    &\quad \times \exp \Bigg[ -i \sum_i \frac{k_i}{K} \int dt \, dt' \, \Big( \hat{q}(t, t') \phi(x_i(t)) \phi(x_i(t')) + \hat{g}(t, t') \phi(x_i(t)) \hat{x}_i(t') + \hat{b}(t, t') \hat{x}_i(t) \hat{x}_i(t') \Big) \Bigg].
\end{align}

In the thermodynamic limit $N\rightarrow \infty$ we can evaluate the generating functional with the saddle-point method
\begin{equation}
    \bar{Z}[\mathbf{\psi}] = \exp \left(-N\mathcal{L}[\Xi^*, \hat{\Xi}^*]\right),
\end{equation}
where $[\Xi^*, \hat{\Xi}^*]$ satisfies the saddle-point equations $\nabla\mathcal{L}[\Xi, \hat{\Xi}]_{[\Xi^*, \hat{\Xi}^*]}=0$. In order to satisfy the optimization constraints the following conditions must hold
\begin{align}
   \hat{q}(t, t')+\frac{1}{2}g^2 b(t, t') &= 0,\\
   \hat{g}(t, t')+\gamma g^2 g(t, t') &= 0,\\
    \hat{b}(t, t')+\frac{1}{2}g^2 q(t, t') &= 0,\\
    q(t, t') - \frac{1}{N}\sum_i \frac{k_i}{K} \langle\phi(x_i(t))\phi(x_i(t')) \rangle_0 &=0,\\
    g(t,t') - \frac{1}{N}\sum_i \frac{k_i}{K} \langle\phi(x_i(t))\hat{x}_i(t') \rangle_0 &=0,\\
    b(t, t') - \frac{1}{N}\sum_i \frac{k_i}{K} \langle\hat{x}_i(t)\hat{x}_i(t') \rangle_0 &=0.
\end{align}

The average $\langle ...\rangle_0$ must be computed on the trajectories of the process itself. Due to the normalization condition $Z[\mathbf{\psi}=0] = 1$ the field $b(t,t')$ vanishes.
Substituting the saddle-point values into $Z_0$, the generating functional factorizes over neurons, each neuron interacting with the rest of the network only through the order parameters. The term quadratic in $\hat{x}_i$, with kernel $\tfrac{1}{2}g^2 (k_i/K)\,q(t,t')$, is the characteristic functional of a Gaussian noise with covariance proportional to $q(t,t')$, while the term coupling $\phi(x_i(t))$ and $\hat{x}_i(t')$ is a linear response to the neuron's own past activity, with kernel proportional to $g(t,t')$, which is identified with the response function $G(t,t')$. Reading the resulting single-neuron functional as an MSRJD integral gives the effective stochastic dynamics below; see Refs.~\cite{helias2020statistical, crisanti2018path} for a pedagogical derivation.
The resulting effective moment-generating functional is
\begin{multline}
    \bar{Z}_{\text{eff}}[\mathbf{\psi}]= \int \mathcal{D}[\mathbf{x}, \mathbf{\hat{x}}] \exp\left(i\sum_i \int dt \left[ \hat{x}_i(t) \left( \dot{x}_i(t) +x_i(t) \right)\right] +\right.\\
    \left. \sum_i \int dt dt' \left[ \frac{1}{2} g^2 q(t,t') \frac{k_i}{K}\hat{x}_i(t)\hat{x}_i(t') + \gamma g^2 g(t, t') \frac{k_i}{K} \phi(x_i(t))\hat{x}_i(t') \right]\right)  \prod_{i=1}^NP(x_i(0)),
\end{multline}
which is the generating functional of the effective dynamics
\begin{equation}
    \label{eq:app_DMFT}
    \dot{x}_k(t) = - x_k(t)+g\sqrt{\frac{k}{K}} \eta(t) +\gamma g^2 \frac{k}{K}\int_0^t \text{d}t' G(t, t') \phi (x_k(t')),
\end{equation}
for all $k$ in the support of $P(k)$. The Gaussian noise $\eta(t)$ must be computed self-consistently knowing that
\begin{equation}
    \langle \eta(t) \rangle = 0,
\end{equation}
and
\begin{equation}
    \langle \eta(t) \eta(t') \rangle  =\sum_k P(k) \frac{k}{K} \langle \phi(x_k(t)) \phi (x_k(t')) \rangle
\end{equation}
together with the response function
\begin{align}
    G(t, t') &= \sum_k P(k) \frac{k}{K} \left\langle \frac{\partial \phi(x_k(t))}{\partial h(t')}\right\rangle.
\end{align}

The averages $\langle \cdots \rangle$ must be computed over both the ensemble of disordered couplings $J$ and the ensemble of adjacency matrices $A$. Note that, in this derivation, all degree-degree correlations between nodes are neglected by construction of the ensemble, and the effective dynamics holds at leading order in $1/K$. We therefore assume that the average degree $K$ satisfies $1 \ll K \ll N$, i.e., it is much larger than one but much smaller than the total number of neurons $N$.

\section{REALIZED VERSUS TARGET DEGREE DISTRIBUTION}\label{app:realized_degree}

The degrees $k_i$ entering $p_{ij}=k_ik_j/(NK)$ are target values drawn from $P(k)$; the
degree actually realized in a given network is $\kappa_i=\sum_j A_{ij}$, a sum of independent
Bernoulli variables. Its generating function is
\begin{equation}
    \big\langle z^{\kappa_i}\big\rangle=\prod_j\big[1+p_{ij}(z-1)\big],
    \qquad
    \log\big\langle z^{\kappa_i}\big\rangle = k_i(z-1)
    - \frac{k_i^2\,\mathbb{E}[k^2]}{2NK^2}(z-1)^2 + \mathcal{O}(N^{-2}),
\end{equation}
where we used $\sum_j p_{ij}=k_i$ and $\sum_j p_{ij}^2 = k_i^2\,\mathbb{E}[k^2]/(NK^2)$. In the
thermodynamic limit $N\to\infty$ at fixed $K$ all terms beyond the first vanish, so that
\begin{equation}
    \kappa_i \,|\, k_i \;\sim\; \mathrm{Poisson}(k_i),
    \qquad
    P_{\mathrm{real}}(\kappa)=\sum_k P(k)\,\frac{e^{-k}k^{\kappa}}{\kappa!}.
\end{equation}
The mean degree is therefore reproduced exactly, $\mathbb{E}[\kappa]=\mathbb{E}[k]=K$, but the
realized distribution is a Poisson mixture of the target one, broadened by the sampling of the
edges. Its second moment is $\mathbb{E}[\kappa^2]=\mathbb{E}[k^2]+K$, so that
\begin{equation}
    \mathrm{CV}^2(\kappa) = \mathrm{CV}^2(k) + \frac{1}{K},
\end{equation}
and $P_{\mathrm{real}}\to P$ only for $K\gg1$, when the relative Poisson fluctuation
$k^{-1/2}$ becomes negligible over the support of $P(k)$. Equivalently, the ratio of moments
entering the critical condition~(\ref{eq:crit_cond_gamma0}) becomes
$\mathbb{E}[\kappa^2]/K^2=\mathbb{E}[k^2]/K^2+1/K$; for a homogeneous target degree
$P(k)=\delta_{k,K}$ this reproduces $\mathbb{E}[\kappa^2]=K+K^2$ and hence
Eq.~(\ref{eq:gc_ER}), showing that the Erd\H{o}s--R\'enyi correction discussed in the main text
is precisely this finite-$K$ sampling effect. The same effect accounts for the deviation between
the sampled degrees and the target lognormal distribution visible in
Fig.~\ref{fig:panel1}b.
\section{LINEAR STABILITY ANALYSIS}\label{app:LSA}
We investigate the stability of the fixed-point solution of Eq.~(\ref{eq:stat_DMFT}) via a linear stability analysis, which has proven to be a robust method for DMFT equations with structured interactions~\cite{poley2025interaction, park2024incorporating}. We begin by adding an external Gaussian input, $h_k(t)$, with zero mean to the DMFT equation for the effective neuron with degree $k$:
\begin{equation}
     \dot{x}_k(t) = - x_k(t) + g \sqrt{\frac{k}{K}} \, \eta(t) + \gamma g^2 \frac{k}{K} \int_0^t \mathrm{d}t' \, G(t, t') \, \phi(x_k(t')) + h_k(t),
\end{equation}
and linearize the dynamics around a fixed point $x^*$. Considering small perturbations, we write
\begin{align}
    x_k(t) &= x^* + \delta x_k(t), \\
    \eta(t) &= \eta^* + \delta \eta(t), \\
    h_k(t) &= \langle h_k \rangle + \delta h_k(t) = \delta h_k(t),
\end{align}
where the mean of the external input vanishes by assumption.

The dynamics linearized around the fixed point is
\begin{align}
    \delta \dot{x}_k(t) &= -x_k^* + g \sqrt{\frac{k}{K}} \, \eta^* + \gamma g^2 \chi x_k^* \notag \\
    &\quad - \delta x_k(t) + g \sqrt{\frac{k}{K}} \, \delta \eta(t)
    + \gamma g^2 \frac{k}{K} \int dt' \, G(t, t') \, \delta \phi(x_k(t'))
    + \delta h_k(t) \notag \\
    &\quad + \gamma g^2 \frac{k}{K} \int dt' \, \delta G(t, t') \, \phi(x_k(t)).
\end{align}

The first line on the right-hand side satisfies the fixed-point condition, Eq.~(\ref{eq:stat_DMFT}), and therefore vanishes. The third term depends on $\delta G(t, t')$, which, from its definition, can be expressed as the average of an unperturbed observable multiplied by $\delta h_k$. Since these two quantities are independent and $\langle h_k \rangle = 0$, we have $\delta G(t, t') = 0$, and thus this term vanishes.

The resulting linearized equation is
\begin{align}
    \delta \dot{x}_k(t) &= - \delta x_k(t) + g \sqrt{\frac{k}{K}} \, \delta \eta(t)
    + \gamma g^2 \frac{k}{K} \int dt' \, G(t, t') \, \delta \phi(x_k(t))
    + \delta h_k(t),
    \label{eq:app_linearized}
\end{align}
where $\delta \eta(t)$ is the self-consistent noise term.

Given the self-consistent definition of the noise correlations,
\begin{equation}
    \langle \eta(t) \eta(t') \rangle = \sum_k P(k) \frac{k}{K} \langle \phi(x_k(t)) \, \phi(x_k(t')) \rangle,
\end{equation}
we can linearize to obtain
\begin{equation}
    \label{eq:app_corr_lin_noise}
    \langle \delta \eta(t) \, \delta \eta (t') \rangle = \sum_k P(k) \frac{k}{K} \langle (\phi_k'^*)^2 \, \delta x_k(t) \, \delta x_k(t') \rangle,
\end{equation}
where we have defined $\phi_k'^* = \partial_x \phi(x_k^*)$.

Applying the Fourier transform to Eq.~(\ref{eq:app_linearized}), we obtain
\begin{equation}
    i \omega \, \delta \tilde{x}_k(\omega) = - \delta \tilde{x}_k(\omega)
    + \frac{k}{K} \, \gamma g^2 \, \phi_k'^* \, \tilde{G}(\omega) \, \delta \tilde{x}_k(\omega)
    + g \sqrt{\frac{k}{K}} \, \delta \tilde{\eta}(\omega)
    + \delta \tilde{h}_k(\omega),
    \label{eq:app_FT_lin_DMFT}
\end{equation}
where the external perturbation $\delta \tilde{h}_k(\omega)$ satisfies
\begin{equation}
    \langle \delta \tilde{h}_k(\omega) \, \delta \tilde{h}_k (\omega') \rangle = 2 \pi \, \epsilon_k^2 \, \delta(\omega + \omega'),
    \label{eq:app_corr_pert}
\end{equation}
since it represents uncorrelated Gaussian noise.

Solving Eq.~(\ref{eq:app_FT_lin_DMFT}) for $\delta \tilde{x}_k(\omega)$, we obtain
\begin{equation}
    \delta \tilde{x}_k(\omega) = \frac{1}{1 + i \omega - \gamma g^2 \frac{k}{K} \tilde{G}(\omega) \phi_k'^*}
    \left[ g \sqrt{\frac{k}{K}} \, \delta \tilde{\eta}(\omega) + \delta \tilde{h}_k(\omega) \right].
\end{equation}
Since the perturbation $h_k(t)$ is assumed to be an uncorrelated Gaussian noise, the following condition must be satisfied:
\begin{equation}
    \langle \delta \tilde{x}_k (\omega) \, \delta \tilde{x}_k(\omega') \rangle = 2 \pi \, \epsilon_k^2 \, \delta(\omega + \omega') \, \delta \tilde{C}_k(\omega).
    \label{eq:app_corr_dyn}
\end{equation}

Since we are interested in the long-time response, we set $\omega = \omega' = 0$ and define $\delta \tilde{C}_k(\omega=0) = \delta \tilde{C}_k$. The amplitude $\delta \tilde{C}_k$ can be obtained from Eqs.~(\ref{eq:app_FT_lin_DMFT}), (\ref{eq:app_corr_lin_noise}), (\ref{eq:app_corr_pert}), and (\ref{eq:app_corr_dyn}), yielding
\begin{equation}
    \delta \tilde{C}_k = \Bigg\langle
    \frac{1}{\big(1 - \gamma g^2 \frac{k}{K} \chi \, \phi_k'^*\big)^2}
    \Bigg[ g^2 \frac{k}{K} \sum_{k'} P(k') \frac{k'}{K} (\phi_{k'}'^*)^2 \, \delta \tilde{C}_{k'} + 1 \Bigg]
    \Bigg\rangle,
    \label{eq:app_eig_equatio}
\end{equation}
for every $k$ in the support of $P(k)$, assuming that the amplitude of the perturbations $\epsilon_k$ is the same for all $k$. Here, we have defined $\chi = \tilde{G}(\omega=0)$, and the average $\langle \cdots \rangle$ is taken over the realizations of the stochastic process $x^*$.

Eq.~(\ref{eq:app_eig_equatio}) is an eigenvalue equation and can be solved using standard linear algebra methods. Alternatively, one can sum Eq.~(\ref{eq:app_eig_equatio}) over all $k$, weighting the sum by $P(k)\frac{k}{K}(\phi_k'^*)^2$, and study the global amplitude of the network,
\begin{equation}
    \delta \tilde{C} = \sum_k P(k) \frac{k}{K} (\phi_k'^*)^2 \, \delta \tilde{C}_k.
\end{equation}
This global amplitude is found to be
\begin{equation}
    \delta \tilde{C} = \frac{\sum_k P(k) \frac{k}{K} \left\langle \left( \frac{1}{1 - \gamma g^2 \phi_k'^* \chi \frac{k}{K}} \right)^2 \right\rangle}{1 -g^2 \sum_k P(k) \frac{k^2}{K^2} \left\langle (\phi_k'^*)^2  \left( \frac{1}{1 - \gamma g^2 \phi_k'^* \chi \frac{k}{K}} \right)^2 \right\rangle},
\end{equation}
which diverges when the following critical condition is met:
\begin{equation}
    1 = g^2 \sum_k P(k) \frac{k^2}{K^2} \left\langle (\phi_k'^*)^2 \left( \frac{1}{1 - \gamma g^2 \chi \frac{k}{K} \phi_k'^*} \right)^2 \right\rangle.
\end{equation}

\section{Numerical Investigation of the Glassy Phase: Lyapunov Exponents}
\label{app:lyapunov}

To characterize the dynamics of networks with partial symmetry $\gamma$, we numerically compute their Lyapunov exponents using the Jacobian method \cite{engelken2023lyapunov}.

\begin{figure}[H]
    \centering
    \includegraphics[width = 0.8\linewidth]{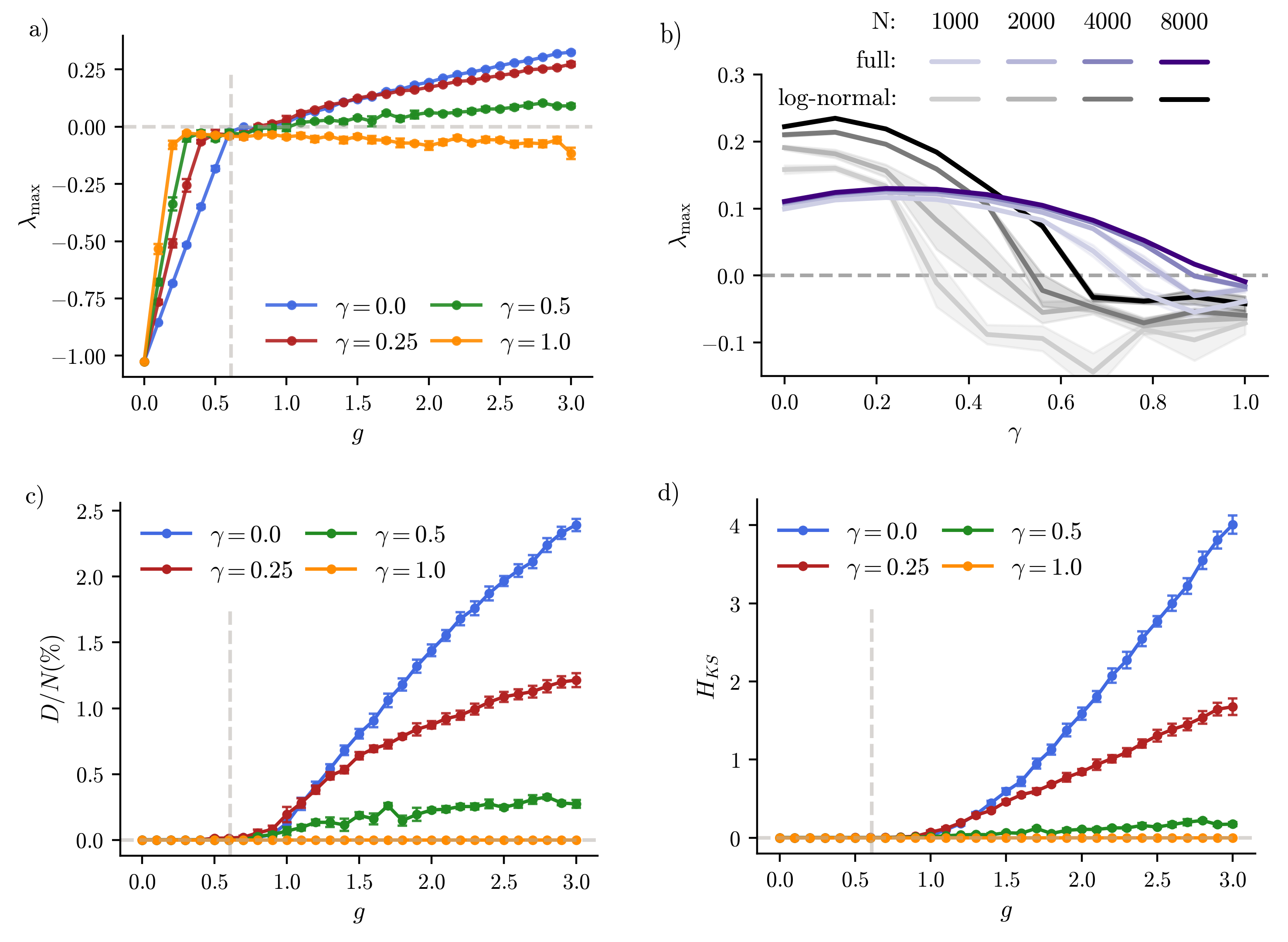}
    \caption{\textbf{Numerical analysis of Lyapunov exponents for a fully connected network and a network with lognormal degree distribution as a function of the partial symmetry parameter $\gamma$}. (a) Largest Lyapunov exponent as a function of $g$ for a network with lognormal degree distribution ($\mu = 3$, $\sigma = 1$) for different values of $\gamma$.
    (b) Comparison of the largest Lyapunov exponent for a fully connected network and a network with lognormal degree distribution as in (a), as a function of $\gamma$, shown for different network sizes $N$.
    (c) Normalized Lyapunov dimension computed using the Kaplan--Yorke formula for a network as in (a), as a function of $g$ for different values of $\gamma$.
    (d) Kolmogorov--Sinai entropy calculated from the Lyapunov exponents for a network as in (a), as a function of $g$ for different values of $\gamma$.}
    \label{fig:supp_lyapunov_exponents}
\end{figure}

\begin{figure}[H]
    \centering
    \includegraphics[width=0.8\linewidth]{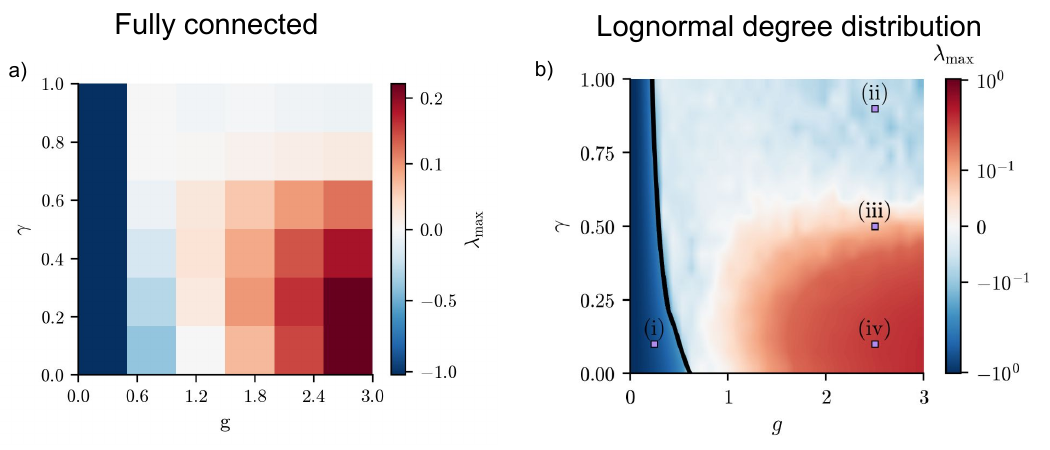}
    \caption{\textbf{Numerical phase diagrams from the largest Lyapunov exponent.} Global largest Lyapunov exponent $\lambda_{\max}$ in the $(g,\gamma)$ plane for (a) a fully connected network and (b) a network with a lognormal degree distribution ($\mu=3$, $\sigma=1$). In (b) the black line is the critical line, Eqs.~(\ref{eq:crit_cond_silent}) and~(\ref{eq:chi}), at which the silent phase loses linear stability, and the color scale is logarithmic away from zero. The markers (i)--(iv) indicate the parameters of the representative dynamics shown in Fig.~\ref{fig:phase_diagram_gamma}a. Simulations use $dt=0.05$, $N=3000$ and $t_{\mathrm{sim}}=3000$; $\lambda_{\max}$ is averaged over 5 network realizations.}
    \label{fig:supp_phase_diagrams}
\end{figure}

The Lyapunov exponents quantify the sensitivity of the network dynamics to infinitesimal perturbations in the state space. In particular, the largest Lyapunov exponent (LLE) indicates whether small perturbations grow or decay over time: a positive LLE signals chaotic dynamics, while a negative LLE corresponds to stable, convergent behavior. By computing the full spectrum of Lyapunov exponents, we can further estimate the Kolmogorov-Sinai (KS) entropy, which measures the rate of information production in the system, and the Lyapunov dimension (or Kaplan-Yorke dimension), which characterizes the effective number of degrees of freedom that contribute to the chaotic dynamics.
In Fig.~\ref{fig:supp_lyapunov_exponents}, we summarize the numerical analysis of networks with lognormal degree distributions and compare them to fully connected networks. Panel (a) shows the largest Lyapunov exponent as a function of the coupling strength $g$ for networks with lognormal degree distribution ($\mu=3$, $\sigma=1$) and different values of the symmetry parameter $\gamma$. Panel (b) directly compares the LLE of fully connected networks with networks with lognormal degree distributions as $\gamma$ varies. Panel (c) displays the normalized Lyapunov dimension computed using the Kaplan-Yorke formula, providing insight into the dimensionality of the attractor associated with each network. Finally, panel (d) shows the Kolmogorov-Sinai entropy calculated from the Lyapunov spectrum, reflecting the overall complexity and unpredictability of the network dynamics.
Figure~\ref{fig:supp_phase_diagrams} summarizes the numerical phase diagrams obtained from the global largest Lyapunov exponent $\lambda_{\max}$, computed along trajectories with the Jacobian method, in the $(g,\gamma)$ plane. Panel (a) shows the phase diagram of a fully connected network, highlighting the transition from stable to chaotic dynamics as the gain increases and the suppression of chaos as the symmetry parameter $\gamma$ grows. Panel (b) extends this analysis to a network with a lognormal degree distribution, together with the critical line obtained from Eqs.~(\ref{eq:crit_cond_silent}) and~(\ref{eq:chi}), at which the silent phase loses stability. In the heterogeneous network, chaotic dynamics ($\lambda_{\max}>0$) is observed at small $\gamma$ and large $g$, while for large $\gamma$ the exponent becomes negative, signaling the multistable phase also identified by the replica analysis of Fig.~\ref{fig:phase_diagram_gamma}c. Between the critical line and the chaotic region, $\lambda_{\max}$ is close to zero, since near the transition it grows slowly with $g-g_c$ (Fig.~\ref{fig:supp_lyapunov_exponents}a) and cannot be resolved for finite $N$ and simulation time. The markers (i)--(iv) indicate the parameters of the representative neuronal traces and stationary distributions shown in Fig.~\ref{fig:phase_diagram_gamma}a, where high-degree nodes exhibit bistable activity. These results collectively reveal how partial symmetry and network heterogeneity influence the emergence of chaotic and glassy dynamics in recurrent networks.


\section{NUMERICAL SOLUTION OF THE DYNAMICAL MEAN FIELD THEORY}\label{app:num_DMFT}

We analyze the temporal properties at the single-node level by solving the hDMFT equations (\ref{eq:DMFT})--(\ref{eq:response}) for a network with a heterogeneous degree distribution $k \in \{100, 1000\}$, occurring with probabilities $P_k = \{0.9, 0.1\}$.

A general and detailed procedure for numerically solving the Dynamical Mean Field Theory (DMFT) equations can be found in Ref.~\cite{roy2019numerical}. The basic idea is to proceed iteratively: one initializes the observables randomly, runs the dynamics, and then uses the resulting trajectories to update the observables with a soft injection to ensure convergence. The code for the solution can be found in the public repository \cite{myrepo}.

\begin{algorithm}[H]
\caption{Numerical solution of the heterogeneous DMFT equations \eqref{eq:app_DMFT}}
\label{alg:dmft_general}
\begin{algorithmic}[1]
\State \textbf{Input:} Network parameters: degree distribution $P(k)$, average degree $K$, coupling strength $g$, partial symmetry $\gamma$, activation function $\phi$, time step $dt$, total simulation time $T_{\rm sim}$
\State Initialize self-consistent fields:
\Statex \quad Gaussian noise $\eta(t) = 0$, response function $G(t,t') = 0$, autocorrelations $C_k(t,t') = 0$ for all $k$
\State Set convergence tolerance $\epsilon$ and maximum number of iterations $N_{\rm iter}$
\For{$n = 1$ to $N_{\rm iter}$}
    \For{each degree $k$ in the support of $P(k)$}
        \State Simulate the effective dynamics for neuron $k$:
        \[
            \dot{x}_k(t) = -x_k(t) + g \sqrt{\frac{k}{K}} \eta(t)
            + \gamma g^2 \frac{k}{K} \int_0^t dt' \, G(t,t') \phi(x_k(t'))
        \]
        using the current estimates of $\eta(t)$ and $G(t,t')$
        \State Compute the activation $\phi(x_k(t))$ along the trajectory
    \EndFor
    \State Update the Gaussian noise correlation self-consistently:
    \[
        \langle \eta(t) \eta(t') \rangle = \sum_k P(k) \frac{k}{K} \langle \phi(x_k(t)) \phi(x_k(t')) \rangle
    \]
    \State Update the response function:
    \[
        G(t,t') = \sum_k P(k) \frac{k}{K} \left\langle \frac{\partial \phi(x_k(t))}{\partial h(t')} \right\rangle
    \]
    \State Check convergence of $\eta(t)$, $G(t,t')$, and $C_k(t,t')$
    \If{changes $< \epsilon$}
        \State \textbf{break}
    \EndIf
\EndFor
\State \textbf{Output:} Self-consistent trajectories $x_k(t)$, autocorrelations $C_k(t,t')$, response function $G(t,t')$, and Gaussian noise $\eta(t)$
\Statex These outputs can be further used to compute Lyapunov exponents, KS entropy, or other observables.
\end{algorithmic}
\end{algorithm}

We note in particular that convergence requires substantial computational resources in the case of partial symmetry $\gamma \neq 0$, as estimating the response function is extremely demanding. We therefore solve this case for a small values of $\gamma$ up to $0.25$, exploiting the Novikov theorem to estimate the kernel $G(t, t')$ (see Ref.~\cite{roy2019numerical} for details). The computational complexity in this case scales as $\mathcal{O}(N_\text{steps}^3)$.

In this section, we first present the numerical solution of the hDMFT equations for a network gain $g = 3$, comparing the cases $\gamma = 0$ and $\gamma = 0.1$. For each degree $k$, we use $N = 10^4$ replicas of the effective system. We then analyze and compare the intrinsic timescales arising in the network with a discrete degree distribution.
Finally, we test the effective self-coupling approximation of the main text, $G(\tau)\simeq\chi\,\delta(\tau)$, by solving the hDMFT with memory kernels of decreasing temporal extent and comparing the results with the full solution and with direct simulations of the network.

\textbf{Results for $\gamma = 0$}\\
In the asymmetric case $\gamma = 0$, there is no need to estimate the response function $G(t, t')$. The system behaves as described in Fig.~\ref{fig:panel1} of the main text, exhibiting a transition from a trivial fixed point to chaos.
In this case, the stationary distribution is unimodal, and the autocorrelation timescale is not influenced by the effective self-coupling term that arises in the presence of partial symmetry.

\begin{figure}[h]
    \centering
    \includegraphics[width=0.9\linewidth]{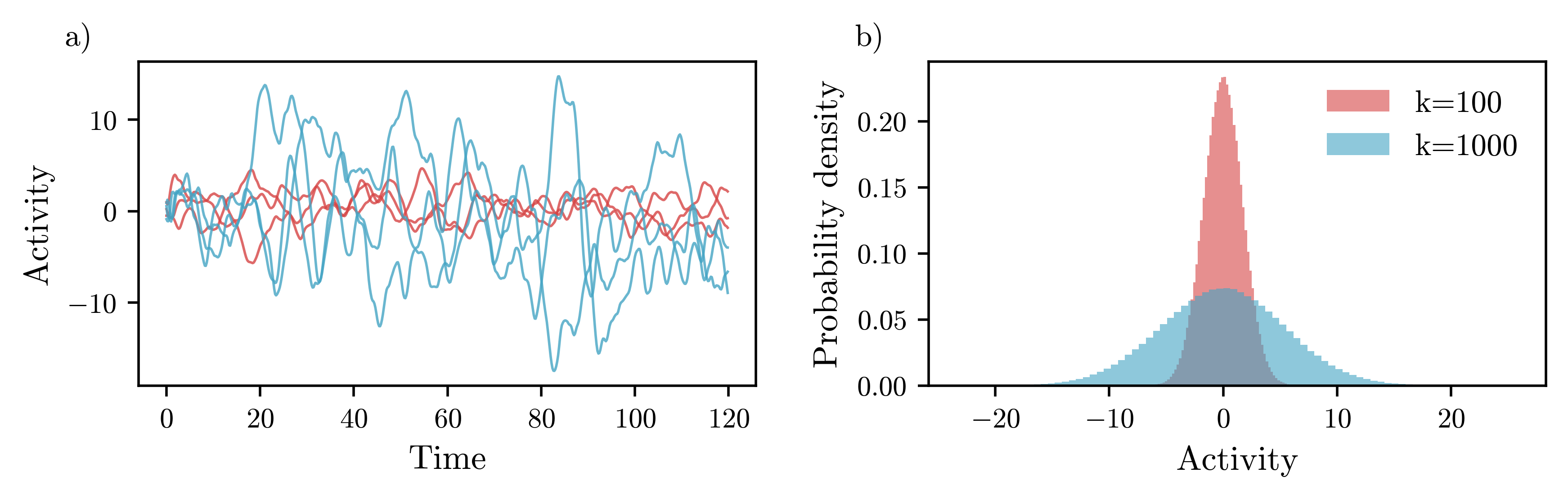}
    \caption{DMFT of a recurrent neural network with $g = 3$, $\gamma = 0$, $k = \{100, 1000\}$, and $P(k) = \{0.9, 0.1\}$. The DMFT equations have been solved using the methods described in Ref.~\cite{roy2019numerical}, with $N = 10^4$ replicas of the units. (a) Example neural activity trajectories, color-coded by the degree of the node. (b) Stationary distributions for the two populations of units with different degrees.}
    \label{fig:study_case_trajectories_gamma0}
\end{figure}

\begin{figure}[h]
    \centering
    \includegraphics[width=0.9\linewidth]{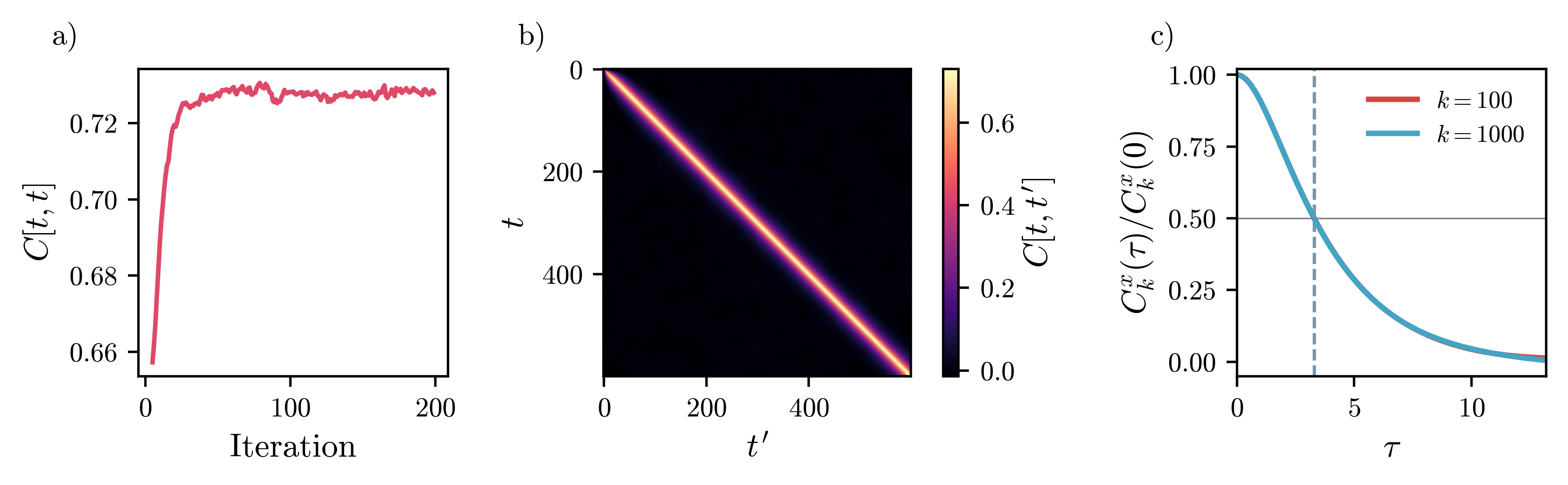}
    \caption{Numerical solution of the DMFT equations for the observables of a recurrent neural network with $g = 3$, $\gamma = 0$, $k = \{100, 1000\}$, and $P(k) = \{0.9, 0.1\}$. The DMFT equations have been solved using the methods described in Ref.~\cite{roy2019numerical}, with $N = 10^4$ replicas of the units.
    (a) Convergence of the variance $C(t, t)$ with the number of iterations.
    (b) Final matrix estimating the noise correlator $C(t, t')$.
    (c) Normalized autocorrelation of $x$ for units with degree $k$, estimated over the stationary second half of the trajectories; dashed vertical lines mark the half width at half maximum (HWHM).
}
    \label{fig:case_study_gamma0}
\end{figure}

\textbf{Results for $\gamma = 0.1$}\\
In the case of partially symmetric couplings, in addition to the noise correlator, it is also necessary to estimate the response kernel $G(t, t')$.

\begin{figure}[h]
    \centering
    \includegraphics[width=0.9\linewidth]{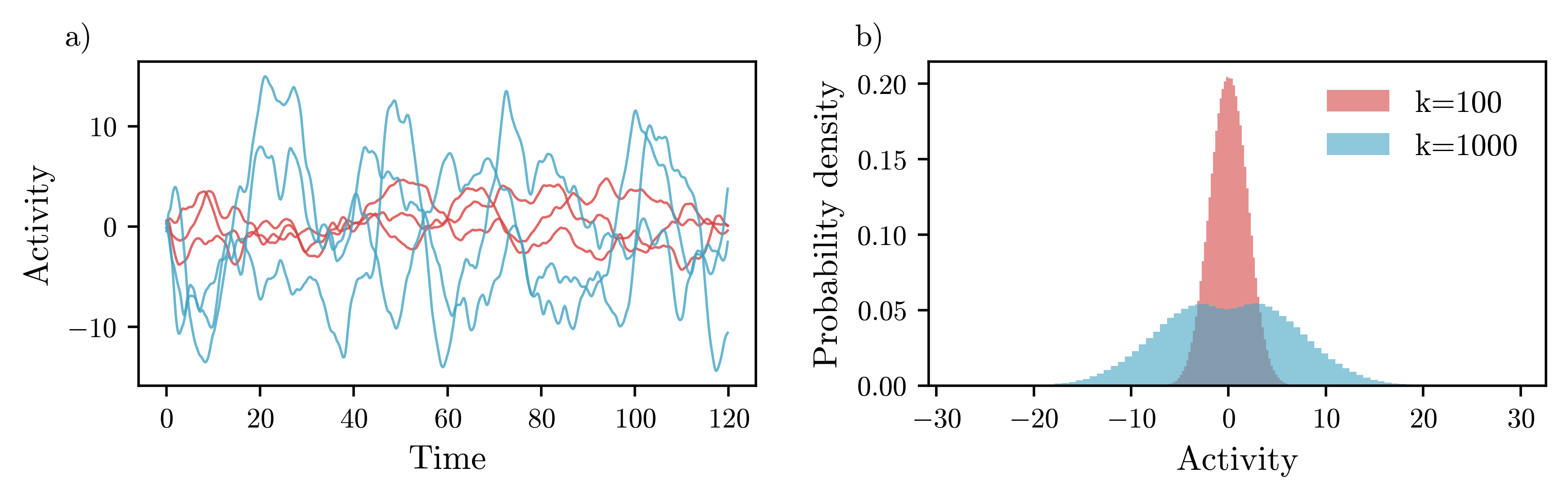}
    \caption{DMFT of a recurrent neural network with $g = 3$, $\gamma = 0.1$, $k = \{100, 1000\}$, and $P(k) = \{0.9, 0.1\}$. The DMFT equations have been solved using the methods described in Ref.~\cite{roy2019numerical}, with $N = 10^4$ replicas of the units. (a) Example neural activity trajectories, color-coded by the degree of the node. (b) Stationary distributions for the two populations of units with different degrees.}
    \label{fig:case_study_trajectories}
\end{figure}

At this value of $\gamma$ the stationary distribution of hubs is much broader than that of low-degree neurons, but it is still unimodal (Fig.~\ref{fig:case_study_trajectories}b); bimodal distributions appear at larger values of $\gamma$ (Appendix~\ref{app:bistability}).

We note in particular that the typical decay timescale of the memory kernel $G(\tau)$ is much shorter than that of the autocorrelations $C(\tau)$.

\begin{figure}[h]
    \centering
    \includegraphics[width=0.9\linewidth]{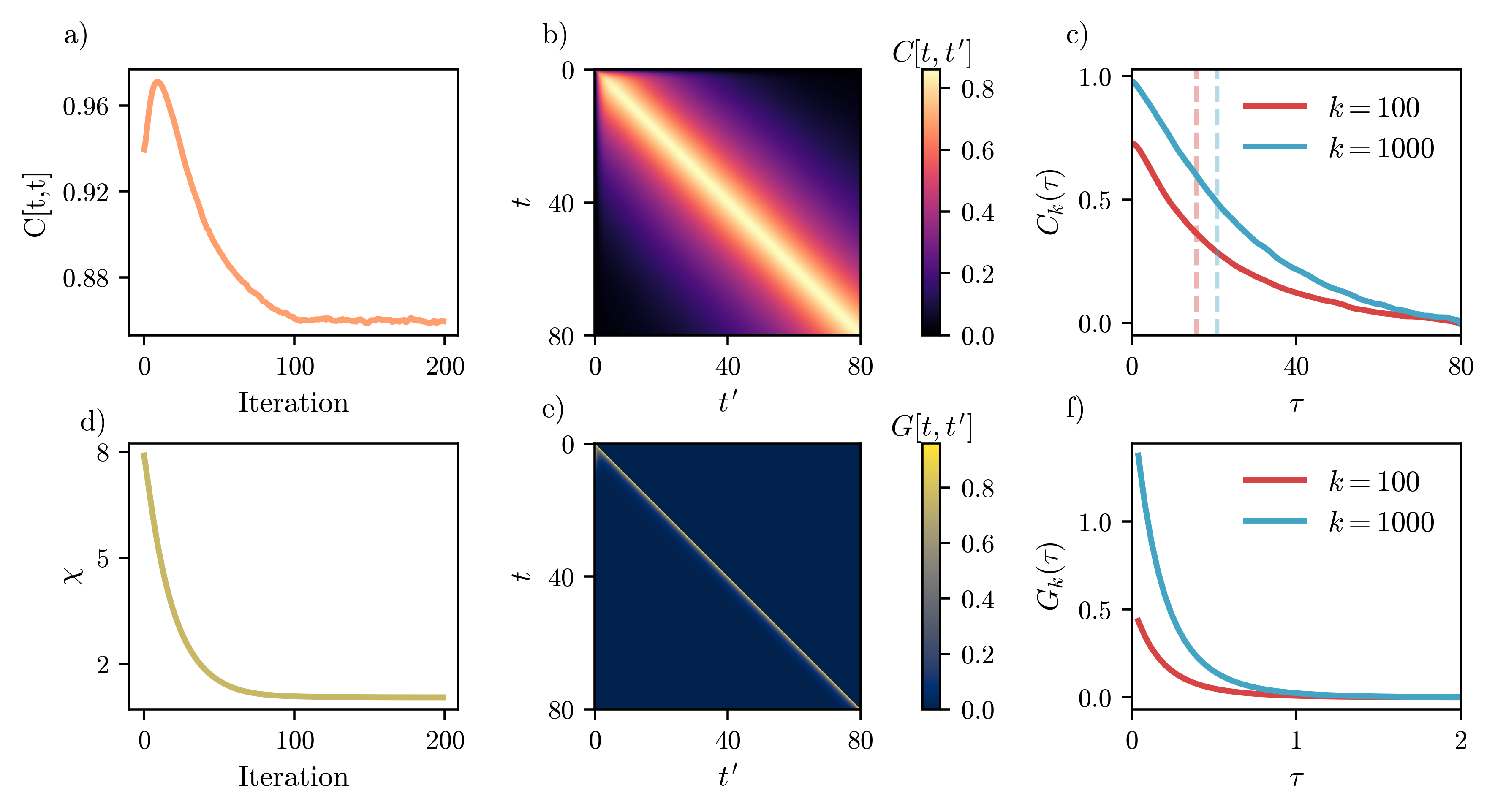}  \caption{Numerical solution of the DMFT equations for the observables of a recurrent neural network with $g = 3$, $\gamma = 0.1$, $k = \{100, 1000\}$, and $P(k) = \{0.9, 0.1\}$. The DMFT equations have been solved using the methods described in Ref.~\cite{roy2019numerical}, with $N = 10^4$ replicas of the units.
(a) Convergence of the variance $C(t, t)$ with the number of iterations.
(b) Final matrix estimating the noise correlator $C(t, t')$.
(c) Normalized autocorrelation of $x$ for units with degree $k$, estimated over the stationary second half of the trajectories; dashed vertical lines mark the HWHM.
(d) Convergence of the observable $\chi = \int d\tau\, G(\tau)$ with the number of iterations.
(e) Final matrix estimating the response kernel $G(t, t')$ (lower triangular due to causality).
(f) Response kernel of units with degree $k$.}
    \label{fig:case_study}
\end{figure}

\textbf{Temporal scales in RNN with discrete degree distribution}\\

\begin{figure}{h}
    \centering
\includegraphics[width=0.9\linewidth]{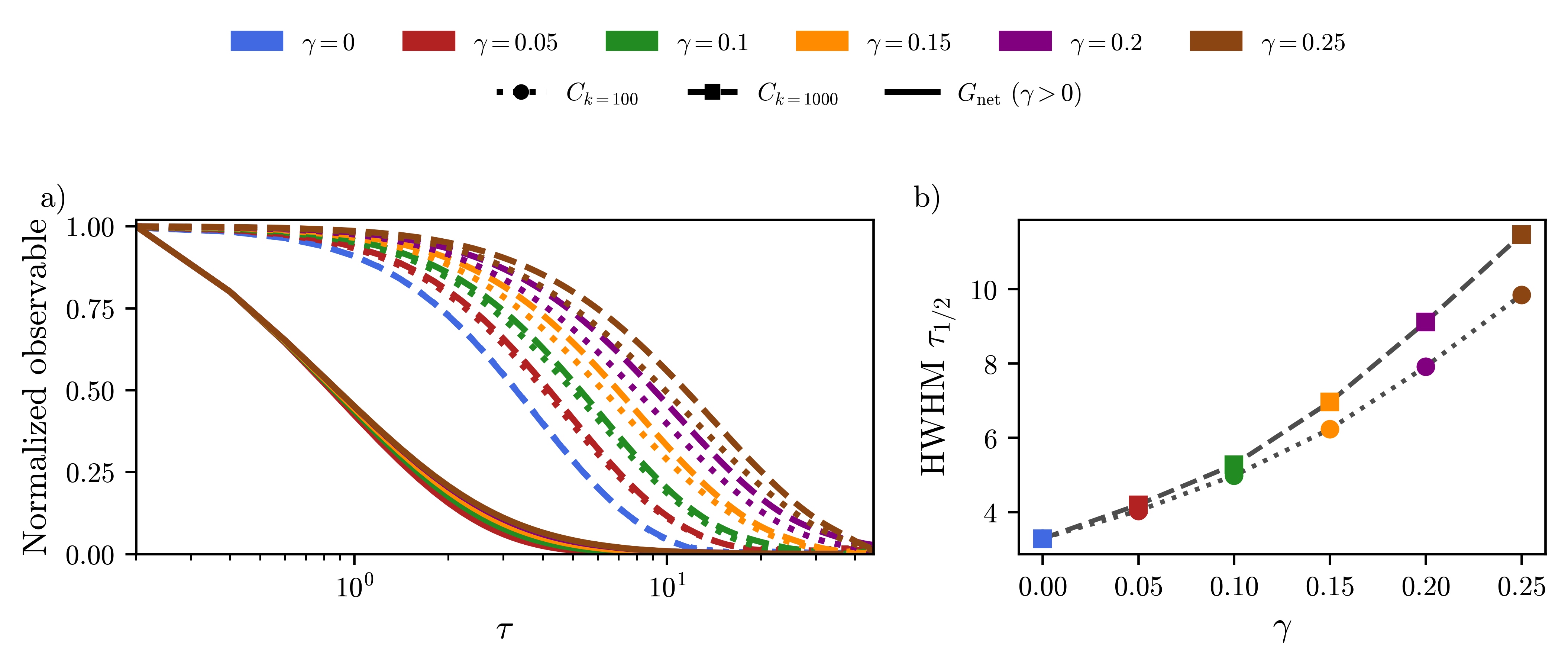}
    \caption{\textbf{Timescales in a network with a discrete degree distribution.}
 Recurrent neural network with $g = 3$, $k = \{100, 1000\}$, and $P(k) = \{0.9, 0.1\}$, for $\gamma = 0, 0.05, 0.1, 0.15, 0.2, 0.25$ (colours). (a) Normalized autocorrelation of $x$ for neurons with $k=100$ (dotted) and $k=1000$ (dashed), and normalized memory kernel $G(\tau)$ of the network for each $\gamma>0$ (solid). The horizontal axis is shown on a logarithmic scale. (b) Half width at half maximum (HWHM) of the autocorrelation as a function of $\gamma$, for $k=100$ (circles) and $k=1000$ (squares).}
    \label{fig:DMFT_solution}
\end{figure}

The neuronal-level timescales of low-degree neurons and hubs are shown in Fig.~\ref{fig:DMFT_solution} for $\gamma = 0, 0.05, 0.1, 0.15, 0.2, 0.25$. In the case of uncorrelated synapses ($\gamma = 0$), the two degree classes have the same timescale (HWHM $\simeq 3.3$ for both) and autocorrelation times are significantly shorter than in the presence of partial symmetry. Increasing $\gamma$ activates the effective self-coupling term, which slows down the dynamics of all units and separates the two classes: between $\gamma=0$ and $\gamma=0.25$ the HWHM grows from $3.3$ to $11.5$ for hubs and from $3.3$ to $9.8$ for low-degree neurons, so that the ratio between the two increases from $1.00$ to $1.17$ (Fig.~\ref{fig:DMFT_solution}b).

Fig.~\ref{fig:DMFT_solution}a also shows the response kernel $G(\tau)$ of the network for each $\gamma>0$. It decays on a timescale of order one (HWHM $\simeq 1$), shorter than that of the autocorrelations. This motivates the approximation $G(\tau) \simeq \chi\,\delta(\tau)$ discussed in the main text, whose accuracy is tested below.

\textbf{Effective self-coupling and finite memory kernel}\\

\begin{figure}[H]
    \centering
    \includegraphics[width=1\linewidth]{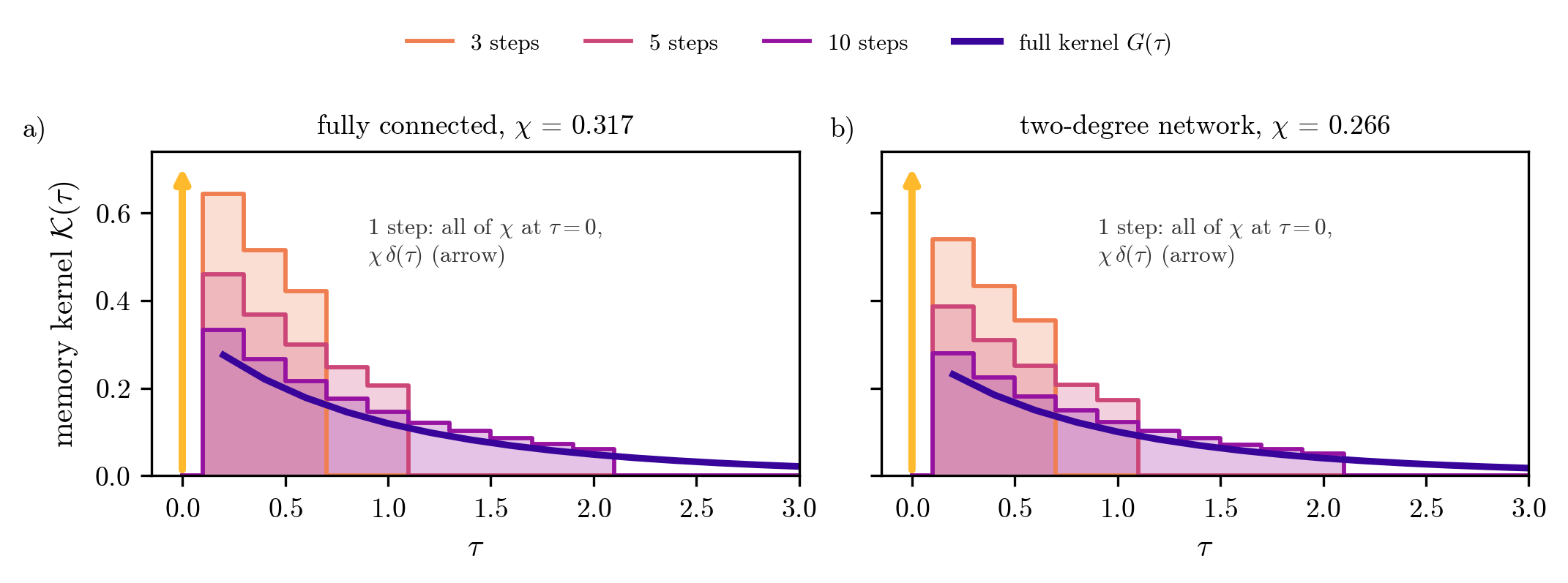}
    \caption{\textbf{Memory kernels.} Kernels used in the memory term of the hDMFT at $g=3$, $\gamma=0.1$, for (a) a fully connected network and (b) the network with $k = \{100, 1000\}$, $P(k) = \{0.9, 0.1\}$. Solid line: full response kernel $G(\tau)$. Shaded steps: $G(\tau)$ restricted to its first 3, 5 and 10 time steps ($\Delta t=0.2$) and rescaled to the same integral $\chi$. The arrow marks the instantaneous kernel $\chi\,\delta(\tau)$ (1 step), which corresponds to the effective self-coupling $s_k=\gamma g^2\chi k/K$.}
    \label{fig:kernels_memory}
\end{figure}

In the main text, the memory term of Eq.~(\ref{eq:DMFT}) is approximated by an instantaneous self-coupling, $G(\tau)\simeq\chi\,\delta(\tau)$, which gives $s_k=\gamma g^2\chi\,k/K$ [Eq.~(\ref{eq:hDMFT})]. For the network above at $\gamma=0.1$, $\chi\simeq0.27$, so that $s_1\simeq0.13$ for $k=100$ and $s_2\simeq1.3$ for $k=1000$. To test this approximation, we solve the stationary hDMFT self-consistently with the same degree-dependent noise, replacing $G(\tau)$ in the memory term by kernels of decreasing temporal extent: $G(\tau)$ restricted to its first $n=10, 5, 3$ time steps ($\Delta t = 0.2$) and rescaled so that its integral remains equal to $\chi$, and, for $n=1$, the instantaneous kernel $\chi\,\delta(\tau)$ (Fig.~\ref{fig:kernels_memory}). All kernels therefore carry the same total feedback and differ only in how it is distributed in time. We do this both for the network with $k \in \{100, 1000\}$ and for a fully connected network with the same $g$ and $\gamma$.

\begin{figure}[H]
    \centering
    \includegraphics[width=1\linewidth]{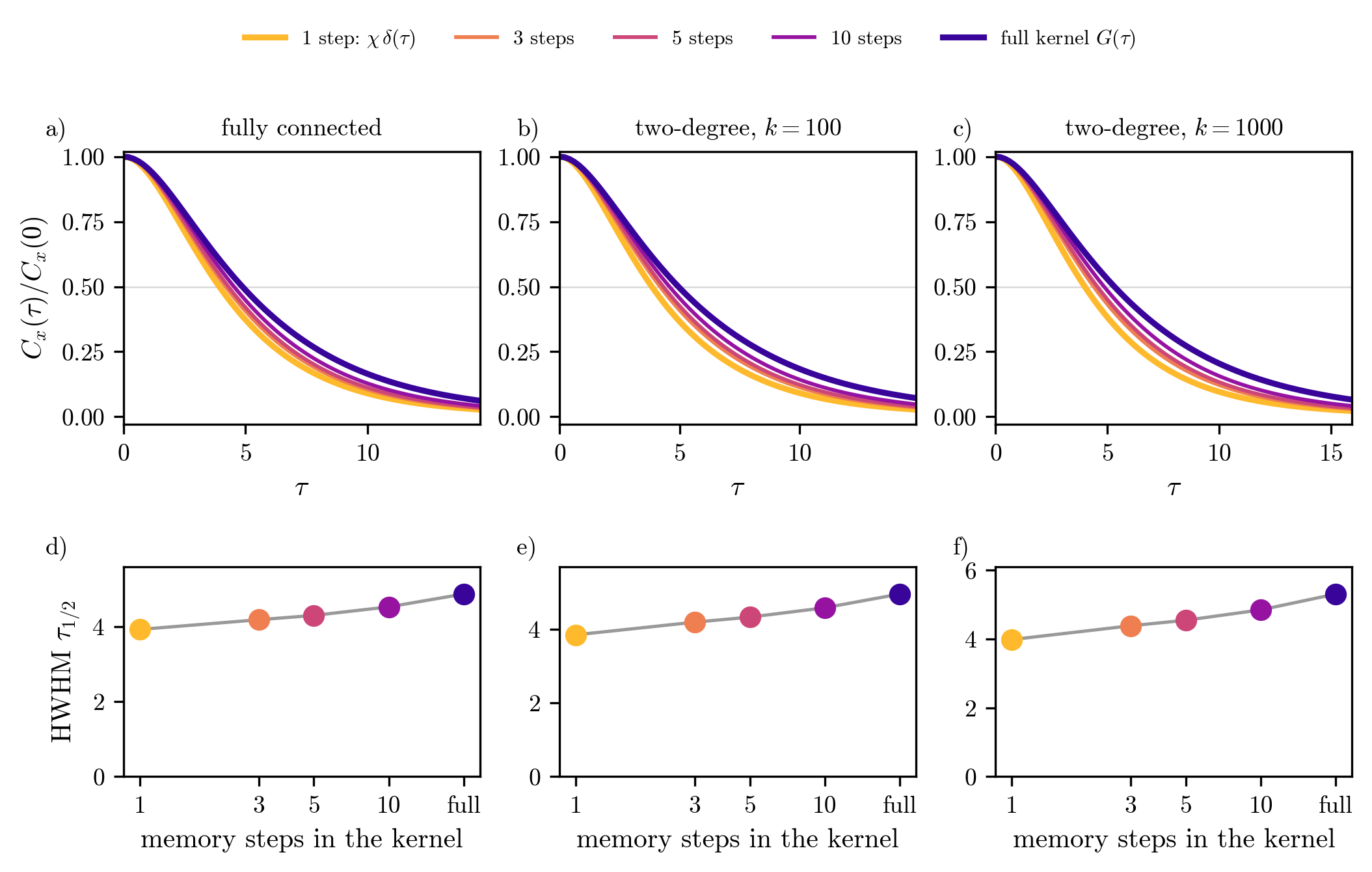}
    \caption{\textbf{Autocorrelations as a function of the memory kernel.} Self-consistent hDMFT solution at $g=3$, $\gamma=0.1$, with the memory kernels of Fig.~\ref{fig:kernels_memory}: 1 step ($\chi\,\delta(\tau)$), 3, 5 and 10 steps, and full $G(\tau)$ (colours). (a) Normalized autocorrelation of $x$ for a fully connected network; (b,c) the same for the network with $k = \{100, 1000\}$, $P(k) = \{0.9, 0.1\}$, for units with $k=100$ (b) and $k=1000$ (c). (d--f) Corresponding HWHM $\tau_{1/2}$ as a function of the number of time steps retained in the kernel.}
    \label{fig:acf_memory_steps}
\end{figure}

Fig.~\ref{fig:acf_memory_steps} shows that shortening the memory kernel monotonically shortens the autocorrelation timescales of all units. With the full kernel, the HWHM of the hDMFT solution is $0.93$--$0.94$ of that measured in direct simulations of the network (fully connected network with $N=2000$; discrete degree distribution with $N=6000$; three realizations each). With the instantaneous self-coupling it drops to $0.70$--$0.75$, and the kernels with 3, 5 and 10 time steps interpolate between the two. A network with independent couplings and explicit self-couplings $s_k$ gives the same timescales as the instantaneous kernel, within $5\%$. The effective self-coupling therefore captures the mechanism, namely a memory term whose strength grows with the degree, which slows down the dynamics and orders the timescales by degree, but underestimates its magnitude: part of the slowing down comes from the finite temporal extent of the response kernel.


\section{BISTABILITY AND AMPLIFICATION OF TIMESCALES}\label{app:bistability}
\begin{figure}[H]
    \centering
    \includegraphics[width=0.7\linewidth]{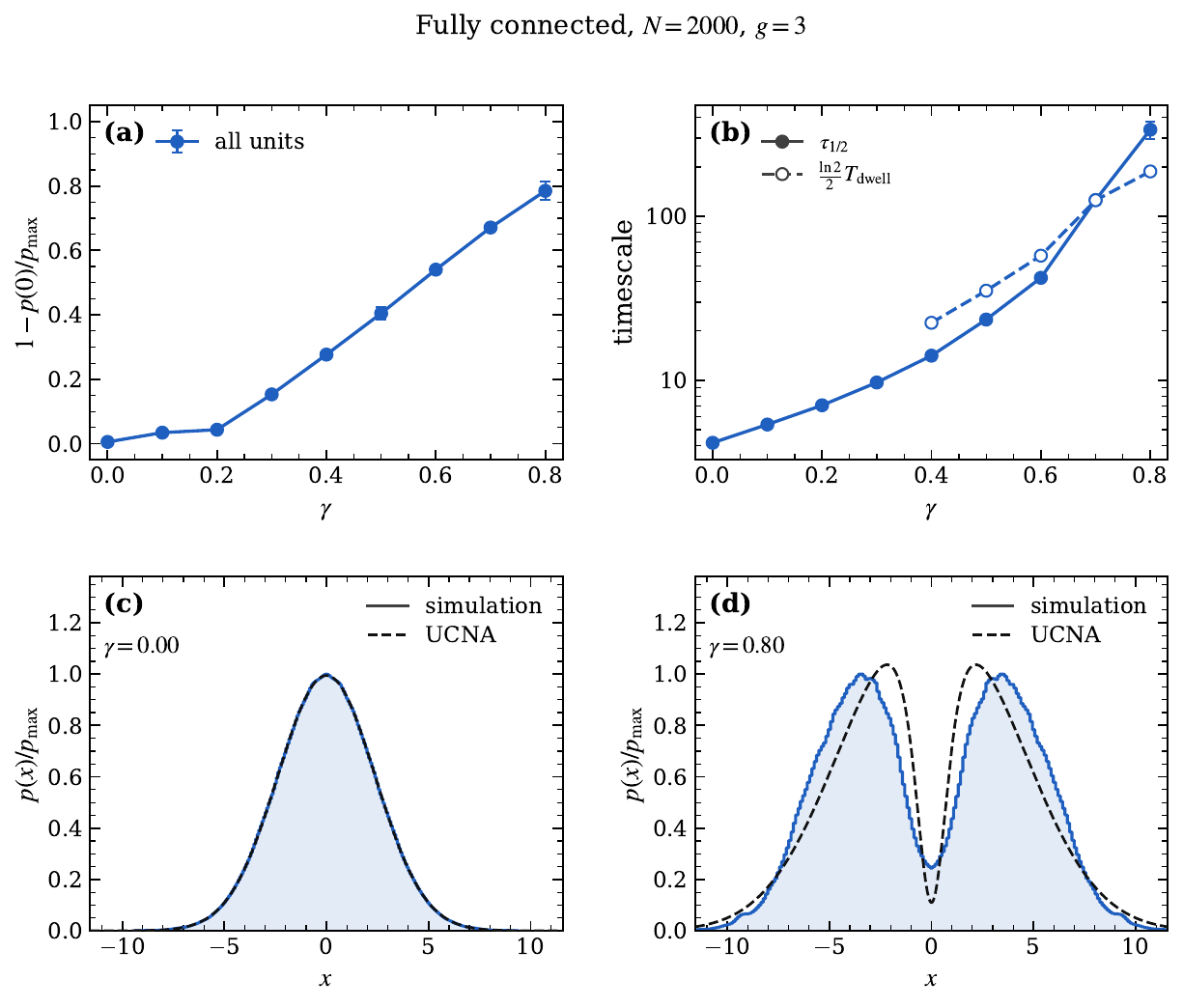}
    \caption{\textbf{Fully connected network}, $N=2000$, $g=3$.
    (a) Bistability parameter $\beta$, Eq.~(\ref{eq:appG_beta}), versus $\gamma$.
    (b) Half width at half maximum $\tau_{1/2}$ of the activity autocorrelation (filled
    symbols, solid) and $\tfrac{\ln2}{2}T_{\rm dwell}$ from the measured mean residence time
    (open symbols, dashed), the latter shown only where the distribution is bimodal.
    (c,d) Stationary distribution at $\gamma=0$ and at the largest $\gamma$ still resolved,
    normalized to its maximum, with the UCNA~(\ref{eq:appG_ucna}) (dashed): $\tau_1$ is
    measured and $(s,D)$ fitted to the histogram. Error bars are the SEM over five
    realizations; distributions are averaged over the same realizations and symmetrized.}
    \label{fig:appG_fully}
\end{figure}

\begin{figure}[H]
    \centering
    \includegraphics[width=0.7\linewidth]{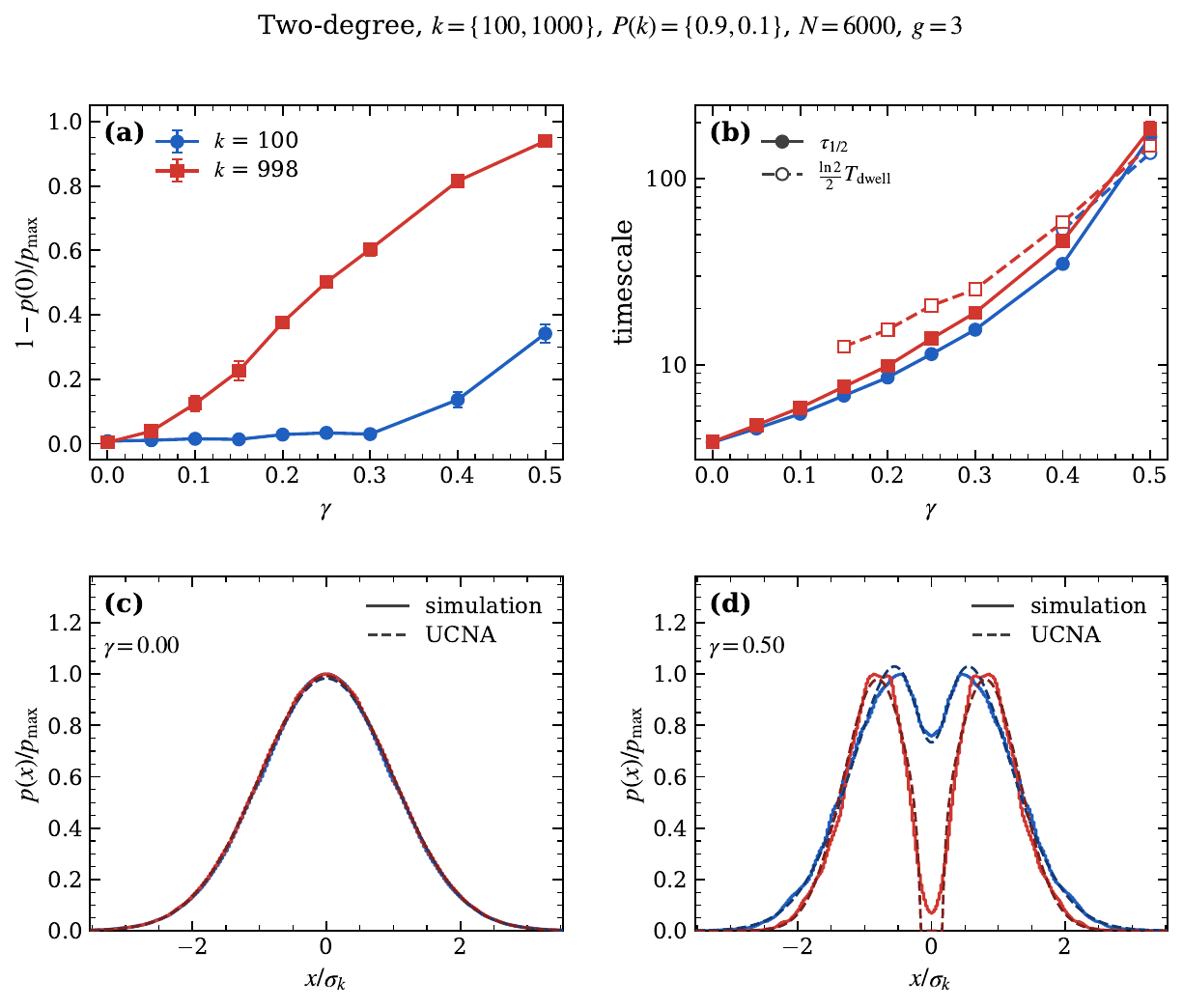}
    \caption{\textbf{Discrete degree distribution}, $k\in\{100,1000\}$,
    $P(k)=\{0.9,0.1\}$, $N=6000$, $g=3$. Panels as in Fig.~\ref{fig:appG_fully}, with the two
    degree classes shown separately. In (c,d) the activity is rescaled by the standard
    deviation $\sigma_k$ of each class, and the UCNA of each class is drawn in a darker shade
    of its colour.}
    \label{fig:appG_twodeg}
\end{figure}

In the $\delta$-kernel approximation the effective dynamics~(\ref{eq:hDMFT}) of a neuron of
degree $k$ is a one-dimensional Langevin equation driven by colored noise, as in
Ref.~\cite{stern2023reservoir},
\begin{equation}
    \dot{x}_k = f(x_k) + g\sqrt{\tfrac{k}{K}}\,\eta(t),
    \quad
    f(x) = -x + s_k\tanh x,
    \quad
    s_k = \gamma g^2 \tfrac{k}{K}\chi ,
    \label{eq:appG_langevin}
\end{equation}
with $\eta$ the self-consistent noise~(\ref{eq:corr}), whose stationary covariance
$C_\eta(\tau)=\sum_k P(k)(k/K)\,C^{\phi}_k(\tau)$ we measure in the simulations below.
Following Ref.~\cite{stern2023reservoir}, we approximate $\eta$ by an exponentially
correlated noise of correlation time $\tau_1$ and intensity
\begin{equation}
    D_k = g^2\,\frac{k}{K}\int_0^\infty \! d\tau\, C_\eta(\tau).
    \label{eq:appG_Dk}
\end{equation}
For $s_k>1$, $f$ has two stable zeros $\pm x^*_k$, which move away from the origin as $s_k$
grows. Both $s_k$ and $D_k$ grow linearly with $k$: the first deepens the wells, the second
drives the escapes between them.
The stationary density of Eq.~(\ref{eq:appG_langevin}) follows from the unified
colored-noise approximation (UCNA)~\cite{jung1987dynamical, hanggi2007colored},
\begin{equation}
    p_k(x) = \mathcal{Z}^{-1}\,h(x)\,e^{-U_{\rm eff}(x)/D_k},
    \label{eq:appG_ucna}
\end{equation}
\begin{equation}
    h(x) = 1 + \tau_1 - \tau_1 s_k\,\mathrm{sech}^2 x,
    \qquad
    U_{\rm eff}(x) = \frac{x^2}{2} - s_k \ln\cosh x + \frac{\tau_1}{2} f(x)^2 ,
    \label{eq:appG_Ueff}
\end{equation}
valid where $h>0$, i.e.\ $|x|>x_c$ with $\tanh^2 x_c = 1-(1+\tau_1)/(\tau_1 s_k)$. The
approximation deteriorates near the barrier, where $f'>0$. The
support splits when $\tau_1 s_k/(1+\tau_1)>1$; the density develops a trough at the origin
already for $s_k>1$, before the support splits, since the prefactor $h$ has its minimum
there. We quantify bistability by the depth of that trough,
\begin{equation}
    \beta_k = 1 - p_k(0)\big/\max_x p_k(x) ,
    \label{eq:appG_beta}
\end{equation}
which is invariant under a rescaling of $x$ and can therefore be compared across degree
classes of very different variance.
We simulate Eq.~(\ref{eq:RNN}) directly at $g=3$, sweeping $\gamma$, for a fully connected
network with $N=2000$ and for the discrete ensemble $k\in\{100,1000\}$, $P(k)=\{0.9,0.1\}$
with $N=6000$, the network of Appendix~\ref{app:num_DMFT}. Integration uses a Heun scheme,
$dt=0.05$, $T_{\rm tr}=300$, $T_{\rm run}=2000$, with five independent realizations per
point. We measure $\beta_k$ from the pooled histogram of each degree class, the timescale
$\tau_{1/2}$ as the half width at half maximum of the class-averaged autocorrelation of $x$,
and $\tau_1$ as the half width of $C_\eta(\tau)$.
Histograms are averaged over realizations and then symmetrized, which is exact because
Eq.~(\ref{eq:RNN}) is invariant under $x\to-x$; a single realization is asymmetric because
each unit switches only $O(T_{\rm run}/T_{\rm dwell})$ times. Realizations that relax onto a
fixed point at finite $N$ are excluded. The UCNA enters only as a check: at fixed measured
$\tau_1$, the pair $(s_k,D_k)$ is fitted to each measured histogram and the fitted $D_k$
compared with Eq.~(\ref{eq:appG_Dk}).
We also measure the mean residence time $T_{\rm dwell}$ between wells. For a symmetric
two-state process the autocorrelation is $e^{-2\tau/T_{\rm dwell}}$, so
$\tau_{1/2}=\tfrac{\ln2}{2}T_{\rm dwell}$ exactly. Each trajectory, however, combines
switching between the wells with fast chaotic fluctuations within a well. If $\rho$ is the
fraction of the variance carried by the switching, and the within-well fluctuations decay
faster than the switching,
\begin{equation}
    \frac{T_{\rm dwell}}{2}\ln(2\rho)
    \;\le\; \tau_{1/2} \;\le\;
    \frac{\ln 2}{2}\,T_{\rm dwell},
    \label{eq:appG_bounds}
\end{equation}
both bounds collapsing onto $\tfrac{\ln2}{2}T_{\rm dwell}$ as $\rho\to1$. We estimate $\rho$ from the ratio $\langle|x|\rangle^2/\langle x^2\rangle$, which equals $2/\pi$ at $\rho=0$ and $1$ at $\rho=1$.
Figures~\ref{fig:appG_fully} and~\ref{fig:appG_twodeg} report the resulting sweeps for the two ensembles.


\section{FROM THE DEGREE DISTRIBUTION TO THE DISTRIBUTION OF TIMESCALES}\label{app:tail}
If the timescale of a neuron is a monotone function $\tau(k)$ of its degree, the distributions of degrees and timescales are related by a change of variables,
\begin{equation}
P(\tau) = \frac{P(k)}{|d\tau/dk|}, \qquad\text{equivalently}\qquad S_\tau\big(\tau(k)\big) = S_k(k),
\label{eq:tail_change}
\end{equation}
where $S_k$ and $S_\tau$ are the survival functions, $S_k(k)=\int_k^\infty P(k')\,dk'$. Any two of $P(k)$, $P(\tau)$ and $\tau(k)$ therefore determine the third. Writing the local tail as $P(\tau) \propto \tau^{-1-\mu}$ gives
\begin{equation}
\mu(k) = \frac{-\,d \ln P(k)/dk}{d \ln \tau / dk} ,
\label{eq:tail_mu}
\end{equation}
i.e.\ the exponent is the ratio between the rate at which high-degree neurons become rare and the rate at which they become slow. $P(\tau)$ is a power law only if this ratio is constant.
The function $\tau(k)$ is set by the dynamics and is in general not known; in our model it is available only within the approximations of Appendix~\ref{app:bistability}. Experimentally, degree distributions are close to lognormal~\cite{piazza2025physical} (Fig.~\ref{fig:microns}c), while distributions of single-neuron timescales are heavy tailed and compatible with power laws~\cite{shi2025brain}. Note that, over the limited range of timescales accessible in simulations or recordings, a power law can be fitted even when Eq.~(\ref{eq:tail_mu}) gives no constant exponent: for the lognormal case of Fig.~\ref{fig:tail}b, a fit over $10<\tau<300$ gives a slope of about $-1.6$, although the local exponent keeps drifting. Below we use Eqs.~(\ref{eq:tail_change})--(\ref{eq:tail_mu}) in both directions: we first take $\tau(k)$ from the effective bistable dynamics and derive $P(\tau)$, and then take a lognormal $P(k)$ and a power-law $P(\tau)$ and derive the $\tau(k)$ that connects them.
\subsection{Timescales set by the bistable dynamics}
\begin{figure}[H]
\centering
\includegraphics[width=0.9\linewidth]{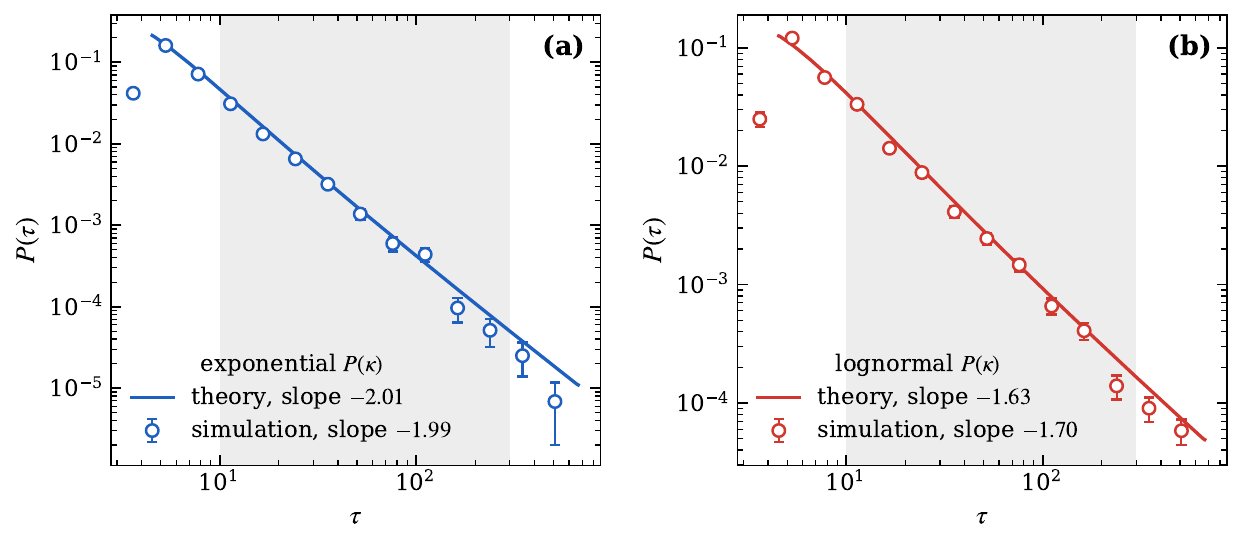}
\caption{\textbf{From the degree distribution to the distribution of timescales.} Distribution of timescales $P(\tau)$ for $1500$ independent bistable units with $s = \kappa$, $D = 0.25\,\kappa$, and $\kappa$ drawn from (a) an exponential distribution with rate $\lambda = 1.6$ and (b) a lognormal distribution with $\sigma = 0.6$, both restricted to $1.5 \le \kappa \le 4.5$. Circles: simulations, with $\tau$ the half width at half maximum of each unit's autocorrelation; lines: theory, $P(\tau) = P(\kappa)/|d\tau/d\kappa|$ with $\tau = \ln 2\, T_0$. The slopes in the legends are fitted over the shaded range, $10 < \tau < 300$.}
\label{fig:tail}
\end{figure}

In the $\delta$-kernel approximation, a neuron of degree $k$ obeys Eq.~(\ref{eq:appG_langevin}): its self-coupling, $s_k = \gamma g^2 \chi\, k/K$, and the intensity of its recurrent input, $D_k \propto k$, both grow linearly with the degree. For $s_k > 1$ the neuron switches between two wells, and its timescale is set by the mean residence time, $\tau = \ln 2\, T_0$, where $T_0$ is the mean first-passage time from a well to the barrier (Appendix~\ref{app:bistability}).
In the limit of deep wells and white noise, $\tau_1\to0$, Kramers' formula gives $T_0 \propto e^{\Delta U/D}$, with a barrier $\Delta U \propto s_k^2 \propto k^2$ and a noise intensity $D_k \propto k$, so that
\begin{equation}
\ln \tau(k) \simeq \mathrm{const} + \beta k , \qquad \beta = \frac{\gamma^2 g^2 \chi^2}{2 K Q},
\label{eq:tail_kramers}
\end{equation}
with $Q = \int_0^\infty C_\eta(\tau)\, d\tau$.
For colored noise, the UCNA leaves this exponent unchanged, since the term $\tau_1 f^2/2$ in $U_{\rm eff}$ vanishes at both extrema, and $\tau_1$ enters only through the prefactor. The UCNA, however, is not reliable near the barrier, where $f'>0$ (Appendix~\ref{app:bistability}), and beyond it a finite correlation time increases the effective barrier~\cite{hanggi2007colored}, so that $\ln\tau$ is expected to grow faster with $k$ than in Eq.~(\ref{eq:tail_kramers}). Equation~(\ref{eq:tail_mu}) still applies, with the corresponding $\tau(k)$.
For an exponential degree distribution, $P(k) \propto e^{-\lambda k}$, Eq.~(\ref{eq:tail_mu}) gives a constant $\mu = \lambda/\beta$, and $P(\tau)$ is a power law, $P(\tau) \propto \tau^{-1-\lambda/\beta}$; in particular, $P(\tau) \propto \tau^{-2}$ for $\lambda = \beta$. For a lognormal degree distribution, $-d\ln P(k)/dk = [1 + (\ln k - \mu_{\rm LN})/\sigma^2]/k$ decreases with $k$, so that $\mu(k)$ is not constant: $P(\tau)$ has no single exponent, and its local slope depends on the range of degrees, and hence of timescales, considered. We tested these predictions on independent bistable units with $s = \kappa$ and $D = 0.25\,\kappa$, $\kappa = k/K$, driven by white noise, for which $T_0$ can be computed exactly, so that the test of Eq.~(\ref{eq:tail_change}) involves no approximation; $\kappa$ is drawn either from an exponential or from a lognormal distribution in the bistable range $1.5 \le \kappa \le 4.5$. The timescale of each unit was measured as the half width at half maximum of its autocorrelation and compared with the theory, $\tau = \ln 2\, T_0$ with $T_0$ computed exactly, which has no free parameters (Fig.~\ref{fig:tail}). For an exponential distribution with $\lambda \simeq \beta$, $P(\tau)$ is a power law with exponent close to $-2$. For a lognormal distribution, the slope fitted over the same range is about $-1.6$, and it decreases in magnitude at larger $\tau$. In the network matched to the MICrONS data, the timescale grows only weakly with the degree, $d\ln\tau/dk \approx 10^{-3}$, so that Eq.~(\ref{eq:tail_mu}) gives $\mu \gg 1$ for the measured lognormal degree distribution: degree heterogeneity produces a broad but not scale-free distribution of timescales. A power-law tail would require high-degree neurons deep in the bistable regime, while the rest of the network remains chaotic, together with a degree distribution whose tail is matched to $\beta$.
\subsection{Degree dependence required by a power-law distribution of timescales}
Conversely, we ask which $\tau(k)$ maps a lognormal degree distribution, with parameters $\mu_{\rm LN}$ and $\sigma$, onto a power-law distribution of timescales, $P(\tau) = \mu\,\tau_{\min}^{\mu}\,\tau^{-1-\mu}$ for $\tau\ge\tau_{\min}$. Since $S_\tau(\tau) = (\tau/\tau_{\min})^{-\mu}$, Eq.~(\ref{eq:tail_change}) gives
\begin{equation}
\tau(k) = \tau_{\min}\left[\tfrac{1}{2}\,\mathrm{erfc}\!\left(\frac{z}{\sqrt{2}}\right)\right]^{-1/\mu},
\qquad z = \frac{\ln k - \mu_{\rm LN}}{\sigma}.
\label{eq:tau_required}
\end{equation}
For $z \gg 1$, $-\ln S_k \simeq z^2/2 + \ln(z\sqrt{2\pi})$, so that
\begin{equation}
\ln \frac{\tau(k)}{\tau_{\min}} \simeq \frac{1}{\mu}\left[\frac{(\ln k - \mu_{\rm LN})^2}{2\sigma^2} + \ln\!\left(\sqrt{2\pi}\,\frac{\ln k - \mu_{\rm LN}}{\sigma}\right)\right],
\qquad
\frac{d\ln\tau}{d\ln k} \simeq \frac{\ln k - \mu_{\rm LN}}{\mu\sigma^2}.
\label{eq:tau_required_asym}
\end{equation}
A model dynamics that reproduces both the lognormal degree distribution and a power-law distribution of timescales must therefore have a timescale that depends on the degree as in Eq.~(\ref{eq:tau_required}): at large degree, $\ln\tau$ grows as $(\ln k)^2$, and the local exponent $d\ln\tau/d\ln k$ increases logarithmically with $k$.

\section{TIMESCALES SCALING WITH NETWORK SIZE }\label{app:timescales_scaling}
In this section we compare different models proposed in the literature for the generation of heterogeneous timescales.
In particular, we distinguish between two classes of models: those in which timescales emerge as an intrinsic property of the network in the thermodynamic limit, and those in which apparent timescale heterogeneity arises only in finite-size networks, for instance when the system is tuned close to a critical point of the corresponding infinite systems.
To discriminate between these two scenarios, we perform a finite-size scaling analysis by computing the coefficient of variation of the timescales,
\begin{equation}
    CV(\tau) = \frac{\mathrm{std}(\mathrm{HWHM})}{\mathrm{mean}(\mathrm{HWHM})},
\end{equation}
where the timescale of each neuron is quantified by the half width at half maximum (HWHM) of its autocorrelation function.
All models share the dynamics of Eq.~(\ref{eq:RNN}) with $\phi(x)=\tanh(x)$ and $I_i=0$, and differ only in the statistics of the coupling matrix $W$. The ensembles are summarized in Table~\ref{tab:scaling_models}. For every model, Eq.~(\ref{eq:RNN}) is integrated with the Euler scheme, $dt=0.1$, for a total time $T=2000$, discarding a transient of $300$. The autocorrelation function of each $x_i$ is computed up to a lag of $1000$. Neurons whose activity variance is below $10^{-4}$, and neurons whose autocorrelation does not decay below $1/2$ within the maximal lag, are excluded from the computation of $\mathrm{CV}(\tau)$. Realizations that converge to a fixed point are discarded. For each size $N\in\{1000,2000,\dots,10000\}$, $\mathrm{CV}(\tau)$ is computed separately for each independent network realization, and we report its mean and standard error across realizations.
\begin{table}[h]
\centering
\begin{tabular}{l l l}
\hline\hline
Model & Coupling ensemble & Parameters \\
\hline
Fully & $W_{ij}\sim\mathcal{N}(0,g^2/N)$ i.i.d., $W_{ii}=0$ & $g=1.5$ \\[1mm]
Heavy tailed & symmetric $\alpha$-stable, scale $g/N^{1/\alpha}$, i.i.d., $W_{ii}=0$ & $\alpha=1.2$, $g=1.8$ \\[1mm]
Heavy tailed, $\gamma$ & same marginals; $(W_{ij},W_{ji})$ i.i.d.\ over pairs, Spearman correlation $\gamma$ & $\alpha=1.2$, $g=1.8$, $\gamma=0.4$ \\[1mm]
Het self-couplings & $W_{ij}\sim\mathcal{N}(0,g^2/N)$ for $i\neq j$, $W_{ii}=s_i$, $\ln s_i\sim\mathcal{N}(\mu,\sigma^2)$ & $g=3$, $\mu=0.6$, $\sigma=0.6$ \\[1mm]
Het degree, $\gamma$ & $W_{ij}=A_{ij}J_{ij}$, lognormal degrees, $J_{ij}\sim\mathcal{N}(0,g^2/K)$, $\mathrm{corr}(J_{ij},J_{ji})=\gamma$ & $g=3$, $\mu=3$, $\sigma=1$, $\gamma=0.4$ \\
\hline\hline
\end{tabular}
\caption{Coupling ensembles used in the finite-size scaling analysis of Fig.~\ref{fig:timescales_scaling}.}
\label{tab:scaling_models}
\end{table}

\textit{Fully connected Gaussian network.}
We first consider a classical random neural network with fully connected architecture and Gaussian-distributed couplings \cite{sompolinsky1988chaos}, tuned close to criticality: $W_{ij}$ are i.i.d.\ Gaussian variables with zero mean and variance $g^2/N$, and $W_{ii}=0$. Tuning near criticality is often proposed as a mechanism for generating heterogeneous timescales. The underlying rationale is that, at criticality, the network exhibits several distinctive features, most notably the divergence of the autocorrelation time. This divergence amplifies the differences between neuronal timescales that arise in finite networks due to finite-size fluctuations. As a result, apparent heterogeneity in timescales can be observed even in otherwise homogeneous systems.
This effect persists as long as the network is tuned sufficiently close to the critical gain. However, numerical demonstrations become increasingly challenging in this regime due to critical slowing down and finite-size effects, such as the presence of fixed points or limit cycles. For this reason, we analyze the network at a coupling strength $g = 1.5$, above the critical gain $g_c=1$ of the infinite network~\cite{sompolinsky1988chaos}. In the thermodynamic limit, the behavior of this model can be characterized analytically using dynamical mean-field theory (DMFT) \cite{sompolinsky1988chaos}.

\begin{figure}[H]
    \centering
\includegraphics[width=0.6\linewidth]{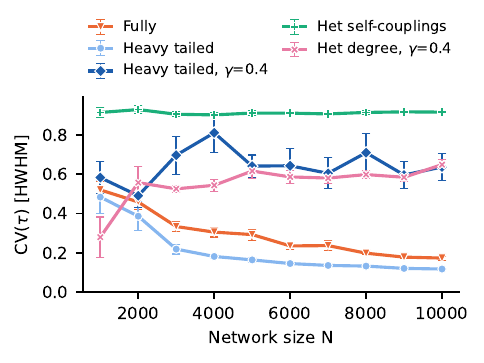}
\caption{\textbf{Comparison of timescale heterogeneity scaling with $N$ across different models.}
We quantify the heterogeneity of timescales in a network by computing the coefficient of variation (standard deviation divided by mean) of the half width at half maximum (HWHM) of the neuronal autocorrelation functions. The coupling ensembles are detailed in Table~\ref{tab:scaling_models}.
Fully: fully connected network with i.i.d.\ Gaussian couplings of variance $g^2/N$, tuned near criticality ($g=1.5$).
Heavy tailed: fully connected network with i.i.d.\ couplings sampled from a symmetric $\alpha$-stable distribution with scale $g/N^{1/\alpha}$, $\alpha=1.2$, $g=1.8$.
Heavy tailed, $\gamma=0.4$: same $\alpha$-stable marginals, with reciprocal pairs $(W_{ij},W_{ji})$ joined by a Gaussian copula with Spearman correlation $\gamma=0.4$.
Het self-couplings: i.i.d.\ Gaussian off-diagonal couplings of variance $g^2/N$ and self-couplings sampled from a lognormal distribution with $\mu=0.6$, $\sigma=0.6$, as in Ref.~\cite{stern2023reservoir}, with $g=3$.
Het degree, $\gamma=0.4$: the model proposed in this paper, with lognormal degree distribution ($\mu=3$, $\sigma=1$), Gaussian weights of variance $g^2/K$ and partial symmetry $\gamma=0.4$, with $g=3$.
Each point is the mean over independent network realizations (100 for fully, 20 for the two heavy-tailed ensembles, 50 for het self-couplings and het degree); error bars denote the standard error of the mean.}
\label{fig:timescales_scaling}
\end{figure}

\textit{Heavy-tailed i.i.d.\ couplings.}
Another interesting proposed model is the extension of recurrent neural networks to include heavy-tailed coupling distributions, as discussed in Refs.~\cite{kusmierz2020edge,xie2025slow, wardak2022extended}.
Here $W_{ij}$ are i.i.d.\ symmetric $\alpha$-stable variables with characteristic function $\langle e^{iqW_{ij}}\rangle=\exp(-g^\alpha|q|^\alpha/N)$, i.e.\ with scale $g/N^{1/\alpha}$, and $W_{ii}=0$. This scaling with $N$ keeps the local field $\sum_jW_{ij}\phi(x_j)$ of order one. We use $\alpha=1.2$ and $g=1.8$.
Surprisingly, introducing heavy-tailed coupling distributions is not sufficient to generate the emergence of heterogeneous timescales (see Fig.~\ref{fig:timescales_scaling}). Intuitively, for $\alpha<2$ the input of each neuron is dominated by a few large couplings, so that its amplitude may vary across neurons. A variation of the input amplitude alone, however, is not expected to change the shape of the autocorrelation, and thus the timescale, which may explain why the timescale heterogeneity decreases with $N$ (Appendix~\ref{app:iid}).

\textit{Heavy-tailed couplings with partial symmetry.}
We also consider the heavy-tailed ensemble with partial symmetry. For $\alpha<2$ the variance of the couplings is infinite and the Pearson correlation between $W_{ij}$ and $W_{ji}$ is undefined. We therefore define the symmetry parameter through the Spearman rank correlation, $\gamma=\rho_{\rm S}(W_{ij},W_{ji})$.
The couplings are generated with a Gaussian copula. For each pair $i<j$ we draw $(z_{ij},z_{ji})$ from a bivariate standard normal distribution with correlation $\rho=2\sin(\pi\gamma/6)$, independently across pairs, and set
\begin{equation}
    W_{ij}=\frac{g}{N^{1/\alpha}}\,F_\alpha^{-1}\!\big(\Phi(z_{ij})\big),
\end{equation}
with $W_{ii}=0$. Here $\Phi$ is the standard normal cumulative distribution and $F_\alpha$ the cumulative distribution of the standard symmetric $\alpha$-stable law. The marginals are therefore identical to those of the i.i.d.\ heavy-tailed ensemble, and $\rho_{\rm S}(W_{ij},W_{ji})=\gamma$. For $\alpha=2$ the construction reduces to Gaussian couplings with Pearson correlation $2\sin(\pi\gamma/6)\simeq\gamma$. The inverse $F_\alpha^{-1}$ is evaluated numerically in the bulk and from the asymptotic series of the stable law in the tails. We use $\alpha=1.2$ and $g=1.8$, as in the i.i.d.\ case, and $\gamma=0.4$, as in our model. In contrast with the i.i.d.\ case, the coefficient of variation of the timescales does not decrease with network size over the range of $N$ tested (Fig.~\ref{fig:timescales_scaling}). Whether the corresponding reciprocal terms remain finite as $N\to\infty$ likely depends on the joint tail of $(W_{ij},W_{ji})$ (Appendix~\ref{app:iid}).

\textit{Heterogeneous self-couplings.}
In order to obtain the emergence of heterogeneity at the network level, a diversity of some feature must be introduced at the neuronal level. This is done, for example, in Ref.~\cite{stern2023reservoir}, where each rate unit has a self-coupling of order one, interpreted as the effect of a neural assembly rather than as an autapse, and these self-couplings are distributed across units according to a lognormal distribution.
The off-diagonal couplings are i.i.d.\ Gaussian with zero mean and variance $g^2/N$, and the self-couplings $W_{ii}=s_i$ are i.i.d.\ lognormal, $\ln s_i\sim\mathcal{N}(\mu,\sigma^2)$, with $\mu=0.6$, $\sigma=0.6$ and $g=3$.
As shown in Fig.~\ref{fig:timescales_scaling}, for this model the coefficient of variation of the timescales remains constant as the network size $N$ increases.

\textit{Degree heterogeneity with partial symmetry.}
Finally, in Fig.~\ref{fig:timescales_scaling} we also report the results obtained for our
model, which combines a lognormal degree distribution that introduces heterogeneity at
the single-neuron level with partial symmetry, which slows down the dynamics and
amplifies this heterogeneity, leading to the emergence of broadly distributed intrinsic
timescales.
Here $W_{ij}=A_{ij}J_{ij}$, where $A$ is an undirected configuration-model graph without self-loops. Its target degrees are drawn from a lognormal distribution with $\mu=3$ and $\sigma=1$, independent of $N$. The weights $J_{ij}$ are Gaussian with zero mean, variance $g^2/K$, where $K$ is the mean degree, and $\mathrm{corr}(J_{ij},J_{ji})=\gamma$, with $g=3$ and $\gamma=0.4$.
Notably, the coefficient of variation of the timescales does not decrease
with network size. At small $N$ it slightly increases with $N$, which we attribute to the lower probability of sampling hubs with very large degree in small networks.

\section{A HEURISTIC ARGUMENT ON TIMESCALE HETEROGENEITY WITH I.I.D. COUPLINGS}
\label{app:iid}
In this appendix we propose a heuristic argument that may help interpret the numerical results of Fig.~\ref{fig:timescales_scaling}. It is not a derivation, but it suggests which single-neuron properties could produce a timescale heterogeneity that persists as $N\to\infty$.

\textit{Input amplitude and timescale.}
Timescales are measured on $x_i$, which obeys $(\partial_t+1)\,x_i = h_i(t)$, with $h_i=\sum_j W_{ij}\,\phi(x_j)$. Since $x_i$ is a linear filter of $h_i$, its normalized autocorrelation depends only on the normalized autocorrelation of $h_i$, and not on its amplitude, as long as $h_i$ contains no feedback of $x_i$ itself. This suggests that neurons whose inputs differ mainly by a constant factor should have similar timescales.

\textit{Degree heterogeneity without symmetry.}
At $\gamma=0$, Eq.~(\ref{eq:DMFT}) reduces to $(\partial_t+1)\,x_k = g\sqrt{k/K}\,\eta(t)$. Because $p_{ij}=k_ik_j/(NK)$ factorizes, all neurons receive the same mixture of presynaptic activities, and the degree only sets the amplitude of $\eta$. Consistently, the numerical solution of the hDMFT gives the same timescale for the two degree classes, $k\in\{100,1000\}$, at $\gamma=0$ (Appendix~\ref{app:num_DMFT}). This indicates that a quenched single-neuron variable, such as the degree, may not be sufficient on its own to generate timescale heterogeneity.

\textit{I.i.d.\ couplings.}
For i.i.d.\ couplings, the reciprocal sum $\sum_j W_{ij}W_{ji}$, which would feed the activity of neuron $i$ back onto itself, has zero mean and is expected to vanish as $N$ grows. For Gaussian couplings the input also self-averages, so that all neurons are expected to receive statistically equivalent inputs. For $\alpha$-stable couplings with $\alpha<2$ the conclusion is expected to be the same: although the input of each neuron is dominated by a few large couplings, so that its amplitude may vary across neurons without averaging out, the argument above suggests that this variability alone should have little effect on the timescale. In our simulations, $\mathrm{CV}(\tau)$ decreases with $N$ over the range tested (Fig.~\ref{fig:timescales_scaling}), which seems consistent with this picture.

\textit{Partial symmetry.}
For $\gamma>0$ the reciprocal sum no longer vanishes. In Eq.~(\ref{eq:DMFT}) it produces the memory term $\gamma g^2 (k/K)\int dt'\,G(t,t')\,\phi(x_k(t'))$, which acts as a feedback of the neuron's own activity. Unlike an input amplitude, a self-feedback can change the shape of the autocorrelation: in the instantaneous approximation of Eq.~(\ref{eq:hDMFT}), for linear $\phi$ it shifts the relaxation rate from $1$ to $1-s_k$, and for $\phi=\tanh$ it can induce the bistability discussed in Appendix~\ref{app:bistability}. Since $s_k\propto k$, this suggests that partial symmetry allows the degree to set the timescale, and that the resulting distribution, inherited from $P(k)$, should not depend on $N$ at leading order in $1/K$. Heterogeneous self-couplings of order one~\cite{stern2023reservoir} would fall in the same class, with the self-feedback postulated rather than derived from the connectivity.
\textit{Summary.}
Taken together, these observations suggest that a timescale heterogeneity persisting as $N\to\infty$ may require a single-neuron variable that (i) does not average out and (ii) enters the dynamics as a self-feedback rather than only as an input amplitude. Within this picture, degree heterogeneity alone and i.i.d.\ couplings would each lack at least one of these ingredients, whereas degree heterogeneity with partial symmetry, and explicit self-couplings, would have both. For heavy-tailed couplings with partial symmetry, $\mathrm{CV}(\tau)$ does not decrease over the range of $N$ tested (Fig.~\ref{fig:timescales_scaling}). Whether the corresponding reciprocal terms remain finite as $N\to\infty$ likely depends on the joint tail of $(W_{ij},W_{ji})$, and we leave this question to future work.

\section{SPIKING NETWORKS WITH HETEROGENEITY IN DEGREE DISTRIBUTION OR  NEURAL ASSEMBLY SIZE}
\label{app:spiking}
The rate description of Eq.~(\ref{eq:RNN}) leaves two questions open. The first is whether the
mechanism of Sec.~\ref{sec:emergence} survives when the units are spiking neurons obeying Dale's
law inside a balanced excitatory--inhibitory circuit~\cite{van1996chaos,
brunel2000dynamics}, where the single-neuron transfer function is not a fixed $\phi$ but is set
self-consistently by the recurrent input. The second is whether it can be told apart from the
other known route to a distribution of intrinsic timescales in spiking circuits, in which the
slow variable is the activity of a neural assembly rather than of a single
neuron~\cite{litwin2012slow, mazzucato2015dynamics, stern2023reservoir}, or in which strong recurrent coupling generates slow, self-consistent network fluctuations~\cite{wieland2015slow}.
We therefore build two spiking networks that share the neuron and synapse model, the population
sizes, the four connection probabilities and the mean excitatory firing rate, and differ only in
which structural variable is heterogeneous: the in-degree of a neuron in one case (M-deg), the
size of the assembly it belongs to in the other (M-clust). In both, every synapse of a block
carries exactly its base weight times its structural multiplier, so that no per-synapse weight
noise contributes to the heterogeneity. The two are built the other way round from one another:
M-deg leaves every multiplier equal to one and puts all of its structure in which $E\!\to\!E$
pairs are wired, while M-clust is wired at random in all four blocks and puts its structure in
the weights (Fig.~\ref{fig:spiking}a,b).
\subsection{Neuron and synapse model}
Neurons are current-based leaky integrate-and-fire units with exponential synapses. The membrane
potential $V_i$ of neuron $i$ obeys
\begin{equation}
    \tau_m \frac{dV_i}{dt} = -V_i + \tau_m \left( I^E_i(t) + I^I_i(t) + I^o_i \right),
    \label{eq:spk_V}
\end{equation}
and the two recurrent synaptic currents are low-pass filtered spike trains,
\begin{equation}
    \tau_s \frac{dI^B_i}{dt} = -I^B_i + \sum_{j\in B}\sum_k W_{ij}\,
    \delta\!\left(t - t^j_k - \delta_d\right), \qquad B = E, I,
    \label{eq:spk_I}
\end{equation}
with $W_{ij}\ge 0$ for excitatory and $W_{ij}\le 0$ for inhibitory presynaptic neurons and
$\delta_d$ the transmission delay. When $V_i$ reaches the threshold $V^{\rm thr}_\alpha$
($\alpha=E,I$) a spike is emitted, $V_i$ is clamped at $V^{\rm reset}=0$ for the refractory period
$\tau_{\rm ref}$, and the integration resumes. The constant current $I^o_i$ carries the external
drive and, in M-deg, the compensation defined in Eq.~(\ref{eq:spk_comp}) below.
Both networks receive a constant external drive, $\tau_m I^{\rm ext}_i = \tau_m r_{\rm ext}
N_{\rm ext} J_{\alpha 0}$, i.e.\ the mean of a Poisson afferent input without its shot noise. The
two architectures differ by a factor $40$ in recurrent coupling, so a drive carrying its own noise
would not enter them on the same footing: the weakly coupled network would be dominated by it
while the strongly coupled one would not.
Equations~(\ref{eq:spk_V})--(\ref{eq:spk_I}) are linear between spike arrivals and are integrated
with the exact propagator of the subthreshold dynamics~\cite{rotter1999exact, brunel1998firing}.
Over one step of length $dt$, with $I(t)=I(0)e^{-t/\tau_s}$,
\begin{equation}
    V(dt) = p_{vv} V(0) + p_{ve} I^E(0) + p_{vi} I^I(0) + p_{vo} I^o ,
\end{equation}
\begin{equation}
    p_{vv} = e^{-dt/\tau_m}, \quad
    p_{ve} = p_{vi} = \frac{\tau_s \tau_m}{\tau_s - \tau_m}
    \left( e^{-dt/\tau_s} - e^{-dt/\tau_m} \right), \quad
    p_{vo} = \tau_m \left( 1 - e^{-dt/\tau_m} \right).
\end{equation}
Nothing is approximated in the subthreshold evolution, so $dt$ limits only the resolution at which
spikes are placed and input spikes delivered, not the accuracy of $V$ between them. Within a step
the potentials are propagated first, the synaptic currents then decay by their exact factor, the
threshold is tested, and the spikes emitted one delay earlier are delivered. Initial potentials
are drawn uniformly in $[V^{\rm reset}, V^{\rm thr}_\alpha]$.
\subsection{M-deg: heterogeneous in-degrees with partial symmetry}
M-deg transposes the model of Sec.~\ref{sec:emergence} to the spiking setting. A sequence of
expected degrees $k_i$ is drawn from a lognormal distribution of coefficient of variation
$c_\kappa$ and rescaled so that $\langle k \rangle = K \equiv p_{EE} N_E$ exactly; the
excitatory-to-excitatory adjacency is then drawn pair by pair with
\begin{equation}
    p_{ij} = \frac{k_i k_j}{N_E K},
\end{equation}
so that $\sum_j p_{ij} = k_i$ and $k_i$ is the expected in- and out-degree of neuron $i$. The
joint state of the two directed edges of each unordered pair is drawn with the rule
\begin{equation}
    P(A_{ij}=A_{ji}=1) = \gamma\,p_{ij} + (1-\gamma)\,p_{ij}^2, \qquad
    P(A_{ij}=1, A_{ji}=0) = (1-\gamma)\,p_{ij}(1-p_{ij}),
    \label{eq:spk_rao}
\end{equation}
and symmetrically for the opposite direction. The marginal of each directed edge remains $p_{ij}$
for every $\gamma$, so the mean and variance of the connectivity are those of the $\gamma=0$
network, while
\begin{equation}
    \mathrm{Corr}\left[A_{ij}, A_{ji}\right] = \gamma
\end{equation}
exactly, per pair; at the value used, $96\%$ of the $E\!\to\!E$ connections are reciprocated
against $20\%$ by chance. Because all excitatory-to-excitatory synapses carry the same weight
$J_{EE}$, the correlation of the weights coincides with that of the adjacency,
$\mathrm{Corr}[W_{ij},W_{ji}] = \gamma$: in the language of Sec.~\ref{sec:model} the
synaptic contribution is $\gamma_{\rm w}=1$ and the reciprocity is purely structural. The
remaining three blocks are Erd\H{o}s--R\'enyi at $p=0.5$ with independent directions, and
$c_\kappa = 0$ is the homogeneous baseline, in which the only degree spread left is the binomial
one of the edge sampling (Appendix~\ref{app:realized_degree}).
Dale's law makes the degree structure explicit in the mean connectivity. Since every
excitatory-to-excitatory synapse carries the same weight, the expected excitatory block is
\begin{equation}
    \left\langle W_{EE} \right\rangle
    = \frac{J_{EE}}{N_E K}\, \mathbf{k}\,\mathbf{k}^{\mathsf{T}},
    \label{eq:spk_rank1}
\end{equation}
a rank-one structure added to a random matrix, the setting in which structured connectivity is
known to organize the fluctuations of a strongly coupled network into a collective
mode~\cite{landau2018coherent}; whether such a structure produces low-dimensional variability in
a balanced network, or is instead absorbed by the balance, is decided by how its row and column
spaces overlap~\cite{landau2021macroscopic}. In the rate model of the main text the weights are
zero mean, so the same degree sequence enters the second moment,
$\mathrm{Var}[W_{ij}] = p_{ij}g^2/K$, rather than the first.
The strength of the recurrent excitation is measured by the balanced coupling
\begin{equation}
    g \equiv \frac{J_{EE}\sqrt{K}}{V^{\rm thr}_E},
\end{equation}
which plays here the role of the gain of the rate model. At the value used, the mean recurrent
excitatory input of a neuron is $2.3\,V^{\rm thr}_E$ per Hz of population rate, so a spread of
in-degrees would translate directly into a spread of mean inputs, silencing the low-degree
neurons and saturating the hubs before any effect of $\gamma$ could be seen. We remove that
first-order difference with a constant current,
\begin{equation}
    \tau_m I^{o}_i = \tau_m I^{\rm ext}_i
    - \tau_m \left( \sum_j W_{ij} - \Big\langle \sum_j W_{ij} \Big\rangle_E \right) r_{\rm comp},
    \label{eq:spk_comp}
\end{equation}
which cancels the excess mean recurrent excitatory input that neuron $i$ would receive if the
excitatory population fired at the reference rate $r_{\rm comp}$, the average being taken over the
excitatory population. The current is a function of the row sums of $W$ alone: it equalizes the
first moment of the input across neurons and leaves the reciprocal structure, which is what
Eq.~(\ref{eq:spk_rao}) controls, untouched. Equivalently, it removes the component of the
rank-one structure of Eq.~(\ref{eq:spk_rank1}) that acts on the mean input, the component a
balanced network would otherwise have to absorb on its own~\cite{landau2021macroscopic}. At the
solution the reference rate must equal the rate the network actually settles at, and we fix it by
that identity, bisecting
\begin{equation}
    r_{\rm comp} = \nu_E(r_{\rm comp})
\end{equation}
independently for every network realization.
\subsection{M-clust: heterogeneous assemblies}
M-clust is the clustered balanced network of
Refs.~\cite{litwin2012slow, mazzucato2015dynamics, stern2023reservoir}. A fraction
$1-n_{\rm bgr}$ of the excitatory population is divided into $Q$ assemblies, each holding on
average a fraction $f = (1-n_{\rm bgr})/Q$ of it, and the remaining $n_{\rm bgr}$ belong to none.
Weights between members of the same assembly are multiplied by the potentiation $J^+_{EE}$ and
weights between members of different assemblies by the depression
\begin{equation}
    J^-_{EE} = 1 - \gamma_{\rm dep} \left( J^+_{EE} - 1 \right), \qquad
    \gamma_{\rm dep} = f \left[ 2 - f(Q+1) \right]^{-1},
\end{equation}
while weights to and from the unclustered neurons are left unchanged. Without a background this
rule leaves the mean excitatory weight a neuron receives at the base weight,
$f J^+_{EE} + (1-f) J^-_{EE} = 1$: what an assembly gains inside, the synapses between assemblies
pay for. Inhibitory assemblies are paired with the excitatory ones, with their own potentiation
and depression factors, so that all four blocks carry assembly structure: the multipliers are
$0.51$, $1$ and $21.9$ in $EE$, $0.73$, $1$ and $10.4$ in $EI$, $0.78$, $1$ and $8.8$ in $IE$,
and $0.86$, $1$ and $7.1$ in $II$. Assembly sizes are drawn uniformly within $\pm r_{\rm size}$
of the mean, and this spread is the structural variable of the comparison, the counterpart of
$c_\kappa$ in M-deg. The adjacency is Erd\H{o}s--R\'enyi in every block, with independent
directions, $\gamma = 0$.
The excitatory input a neuron receives grows with the size of its assembly, by
$J_{EE} p_{EE}(J^+_{EE}-J^-_{EE})$ per member, and this difference is not removed: the
compensation of Eq.~(\ref{eq:spk_comp}) is defined for any weight matrix but is not applied here,
and its structural alternative, a per-neuron budget that rescales each neuron's weights from
outside its assembly, cannot be met for the largest assemblies at this potentiation. The two
networks are matched on their output instead, by a single scalar multiplying the constant external
drive of M-clust, bisected until its mean excitatory rate equals that of M-deg. Each network is
therefore compared at equal mean output and at its own coupling.
\subsection{Parameters}
All parameters are listed in Table~\ref{tab:spiking_params}. The two networks share the neuron and synapse model, the population sizes and the connection probabilities; they differ in the weights, which set the structure of M-clust, and in the thresholds, which are calibrated separately for each network.
\begin{table}[h]
\centering
\caption{Parameters of the two spiking networks. $K \equiv p_{EE}N_E$ is the mean excitatory
in-degree and $g \equiv J_{EE}\sqrt{K}/V^{\rm thr}_E$ the balanced coupling. Thresholds are
obtained, for each network, by calibrating the corresponding homogeneous network to a chosen pair
of population rates.}
\label{tab:spiking_params}
\begin{ruledtabular}
\begin{tabular}{llcc}
symbol & meaning & M-deg & M-clust \\
\hline
$N$, $n_E$ & neurons, excitatory fraction & 4000, 0.8 & 4000, 0.8 \\
$p_{EE}$ & $E\!\to\!E$ connection probability & 0.2 & 0.2 \\
$p_{EI}$, $p_{IE}$, $p_{II}$ & other connection probabilities & 0.5 & 0.5 \\
$J_{EE}$, $J_{EI}$ & $E\!\to\!E$, $I\!\to\!E$ weight magnitude & 0.5657, 0.5657 & 0.01414, 0.04243 \\
$J_{IE}$, $J_{II}$ & $E\!\to\!I$, $I\!\to\!I$ weight magnitude & 1.6971, 3.3941 & 0.01414, 0.08485 \\
$K$, $g$ & mean $E$ in-degree, balanced coupling & 640, 4.570 & 640, 0.117 \\
$J^+_{EE}$, $J^-_{EE}$ & assembly potentiation, depression & 1, 1 (none) & 21.92, 0.5135 \\
$c_J$ & weight coefficient of variation & 0 & 0 \\
$Q$, $n_{\rm bgr}$, $r_{\rm size}$ & assemblies, unclustered, size range & --- & 36, 0.1, $\pm40\%$ \\
$c_\kappa$, $\gamma$ & degree spread, reciprocity & $0$ and $0.10$, $0 \ldots 1$ & --- , 0 \\
$V^{\rm thr}_E$, $V^{\rm thr}_I$ & thresholds & 3.1312, 1.5863 & 3.0618, 1.5529 \\
$V^{\rm reset}$ & reset potential & 0 & 0 \\
$\tau_m$, $\tau_s$, $\tau_{\rm ref}$ & membrane, synaptic, refractory & 20, 5, 5 ms & same \\
$\delta_d$, $dt$ & delay, integration step & 0.1, 0.1 ms & same \\
$r_{\rm ext}$ & afferent rate (constant drive) & 6.532 Hz & matched \\
$N_{\rm ext}$, $J_{E0}$, $J_{I0}$ & afferents, external weights & 640, 0.0581, 0.0518 & same \\
\end{tabular}
\end{ruledtabular}
\end{table}
\subsection{Observables}
All measures are taken on the excitatory population after a warm-up of $1$~s.
\emph{Timescale.} Spike trains are binned at $\Delta = 20$~ms and the autocorrelation $A[n]$ of
the counts is computed up to a lag of $5$~s. The timescale $\tau_{1/2}$ is the absolute lag at
which $A$ first falls below $A[k_0]/2$, linearly interpolated between bins. The start lag
$k_0 = 2$ bins skips the trivial peak at lag zero, which a spike count carries but a continuous
rate does not, and the refractory dip that follows it; for an exponential decay of constant
$\tau_c$ this definition returns $k_0\Delta + \tau_c \ln 2$.
\emph{Burst and silence.} An interspike interval longer than $100$~ms is counted as a silence and
a maximal run of shorter intervals as one burst, which gives, per neuron, the number of spikes per
burst and the mean silence duration. These are filter-free counterparts of the autocorrelation
timescale, measured directly on the spikes.
\emph{Structural groups.} The neurons of each network are split into eight equally populated
groups of their own structural variable, with the unclustered background of M-clust excluded, and
each group is located at its structural variable divided by the mean over that network's
population. Two conventions are used, for two different comparisons. Where the two architectures
are compared, the structural variable is the expected relative in-degree $\kappa_i = k_i/K$ for
M-deg and the assembly size for M-clust, and a group is given the timescale read off its mean
autocorrelation; this is necessary because most neurons of the clustered network fire too little
for an individual autocorrelation to resolve at the start lag. Where only M-deg is considered, the
structural variable is instead the \emph{realized} in-degree $\sum_j A_{ij}$, and a group is given
the median of its neurons' own $\tau_{1/2}$. The realized degree is the only degree heterogeneity
an Erd\H{o}s--R\'enyi network has, with $\mathrm{CV}(k) = \sqrt{(1-p_{EE})/K} = 0.035$, so it is
what puts the two degree distributions on one axis.
\subsection{Protocol}
Each configuration is five independent networks, each a different connectivity draw, simulated for
four trials of $500$~s with independent initial conditions; the first second of every trial is
discarded. Trials whose mean excitatory rate differs by more than $20\%$ from the median trial of
their network are dropped from every measure, which isolates the low-rate state of a network that
under a constant drive can also settle into a second, higher one. M-deg is run at
$c_\kappa \in \{0, 0.10\}$, the Erd\H{o}s--R\'enyi baseline and the lognormal degree
distribution, across $\gamma \in \{0, 0.25, 0.5, 0.7, 0.8, 0.9, 0.95, 1\}$, and M-clust once, at
$\gamma = 0$. A network is excluded when its excitatory population saturates, that is when a
finite fraction of its neurons never falls silent for longer than the burst-segmentation
threshold, or when the compensation of Eq.~(\ref{eq:spk_comp}) cannot be made self-consistent.
Summary quantities are quoted as the mean over networks with the standard error of that mean.
\begin{figure*}[p]
\centering
\includegraphics[width=0.86\linewidth]{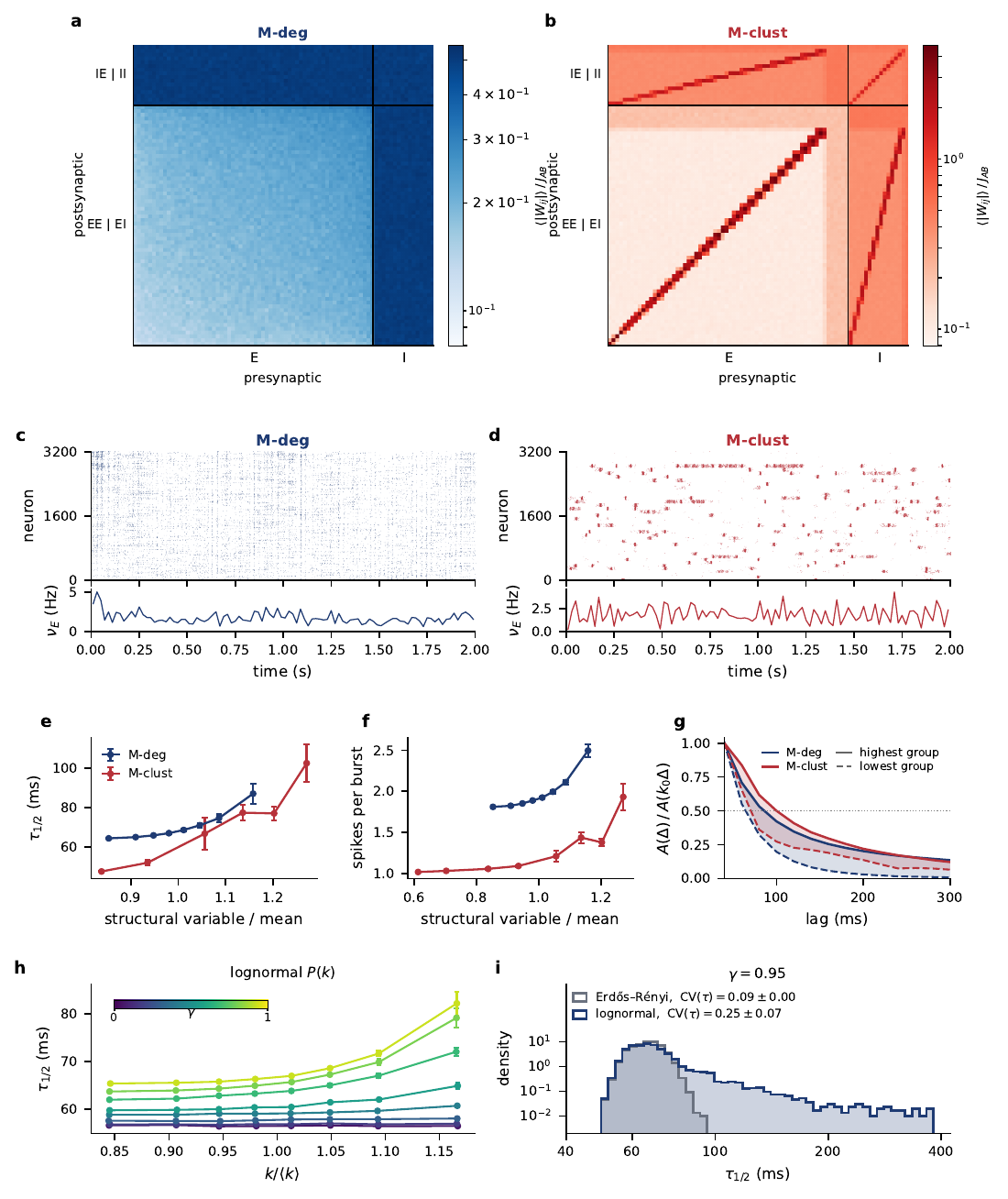}
\caption{\textbf{Degree heterogeneity against neural assemblies in a balanced spiking network.}
(a, b) The coupling matrix of each network, coarse grained into blocks of $50$ neurons and divided
block by block by the base weight $J_{AB}$ of that block, so that the four synaptic blocks share
one scale and an unstructured block reads as its own connection probability. Rows are
postsynaptic and columns presynaptic; the row labels name the two blocks of each row, written
post-pre. Excitatory neurons are ordered by realized in-degree in (a) and by assembly size in (b),
the unclustered neurons last. Every multiplier of M-deg equals one, so its only structure is the
degree gradient of the $E\!\to\!E$ wiring, $\langle W_{EE}\rangle \propto
\mathbf{k}\mathbf{k}^{\mathsf{T}}$, the other three blocks being Erd\H{o}s--R\'enyi; M-clust is
Erd\H{o}s--R\'enyi in every block and carries assembly multipliers in all four.
(c, d) Two seconds of excitatory spiking in each network, in the same neuron order, with the
population rate in $20$~ms bins below.
(e, f) Timescale $\tau_{1/2}$ and spikes per burst of the eight structural groups of each network,
against the group's structural variable divided by its network mean.
(g) Autocorrelation of the $20$~ms spike counts of the lowest (dashed) and highest (solid)
structural group of each network, normalized at the start lag $k_0\Delta = 40$~ms, with the band
between them shaded; the lowest group shown is the lowest one resolvable at the start lag in every
network. The dotted line marks the half level at which $\tau_{1/2}$ is read.
(h) Timescale of the eight octiles of the realized $E\!\to\!E$ in-degree of M-deg at
$c_\kappa = 0.10$, one curve per value of the reciprocity $\gamma$ (colour bar).
(i) Distribution of the single-neuron $\tau_{1/2}$ at $\gamma = 0.95$, pooled over networks, for
the lognormal degree distribution and for the Erd\H{o}s--R\'enyi connectivity, on logarithmic
axes; the legend gives the coefficient of variation of $\tau_{1/2}$ of each.
Panels (a--d) come from one network; elsewhere points are means over the five networks and bars
the standard error of that mean. Parameters as in Table~\ref{tab:spiking_params}, with
$c_\kappa = 0.10$ and $\gamma = 0.95$ for M-deg in (a, c, e--g).}
\label{fig:spiking}
\end{figure*}
\subsection{Results}
\emph{The mechanism in the spiking network.} Without reciprocity, the timescale of M-deg does not depend on the in-degree: at $\gamma = 0$ all octiles have the same $\tau_{1/2}$ (Fig.~\ref{fig:spiking}h), as in the rate model, where the degree alone only rescales the input (Appendix~\ref{app:num_DMFT}). As $\gamma$ increases, all neurons slow down and the timescale starts to increase with the in-degree, an effect that becomes clearly visible only at large reciprocity and is strongest for the highest-degree octiles. Degree heterogeneity is required as well: at strong reciprocity the single-neuron timescales of the lognormal network have a broad, right-skewed distribution with a long tail of slow neurons, whereas the Erd\H{o}s--R\'enyi network, whose only degree spread is the binomial one, gives a narrow distribution (Fig.~\ref{fig:spiking}i). As in the rate model (Fig.~\ref{fig:panel_timescale}b), reciprocity and degree heterogeneity are jointly necessary. \emph{Degree against assemblies.} The dependence of the timescale on the structural variable does not by itself separate the two architectures: in both networks $\tau_{1/2}$ and the number of spikes per burst increase with the structural variable, the in-degree in M-deg and the assembly size in M-clust (Fig.~\ref{fig:spiking}e--g). The two differ in which variable is slow. In M-clust it is the activity of an assembly: its members fire sparsely and become active together, and these collective switches produce large fluctuations of the population rate (Fig.~\ref{fig:spiking}d). In M-deg the activity is asynchronous and the population rate fluctuates little (Fig.~\ref{fig:spiking}c): the slow component belongs to single neurons, as expected from a degree-dependent self-coupling. Two observables could therefore separate the mechanisms in data: the structural variable that correlates with the timescale, in-degree or membership in a strongly coupled assembly, and whether the slow fluctuations of a neuron are shared with those of the neurons in its structural group. \emph{Scope.} These simulations are a proof of principle rather than a quantitative model of the data. (i) The degree dependence requires a reciprocity much stronger than the structural reciprocity measured in MICrONS (Sec.~\ref{sec:microns}). (ii) All $E\!\to\!E$ synapses carry the same weight, so the reciprocity is purely structural and there is no synaptic contribution, $\gamma_{\rm eff} = \gamma$ in the notation of Sec.~\ref{sec:model}; combining the measured structural reciprocity with correlated synaptic strengths, as in the rate model, is left to future work. (iii) Under Dale's law a spread of in-degrees produces a spread of mean inputs, which we remove with the compensating current of Eq.~(\ref{eq:spk_comp}); this isolates the effect of reciprocity, but the compensation is a modelling device rather than a biological mechanism. Finally, the two networks are compared at equal mean rate but at very different couplings (Table~\ref{tab:spiking_params}).

\section{MICrONS DATASET ANALYSIS}
\label{app:microns}

Here we describe the preprocessing and analysis of the MICrONS Cubic Millimeter dataset
\cite{microns2025functional}, which provides large-scale structural connectivity together
with calcium imaging recordings from mouse visual cortex.

\subsection{Dataset selection and preprocessing}

We used public data release \#1412 of the MICrONS dataset. Neurons were selected based on
functional--structural coregistration using the table
coregistration\_manual\_v3, which identifies neurons with reliable matching
between calcium imaging traces and reconstructed neuronal morphologies.
This procedure yielded $15\,434$ matched neurons and a total of $19\,181$ calcium traces,
as some neurons were recorded across multiple imaging sessions.
We excluded six traces that lacked a valid \texttt{root\_id}, which uniquely identifies
each neuron in the structural reconstruction.

To ensure high confidence in the structural connectivity, all subsequent analyses were
restricted to proofread neurons, for which synaptic connections were manually validated
by the MICrONS consortium.

\subsection{Structural connectivity reconstruction}

Structural connectivity was reconstructed starting from the \texttt{root\_id} identifiers
of the matched neurons. For each neuron, we identified all outgoing synapses and the
corresponding postsynaptic targets, following the official MICrONS analysis pipeline
\cite{microns2025functional}. Autapses were excluded from the analysis.

Synaptic contacts were aggregated into a directed adjacency matrix
$A \in \{0,1\}^{N \times N}$, where
\[
A_{ij} = 1 \quad \text{if neuron } j \text{ is presynaptic to neuron } i,
\]
and $A_{ij} = 0$ otherwise.
Multiple synapses between the same ordered pair of neurons were binarized, while their
counts were retained separately to analyze reciprocal connectivity statistics.
Self-connections were removed by setting $A_{ii} = 0$.

\subsection{In-degree distribution and heterogeneity}

The in-degree of each neuron was computed as
\[
k_i = \sum_j A_{ij}.
\]
The resulting in-degree distribution exhibited strong heterogeneity, consistent with a
heavy-tailed form.
We modeled the in-degree distribution using a lognormal distribution,
\begin{equation}
p(k \mid \mu, \sigma) =
\frac{1}{k \sigma \sqrt{2\pi}}
\exp\!\left(
-\frac{(\ln k - \mu)^2}{2\sigma^2}
\right),
\qquad k > 0 ,
\end{equation}
where $\mu$ and $\sigma$ denote the mean and standard deviation of $\ln k$.

To mitigate biases introduced by truncated connectivity near dataset boundaries, we
fitted a truncated lognormal distribution for $k \ge k_{\min}$,
\begin{equation}
p(k \mid \mu, \sigma, k_{\min}) =
\frac{p(k \mid \mu, \sigma)}
{1 - \Phi\!\left( \frac{\ln k_{\min} - \mu}{\sigma} \right)},
\end{equation}
where $\Phi$ is the cumulative distribution function of the standard normal distribution.
Parameters $(\mu,\sigma)$ were estimated by maximum likelihood with the location parameter
fixed to zero.
The cutoff $k_{\min}$ was selected by minimizing the Kolmogorov--Smirnov distance between
the empirical and fitted cumulative distributions.
The fitted parameters were $\mu_{\text{exp}} \approx 3.83$ and
$\sigma_{\text{exp}} \approx 0.69$.

\subsection{Reciprocal connectivity and partial symmetry}

To quantify the degree of reciprocity in the structural network, we computed the global
connection probability
\[
p = \frac{\sum_{i \neq j} A_{ij}}{N(N-1)},
\]
and the fraction of reciprocally connected neuron pairs,
\[
p_{\mathrm{recip}} =
\frac{\sum_{i<j} A_{ij} A_{ji}}{N(N-1)/2}.
\]
Reciprocity beyond chance was quantified using the correlation parameter
\[
r = \frac{p_{\mathrm{recip}} - p^2}{p(1-p)},
\]
which measures the excess of bidirectional connections relative to a null model of
independent connections.
Statistical significance was assessed by generating random networks with identical
connection probability $p$ and comparing their reciprocal fractions to the empirical
value.

\subsection{Functional data analysis}

Functional activity was obtained from calcium imaging recordings matched to the structural
neurons via the \texttt{pt\_root\_id} identifier.
We restricted the analysis to spontaneous (resting-state) activity, excluding periods of
sensory stimulation.
Frame rates varied across recording sessions and were accounted for explicitly when
computing temporal statistics.

For each neuron, we computed the autocorrelation function of its calcium trace during
rest.
The intrinsic timescale was defined as the half-width at half-maximum (HWHM) of the
autocorrelation function.
Although calcium dynamics introduce additional temporal filtering, this measure provides
a robust proxy for comparing relative timescales across neurons.

Finally, we quantified the relationship between structural and dynamical properties by
computing the Spearman rank correlation between neuronal in-degree and intrinsic
timescale. We found a statistically significant positive correlation,
$\rho = 0.164$ with $p = 10^{-4}$, indicating that neurons with higher in-degree
tend to exhibit longer intrinsic timescales.

\begin{figure}[h]
    \centering
    \includegraphics[width=0.9\linewidth]{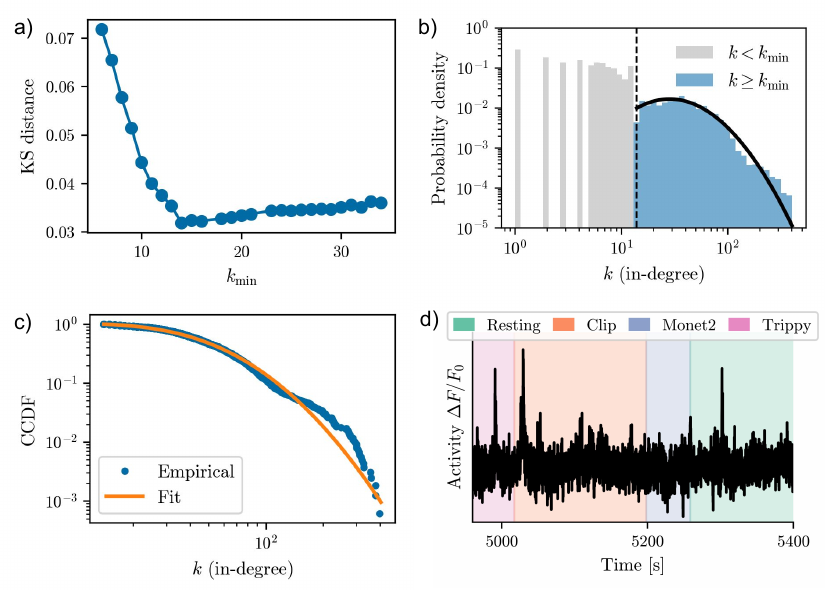}
    \caption{
\textbf{Structural and functional properties of the MICrONS dataset~\cite{microns2025functional}.}
(a) Kolmogorov--Smirnov (KS) distance between the empirical in-degree distribution and the fitted truncated lognormal model as a function of the lower cutoff degree $k_{\min}$. The minimum KS distance is obtained at $k_{\min}=14$.
(b) In-degree distribution after applying the cutoff $k_{\min}=14$, together with the fitted lognormal model. The fitted parameters are $\mu = 3.83$ and $\sigma = 0.69$.
(c) Comparison between the empirical complementary cumulative distribution function (CCDF) of the in-degree and the corresponding lognormal fit.
(d) Example calcium activity trace from a representative neuron in the MICrONS Cubic Millimeter dataset, shown as normalized fluorescence $\Delta F/F_0$. Colored shaded intervals denote different task epochs corresponding to distinct visual stimuli (Clip, Monet2, Trippy). The final green-shaded interval corresponds to spontaneous activity recorded at the end of the scan, with the monitor turned off, which is the condition analyzed in this work.
}
    \label{fig:microns_supplementary}
\end{figure}

\subsection{In-degree, out-degree and timescales}
\label{app:microns_inout}
\begin{figure}[h]
    \centering
    \includegraphics[width=0.7\linewidth]{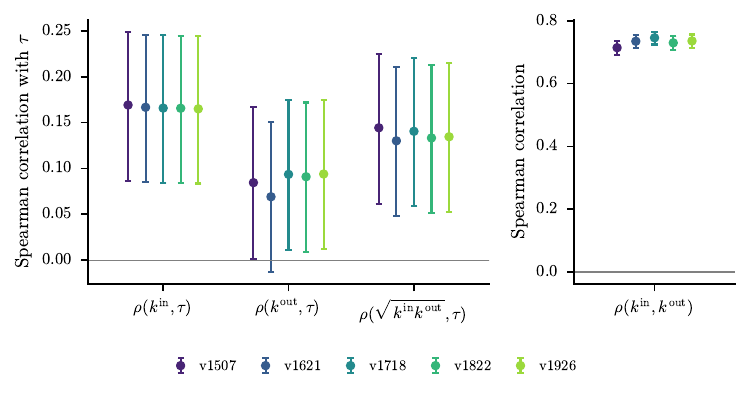}
    \caption{\textbf{In-degree, out-degree and timescales across MICrONS releases.} Left: Spearman correlation between the intrinsic timescale $\tau$ and the in-degree, the out-degree and the geometric degree $\sqrt{k^{\rm in}k^{\rm out}}$. Right: Spearman correlation between in- and out-degree. For each quantity, one point per data release (colours, from v1507 to v1926), with error bars giving the $95\%$ confidence interval. Values are reported in Table~\ref{tab:microns_inout}.}
    \label{fig:microns_inout}
\end{figure}

In the model each neuron carries a single degree, which sets both its in- and its out-degree. In the reconstructed connectome the two can differ, and we therefore repeated the degree--timescale analysis with the out-degree and with the geometric degree, and checked how strongly the two are related.
Degrees were counted within the network of proofread neurons, with autapses removed: the in-degree of neuron $i$ is $k^{\rm in}_i=\sum_j A_{ij}$, its out-degree $k^{\rm out}_i=\sum_j A_{ji}$, and its geometric degree $\sqrt{k^{\rm in}_i k^{\rm out}_i}$. Timescales were computed as described above, from the resting-state calcium traces recorded at $6.3$~fps, using one point per recording and retaining recordings whose HWHM exceeds one frame. For each degree we computed the Spearman rank correlation with the timescale, and we computed the Spearman correlation between in- and out-degree over all proofread neurons. Confidence intervals at $95\%$ were obtained from the Fisher transformation with the Bonett--Wright variance $(1+\rho^2/2)/(n-3)$.
To assess the robustness of the results to the ongoing proofreading of the dataset, the analysis was repeated on five successive public releases of the MICrONS connectome, from v1507 to v1926, which include between $2141$ and $2304$ proofread neurons and between $558$ and $572$ recordings (Table~\ref{tab:microns_inout} and Fig.~\ref{fig:microns_inout}). The analysis in the main text (Figs.\ref{fig:microns} and \ref{fig:microns_matched}) is based on the earlier release v1412, which contains 1965 proofread neurons and gives $\rho(k^{\mathrm{in}},\tau)=0.164$ ($p\approx10^{-4}$).
The correlation between in-degree and timescale is the same in all later releases and matches the v1412 value, $\rho(k^{\mathrm{in}},\tau)=0.165$--$0.169$ ($p<10^{-4}$). In- and out-degree are strongly correlated, $\rho(k^{\rm in},k^{\rm out})=0.71$--$0.75$. The out-degree correlates more weakly with the timescale, $\rho(k^{\rm out},\tau)=0.07$--$0.09$, significant at $p<0.05$ in four of the five releases, and the geometric degree takes intermediate values, $\rho(\sqrt{k^{\rm in}k^{\rm out}},\tau)=0.13$--$0.14$.
The two degrees are not measured with the same accuracy. The in-degree of a neuron counts the proofread axons that contact its dendrites, which are compact and largely contained in the volume. The out-degree instead depends on how much of the neuron's own axon is reconstructed: axons extend well beyond the volume and are proofread to different extents across neurons. Consistently, the out-degree distribution is broader than the in-degree distribution, and its width decreases across successive releases as proofreading progresses, whereas that of the in-degree is stable. A noisier measurement of the out-degree, combined with its strong correlation with the in-degree, is expected to yield a weaker correlation with the timescale. These data do not, however, allow us to separate the contributions of in- and out-degree, and we therefore report the in-degree, which is the more reliably measured quantity, in the main text.
\begin{table}[H]
\centering
\caption{Spearman correlations between degree and intrinsic timescale, and between in- and out-degree, for five public releases of the MICrONS connectome. Each entry gives the correlation and its $95\%$ confidence interval; $^{*}p<0.05$, $^{**}p<0.01$, $^{***}p<0.001$. Correlations with $\tau$ are computed over recordings ($n=558$ for v1507 and $n=572$ for the other releases), the in--out correlation over proofread neurons ($n=2141$, $2273$, $2275$, $2293$ and $2304$).}
\label{tab:microns_inout}
\resizebox{\linewidth}{!}{%
\begin{tabular}{lccccc}
\hline\hline
 & v1507 & v1621 & v1718 & v1822 & v1926 \\
\hline
$\rho(k^{\rm in},\tau)$ & 0.169$^{***}$ [0.087, 0.249] & 0.167$^{***}$ [0.085, 0.246] & 0.166$^{***}$ [0.084, 0.245] & 0.166$^{***}$ [0.084, 0.245] & 0.165$^{***}$ [0.084, 0.244] \\
$\rho(k^{\rm out},\tau)$ & 0.085$^{*}$ [0.001, 0.167] & 0.069 [$-$0.013, 0.150] & 0.094$^{*}$ [0.011, 0.174] & 0.091$^{*}$ [0.009, 0.172] & 0.094$^{*}$ [0.012, 0.175] \\
$\rho(\sqrt{k^{\rm in}k^{\rm out}},\tau)$ & 0.144$^{***}$ [0.062, 0.225] & 0.130$^{**}$ [0.048, 0.210] & 0.140$^{***}$ [0.059, 0.220] & 0.133$^{**}$ [0.051, 0.213] & 0.135$^{**}$ [0.053, 0.215] \\
$\rho(k^{\rm in},k^{\rm out})$ & 0.714$^{***}$ [0.690, 0.737] & 0.735$^{***}$ [0.713, 0.755] & 0.746$^{***}$ [0.724, 0.766] & 0.730$^{***}$ [0.708, 0.751] & 0.736$^{***}$ [0.714, 0.756] \\
\hline\hline
\end{tabular}%
}
\end{table}

\subsection{Cluster size and timescale}
\label{app:dcsbm}
\begin{figure}[h]
\centering
\includegraphics[width=0.5\linewidth]{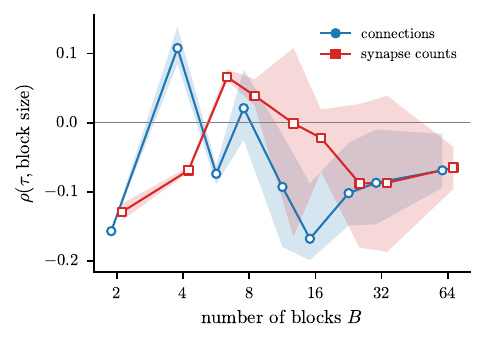}
\caption{Spearman correlation between neuronal timescale and block size against the number of blocks $B$ of a
degree-corrected stochastic block model fitted to the binary connectivity (blue) and to synapse counts (red).
Lines: mean over five fits; shaded areas: range across fits. Open markers: median permutation $p\ge0.05$.}
\label{fig:dcsbm}
\end{figure}

In networks with clustered connectivity, neuronal timescales grow with the size of the cluster a neuron belongs
to~\cite{stern2023reservoir}. To test this in the MICrONS data, we partitioned the proofread network v1926 into $B$ blocks with a
degree-corrected stochastic block model (DC-SBM)~\cite{karrer2011stochastic}, which groups neurons with similar connectivity
while accounting for degree heterogeneity (connections treated as undirected). Since in clustered networks
assemblies arise from stronger coupling, we fitted the model both to the binary connectivity and to the number of
synapses between neurons. As $B$ is a free parameter, we varied it from 2 to 64, with five random initializations
each. For all $B$, both networks show block structure far beyond that of degree-preserving random networks
($z>100$). For each partition, we computed the Spearman correlation between the timescale of the 490 recorded
neurons and the size of their block, assessing significance by permuting sizes among blocks. The correlation was
weak, changed sign across $B$ ($|\rho|\le0.2$), and was significant in only 3 of 90 fits, consistent with chance
(Fig.~\ref{fig:dcsbm}). Thus, the size of structural clusters does not predict neuronal timescales in this network.

\section{DETAILS ON SIMULATIONS COMPARED WITH DATA}\label{app:sim_details}
Here we detail the simulations of Fig.~\ref{fig:microns_matched} and test the robustness of the results with respect to the two free parameters of the model, the synaptic gain $g$ and the reciprocity $\gamma$, and with respect to directed connectivity.
In all simulations the in-degrees are drawn from the lognormal distribution fitted to the MICrONS connectome, $\mu=3.83$, $\sigma=0.69$ (Fig.~\ref{fig:microns}c), with $N=4000$ neurons; the adjacency matrix is generated from the configuration model, $p_{ij}=k_ik_j/(NK)$, and the weights are Gaussian with variance $g^2/K$ and reciprocal correlation $\gamma$, as in Appendix~\ref{app:DMFT}. Integration parameters, stationary window, and number of lags used to estimate the autocorrelation functions are identical to those of Fig.~\ref{fig:microns_matched}; the timescale $\tau_i$ of each unit is the half-width at half maximum of its activity autocorrelation. Each parameter set is averaged over $10$ network realizations.
We characterize each simulation by the two observables compared with the data: the coefficient of variation of the timescale distribution, $\mathrm{CV}(\tau)$, and the Spearman correlation $\rho(k,\tau)$ between in-degree and timescale.
\subsection{Sweep in $g$ and $\gamma$}\label{app:sim_sweep}
\begin{figure}[h]
\includegraphics[width=0.9\linewidth]{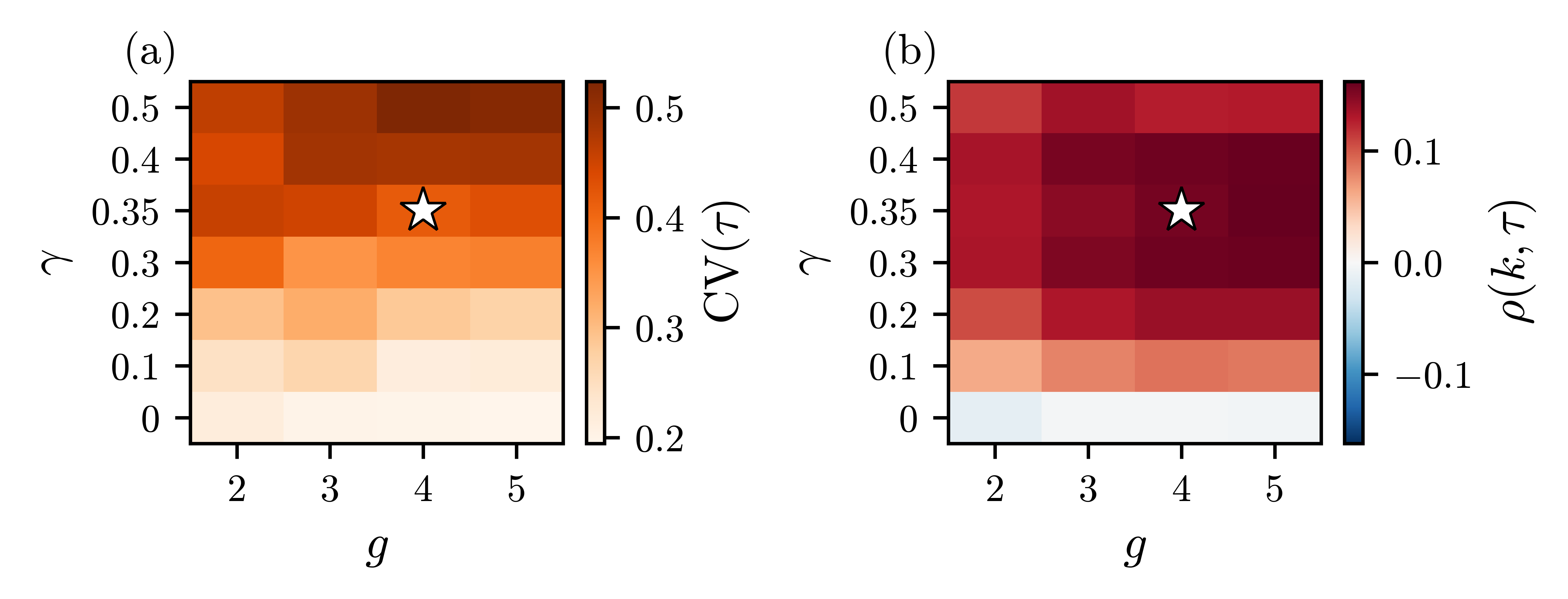}
\caption{Coefficient of variation of the timescales (a) and Spearman correlation between in-degree and timescale (b) as a function of the gain $g$ and the reciprocity $\gamma$, for the MICrONS-matched lognormal degree distribution ($\mu=3.83$, $\sigma=0.69$). The star marks the parameters used in Fig.~\ref{fig:microns_matched}. Each point is an average over $10$ realizations with $N=4000$.}
\label{fig:sweep_heatmaps}
\end{figure}

\begin{figure}[h]
\includegraphics[width=0.9\linewidth]{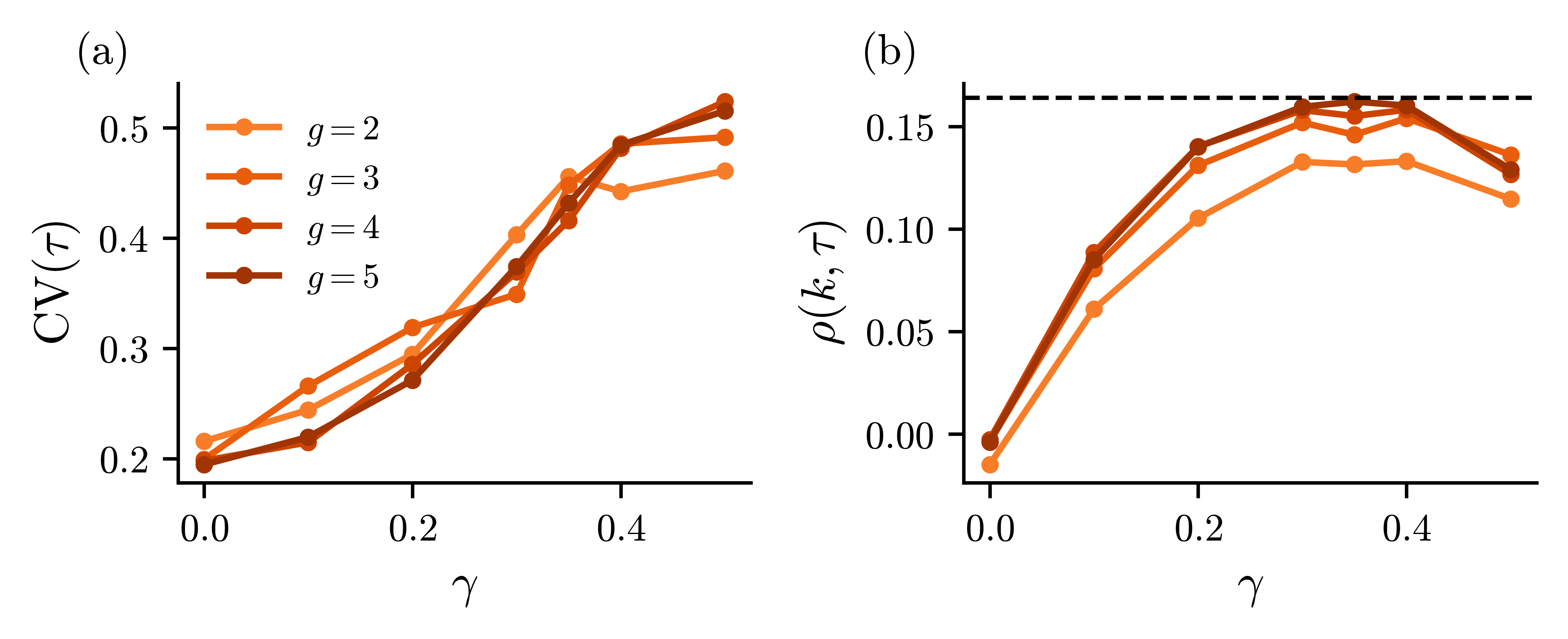}
\caption{Same quantities as in Fig.~\ref{fig:sweep_heatmaps}, plotted as a function of $\gamma$ for each value of $g$. (a) $\mathrm{CV}(\tau)$; (b) $\rho(k,\tau)$, with the value measured in the MICrONS data, $\rho_{\mathrm{data}}=0.164$, indicated by the dashed line.}
\label{fig:sweep_profiles}
\end{figure}

We scan $g\in[2,5]$ and $\gamma\in[0,0.5]$. Figure~\ref{fig:sweep_heatmaps} shows that, once the network is deep in the chaotic regime ($g\gg g_c$), both $\mathrm{CV}(\tau)$ and $\rho(k,\tau)$ vary weakly with $g$ and are controlled mainly by $\gamma$.
Figure~\ref{fig:sweep_profiles} shows the same data as profiles in $\gamma$ at fixed $g$. In the chaotic regime the curves collapse, and the value of $\rho$ measured in the data, $\rho_{\mathrm{data}}=0.164$, is reproduced over a broad range of gains for $\gamma\simeq 0.3\text{--}0.4$, showing that the comparison with the data does not rely on a fine tuning of $g$.
\subsection{Directed networks at fixed effective reciprocity}\label{app:sim_directed}
\begin{figure}[h]
\includegraphics[width=0.9\linewidth]{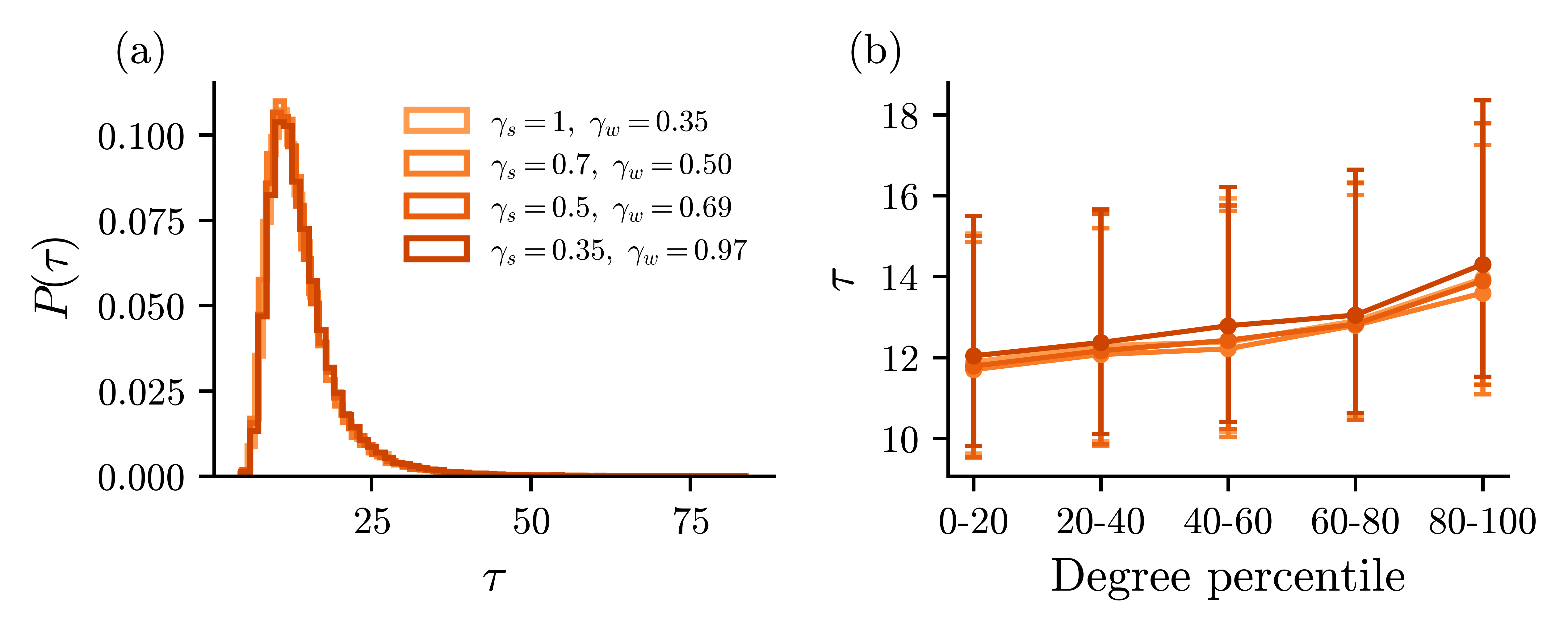}
\caption{Directed networks with different structural reciprocity $\gamma_{\rm s}$ and
synaptic correlation $\gamma_{\rm w}$ at fixed effective reciprocity
$\gamma_{\rm eff}=\gamma_{\rm w}[\gamma_{\rm s}(1-p)+p]=0.35$ and $g=4$.
(a) Distribution of the single-neuron timescales; (b) median timescale as a function of the
in-degree percentile, with bars indicating the interquartile range. Curves for different
$(\gamma_{\rm s},\gamma_{\rm w})$ collapse.}
\label{fig:isoproduct}
\end{figure}

Finally we check that the results depend on the structural reciprocity $\gamma_{\rm s}$ and
on the synaptic correlation $\gamma_{\rm w}$ only through the combination $\gamma_{\rm eff}=\gamma_w[\gamma_s(1-p)+p]$, which reduces to $\gamma_s\gamma_w$ for $p\ll1$ (Sec.~\ref{sec:model}). Since at
leading order in $1/K$ the effective dynamics depends on the couplings only through their second
moments (Appendix~\ref{app:DMFT}), this is the combination expected to control the dynamics.
We generate directed networks in which each neuron keeps a single expected degree $k_i$, drawn from the same distribution as above ($K=58$, $p=0.014$), which sets both its expected in- and out-degree; the two directed edges of each pair are then drawn separately, with correlation $\gamma_s$.
The weights have correlation $\gamma_{\rm w}$, chosen so that
$\gamma_{\rm eff}=0.35$ for each $\gamma_{\rm s}$, namely
$(\gamma_{\rm s},\gamma_{\rm w}) = (1,0.35)$, $(0.7,0.50)$, $(0.5,0.69)$ and $(0.35,0.97)$,
at $g=4$. The correlation between reciprocal weights measured in the generated matrices is
$0.351$, $0.354$, $0.361$ and $0.365$, in agreement with the predicted value $0.350$.
Figure~\ref{fig:isoproduct} shows that the timescale distributions and the dependence of
$\tau$ on the in-degree are essentially unchanged across the four networks:
$\mathrm{CV}(\tau)=0.419\pm0.009$ and $\rho(k,\tau)=0.164\pm0.006$, with a median timescale
varying between $12.5$ and $12.9$. The largest Kolmogorov--Smirnov distance between any pair
of timescale distributions is $D=0.044$. By contrast, a symmetric topology with uncorrelated
weights ($\gamma_{\rm s}=1$, $\gamma_{\rm w}=0$, hence $\gamma_{\rm eff}=0$) gives
$\mathrm{CV}(\tau)=0.194$, $\rho=0.000$ and a median timescale of $3.5$: structural
reciprocity alone produces no slowing down and no degree--timescale correlation. The
undirected model with a single parameter $\gamma$ used in the main text therefore captures
the directed case with the identification $\gamma\to\gamma_{\rm eff}$.

\end{document}